\documentclass[prb,showkeys,showpacs,footnoteinbib,twocolumn,
unsortedaddress]{revtex4}
\usepackage{amsmath,amssymb,epic,eepic}
\usepackage{siunitx}
\usepackage[utf8]{inputenc}
\usepackage{graphicx}
\usepackage{todonotes}
\usepackage{tikz}
\usepackage{circuitikz}
\usetikzlibrary{%
	shapes.geometric,calc,intersections,%
	decorations.markings,decorations.text,decorations,%
	patterns,decorations.pathmorphing,fpu,%
	positioning,automata,shapes.multipart,circuits,%
	circuits.ee, circuits.ee.IEC
}

\newcommand{\normord}[1]{\vcentcolon\mathrel{#1}\vcentcolon}
\providecommand{\vcentcolon}{\mathrel{\mathop{:}}}
\newcommand{\mytodo}[1]%
{{\todo[inline,backgroundcolor=blue!10!white]{#1}
}}
\newcommand{\me}{\mathrm{e}}
\newcommand{\mi}{\mathrm{i}}
\newcommand{\md}{\mathrm{d}}
\DeclareMathOperator{\sinc}{sinc}

\begin{document}

\title{Taming electronic decoherence in 1D chiral ballistic quantum conductors}

\author{C. Cabart$^1$} 
\author{B. Roussel$^1$}
\author{G. F\`eve$^2$}
\author{P. Degiovanni$^1$}

\affiliation{(1) Univ Lyon, Ens de Lyon, Université Claude Bernard 
Lyon 1, CNRS, 
Laboratoire de Physique, F-69342 Lyon, France}

\affiliation{(2) Laboratoire Pierre Aigrain, Ecole normale supérieure,
PSL University, Sorbonne Université, Université Paris Diderot,
Sorbonne Paris Cité, CNRS, 24 rue Lhomond, 75005 Paris France.}

\begin{abstract}
Although interesting per se, decoherence and relaxation of single-electron
excitations induced 
by strong effective screened Coulomb
interactions in Quantum Hall edge channels 
are an important challenge for the applications of electron quantum optics
in quantum information and quantum sensing. In this paper, we study
intrinsic single-electron decoherence within an ideal single-electron channel with
long-range effective Coulomb interactions to determine the influence of the material and sample
properties. We find that weak-coupling materials characterized by a
high velocity of hot-electron excitations may offer interesting
perspectives for limiting intrinsic decoherence due to
electron/electron interactions. We discuss quantitively how extrinsic
decoherence due to the coupling with the channel's electromagnetic
environment can be efficiently inhibited 
in specially designed samples at $\nu=2$ with one closed
edge channel and we propose a realistic geometry for testing decoherence
control in 
an Hong Ou Mandel experiment.
\end{abstract}



\maketitle


\section{Introduction}

Over the last decade, a considerable effort has been devoted 
to the development
of quantum coherent nanoelectronics with the aim of controlling
electronic quantum transport down to the single particle 
level\cite{Bauerle:2018-1,Splettstoesser:2017-1}.
This has led to
the development of electron quantum optics\cite{Bocquillon:2014-1}, 
an emerging field which aims at manipulating electrons in a ballistic
quantum conductor 
just as photons in quantum optical setups. This perspective 
had initially risen strong hopes for on-chip quantum information 
processing using single electrons as quantum information
carriers\cite{Bertoni:2000-1,Ionicioiu:2001-1,Bertoni:2007-1}. 

However, electron
quantum optics differs from quantum optics because electrons, being
charged, interact via effective screened Coulomb interactions.
This leads to electronic decoherence and
relaxation\cite{Roussel:2016-2}. These effects are strong enough
to destroy the electronic quasi-particle 
in the $\nu=2$ quantum Hall edge channel system, a
fact first evidenced by non-equilibrium distribution 
relaxation studies\cite{LeSueur:2010-1} and later confirmed
by recent studies of 
single-electron decoherence through
Hong Ou Mandel (HOM) experiments\cite{Marguerite:2016-1,Freulon:2015-1}. Recent Mach-Zehnder
interferometry (MZI) experiments\cite{Tewari:2016-1} have also confirmed the
plausibility of this
scenario although the most commonly used model based on effective
screened short-range interactions\cite{Levkivskyi:2008-1} fails to
reproduce the observed saturation of the decoherence
scenario\cite{Slobodeniuk:2016-1}. These recent results
suggest that our understanding
of quantitative models of electronic decoherence still needs to be
sharpened. 

On the other hand, using single-electron excitation as carriers of
quantum information requires a high degree of control 
from their generation to their detection, and of course during their
propagation. 
Several single-electron sources have been developed over the
years, from the mesoscopic capacitor \cite{Feve:2007-1} to 
single-electron
pumps \cite{Hohls:2012-1,Waldie:2015-1} and more recently the Leviton 
source
\cite{Dubois:2013-1}. Other systems aims at injecting electrons at very 
high energies \cite{Fletcher:2013-1} using dynamically driven dots 
or at transporting them using surface acoustic waves
\cite{Hermelin:2011-1}. The maturation of technology may lead to
the development of controlled sources able to emit specifically 
tailored electronic wavepackets 
\cite{Ott:2014-1,Kashcheyevs:2017-1,Misiorny:2018-1}.

On the detection side, a full quantum
current analyzer has been developed to extract the single-electron wave
functions present within a time-periodic electric
current\cite{Marguerite:2017-1}. Dynamical quantum dots are envisioned
to  probe single-electron coherence in a time-dependent and 
energy-selective
way\cite{Waldie:2015-1,Johnson:2017-1}.
But controlling the dynamics of propagating single
to few electron excitations is still a challenge.

Understanding single to few electron decoherence is therefore 
crucial both for our understanding of electronic quantum transport and
for the most promising applications of
electron quantum optics such as quantum information processing and the 
quantum
metrology of charge and electric currents. 
It is thus time to ask to what extent electronic decoherence can be
tamed in experimentally relevant systems. 

In this paper, we address this question within our recently developed
non-perturbative framework for studying single-electron decoherence in a
chiral 1D conductor\cite{Degio:2009-1,Ferraro:2014-1}. 
More precisely, we will discuss the influence of the material properties
(intrinsic and induced by its fabrication and 
gating)
by considering single-electron
decoherence induced by effective screened Coulomb interactions 
within an ideal dissipationless single chiral edge channel. Our study
suggests that materials such as exfolliated graphene and AsGa 
respectively
correspond to weak and strong coupling materials, the former being more
favorable for preserving electronic decoherence than the latter. Beyond
the specific example, we think that this shows 
the importance of 
investigating electron quantum optics in various
materials.

We then apply
our approach to the question of passive decoherence control, that is 
through sample
design. We present an in depth discussion of various geometries which 
have been used
in recent experiments\cite{Altimiras:2010-2,Huynh:2012-1}. Our results
suggest that an efficient control of single-electron decoherence  
could be achieved in realistic samples based on edge channels of an 
AsGa 2D
electron gas in the integer quantum Hall regime at $\nu=2$. A new
sample design is proposed for testing our approach in a HOM
interferometer. Let us stress that our work also points out to the
possibility of discriminating among various models of effective screened
electronic Coulomb interactions using HOM interferometry experiments.
As a bonus, we will see that such devices offer
interesting perspective for single edge magnetoplasmon generation, thus 
connecting
electron quantum optics to quantum plasmonics and microwave quantum
optics.

This paper is structured as follows: in Sec.~\ref{sec:review}, we 
briefly review the basic concepts of electron quantum optics and the 
physics of single-electron decoherence in quantum Hall edge channels. 
Then, analytical models of
screened Coulomb interactions for the physical situations 
relevant for
the present paper will be introduced and the corresponding 
edge-magnetoplasmon scattering 
will be discussed. 
Section~\ref{sec:decoherence} is devoted to electronic decoherence.
Decoherence at filling fraction $\nu=1$ in the dissipationless case will
enable us to discuss the influence of the material. We will also
discuss to what extent an HOM experiment could help discriminate between
short and long-range effective interactions in the $\nu=2$ system.
Finally, section~\ref{sec:decoherence-control} is devoted to 
decoherence control for single-electron excitations by sample design. 

\section{Electron quantum optics and finite-frequency quantum transport}
\label{sec:review}

\subsection{Electron quantum optics}
\label{sec:review:eqo}

The key concepts of electron quantum optics are electronic coherences
defined by analogy with photon coherences introduced by Glauber for
photons\cite{Glauber:1963-1}. The first order electronic
coherence at position $x$\cite{Degio:2010-4,Haack:2011-1,Haack:2012-2}
$\mathcal{G}^{(e)}_{\rho,x}(t|t')
=\mathrm{Tr}(\psi^\dagger(x,t)\rho\psi(x,t))$, where $\psi$ is the 
electronic annihilation operator,  
 contains all information on the single-electron 
wavefunctions that can be extracted from the system at position $x$. 
To simplify notation, because our detection setup is at a fixed 
position $x$, we will 
drop it from all equations in the following.
Electronic coherence is most conveniently visualized using a real 
valued 
time/frequency representation called the electronic Wigner function, 
defined as\cite{Ferraro:2013-1}:
\begin{equation}
\mathcal{W}_{\rho}^{(e)}(t,\omega)=
\int v_F
	\mathcal{G}_{\rho}^{(e)}
	\left(t+\frac{\tau}{2},t-\frac{\tau}{2}\right)
	\,\me^{\mi\omega\tau}
\md\tau
\end{equation}
The electronic Wigner function is directly related to physically
relevant quantities: first of all, integrating over $\omega$ leads to 
the average time-dependent current and time averaging gives the 
electronic distribution function. Moreover, the low-frequency 
Hong-Ou-Mandel noise signal for two electronic sources is 
directly proportional to the overlap of the excess Wigner functions of
the two sources\cite{Ferraro:2013-1}, a fact directly exploited in 
electronic tomography
protocols\cite{Degio:2010-4,Marguerite:2017-1,Jullien:2014-1} 
and recent studies of electronic 
decoherence\cite{Freulon:2015-1,Marguerite:2016-1}.

\begin{figure}
\includegraphics{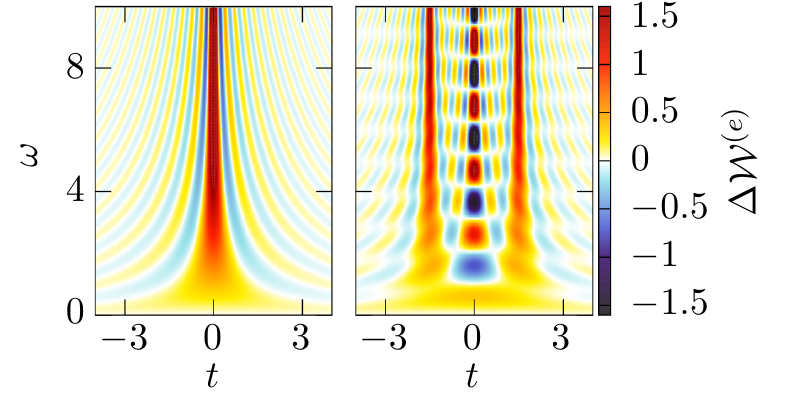}
	\caption{\label{fig:Wigner:1} (Color online) Left panel: excess 
electronic Wigner 
function for a single electronic state of the form 
$\psi^\dagger(t=0)|F\rangle$ measured at $x=0$. Right panel: excess 
electronic Wigner function for a quantum superposition
$(\psi^{\dagger}(-\tau/2)+\psi^\dagger(\tau/2))|F\rangle/\sqrt{2}$
($\tau=3$ on this specific example). The interference contribution is 
clearly visible and overlaps with each localized excitation 
contribution 
for $\omega\tau\lesssim 1$.}
\end{figure}

Within the bosonization framework briefly reviewed in Appendix 
\ref{appendix:bosonization}, 
a single-electron excitation with
wavepacket $\varphi_{\text{e}}$ above the Fermi sea
\begin{equation}
|\varphi_\text{e},F\rangle = \int_{-\infty}^{+\infty}
\varphi_{\mathrm{e}}(t)\psi^\dagger(t)|F\rangle\,\md t
\end{equation}
is a quantum superposition of coherent edge-magnetoplasmon
states. The electronic Wigner function for a perfectly localized 
electronic excitation above the Fermi sea $\psi^\dagger(0)|F\rangle$ 
is, up to normalisation, depicted on the left panel of 
Fig.~\ref{fig:Wigner:1}. As 
expected from the Heisenberg uncertainty principle, such an excitation 
is not limited in energy and, when looked at energy $\varepsilon>0$ 
above the Fermi level, the Wigner function tends to spread over a time 
scale $\hbar/\varepsilon$. 
The Wigner function of a quantum superposition of two such excitations 
at times $t_1$ and $t_2$ contains a contribution for each of the 
excitations within the superposition and an interference contribution 
located at time $(t_1+t_2)/2$ as depicted on the right panel of
Fig.~\ref{fig:Wigner:1}. 

\begin{figure*}
\includegraphics{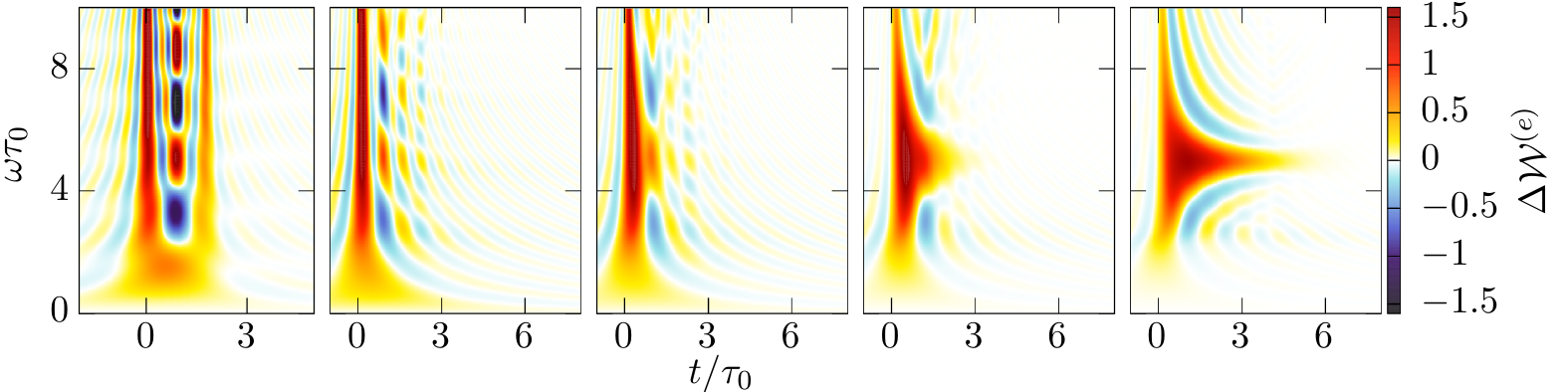}
\caption{\label{fig:Wigner:2} 
Reconstruction of the excess Wigner function for a Landau excitation
with emission energy $\hbar\omega_0$ and duration
$\tau_0$ with $\omega_0\tau_0=5$. Each
panels depicts the Wigner function associated with a finite sum
$\sum_{j=1}^N\varphi_{\mathrm{e}}(t_j)\psi^\dagger(t_j)|F\rangle$ where
the times $t_j$ are sampled randomly using the probability distribution
$|\varphi_{\mathrm{e}}(t)|^2$. From left to right, panels show the 
results corresponding to $N=2$, $N=10$, $N=25$, $N=100$ and $N=500$. 
The specific form of any wavepacket can thus be seen as arising from 
the interference pattern between its different 
time-localized contributions. 
}
\end{figure*}

When considering an arbitrary electronic wavepacket 
$\varphi_{\mathrm{e}}$, these interference contributions are
responsible for cancellations which, in the case of the Landau
excitation emitted at energy $\hbar\omega_0$ above the Fermi
level, localize the main contribution to the excess electronic
Wigner function close to $\omega_0$.
This process is depicted on Fig.~\ref{fig:Wigner:2}, in which the full
excess electronic Wigner function is reconstructed from the excess 
Wigner function of a quantum superposition of more and more localized
electronic excitations at times $t_j$, each of them weighted by the 
value of the electronic wavefunction $\varphi_{\mathrm{e}}(t_j)$. 
This specific Landau wavepacket, given by 
\cite{Degio:2010-4,Ferraro:2013-1}
\begin{equation}
\tilde{\varphi}_{\mathrm{e}}(\omega)=
\frac{\mathcal{N}_{0}\Theta(\omega)}{\omega-\omega_0-i/2\tau_0}
\end{equation}
where $\tau_0$ denotes the excitation lifetime, will be used as our 
main example through this whole text because of its experimental 
relevance for the mesoscopic capacitor in the ideal single-electron 
source regime \cite{Feve:2007-1,Mahe:2008-1,Mahe:2010-1}. Note that the
methods we have developped could also be used for making predictions for
arbitrary injected single electron wavepackets and can therefore be
combined with our recently developped quantum current
analyzis\cite{Marguerite:2017-1,Roussel:2017-1} which enables us to
characterize possible single electron emission regimes from a Floquet
modeling of the source and to extract the corresponding electronic
wavefunction.

\begin{figure}
 \includegraphics{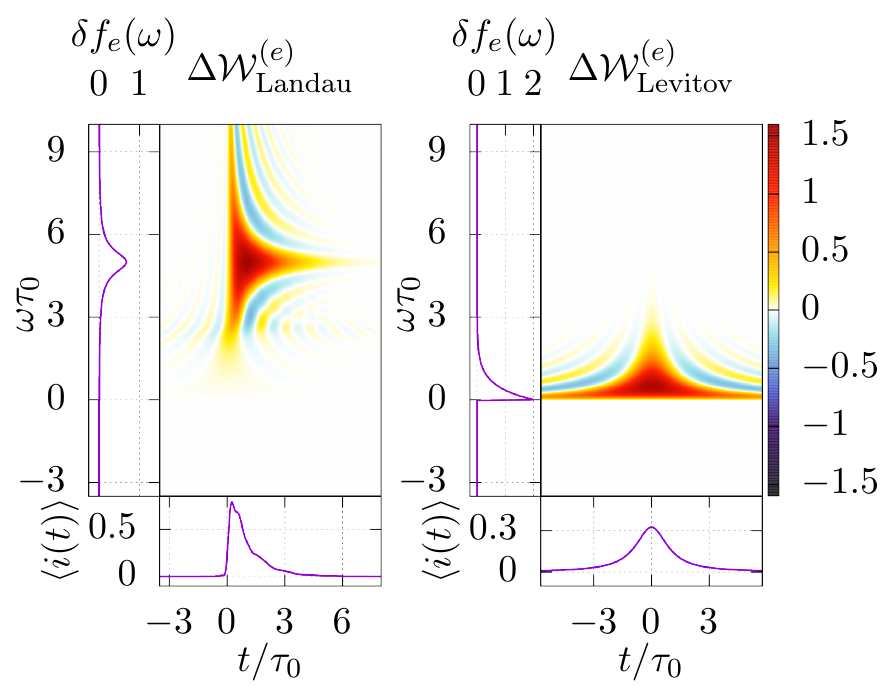}
	\caption{\label{fig:firstwigners} (Color online) Density plot of the 
Wigner function
 of a Landau excitation with parameters $\omega_0\tau_0=5$, left, and 
an $n=1$ Leviton excitation, right, as a function of $t/\tau_0$ and
 $\omega\tau_0$. Marginals are also plotted, giving 
access to the average current as a function of time (bottom of each
plot) and the excess occupation number as a function of energy (left of 
each plot). These two excitations are single-electronic and are 
respectively energy- and time-resolved, with a lorentzian profile. In 
the case of the Landau excitation, we recover the form given by the 
superposition depicted on Fig.~\ref{fig:Wigner:2}.}
\end{figure}

The other important example we use in this article is the recently
observed \cite{Dubois:2013-2} Leviton 
excitation introduced by Levitov, Lee and Lesovik
\cite{Levitov:1996-1} and
whose wavepacket is given by
\begin{equation}
 \varphi_{\mathrm{e}}(t)
	=\sqrt{\frac{\tau_0}{2\pi}}\frac{1}{t+\mi\tau_0}\,.
\end{equation}
This excitation is in fact quite different from other arbitrary 
wavepackets, as it is the only mono-electronic excitation that can be 
created by applying a carefully designed classical voltage drive to an 
ohmic contact \cite{Keeling:2006-1}. Consequently, 
a Leviton is a coherent state 
of edge magnetoplasmons, an essential feature for
understanding the effect of interactions on this 
state\cite{Grenier:2013-1}.
The Wigner functions for both 
types of single-electronic excitations used in this paper are depicted 
on Fig.~\ref{fig:firstwigners}.

\subsection{The physics of single-electron decoherence}
\label{sec:decoherence-physics}

In the original discussion of the decay of an electronic quasi-particle
by Landau\cite{Landau:1957-1}, electronic decoherence arises from 
electron/hole pair
creation by the time and space dependent electric potential generated by
the bare charged injected at a given energy above the Fermi sea. More
than 50 years later, the discovery of dynamical Coulomb blockade
\cite{Devoret:1990-1,Girvin:1990-1}
showed us that electronic relaxation could also arise from the emission 
of
photons within the electromagnetic environment of the conductor.

Our present understanding of single-electron decoherence in quantum Hall
edge channels\cite{Degio:2009-1,Ferraro:2014-2} appears as 
a combination of these two effects: (1) the 
many-body decoherence of the electronic fluid that arises from
the capacitive coupling to external degrees of freedom such as the
second edge channel in the $\nu=2$ system or charge modes of a
neighbouring circuit and (2) the generation of electron/hole pairs in
the same channel, induced by voltage fluctuations within the interacting
region. These fluctuations are due to electron/electron screened Coulomb
interactions within the edge channel as well as from the backaction of
Coulomb induced charge fluctuations from neighboring conductors.

Many-body decoherence arises from the entanglement between the
charge degrees of freedom of the edge channel under consideration and
external degrees of freedom. For example,  
at $\nu=2$, Coulomb interactions induce 
entanglement between the two edge channels. It is responsible for
the fast relaxation of Landau electronic excitations compared to
the Levitov excitations \cite{Ferraro:2014-2}. This striking difference
between these two excitations can be traced back to the fact
that, Levitov excitations being edge-magnetoplasmon coherent states,
they are pointer states\cite{Zurek:1993-1}
with respect to Coulomb interaction induced decoherence.
On the other hand, all other single-electron
excitation being quantum superpositions of such edge-magnetoplasmon
coherent states, many-body decoherence kills interferences 
between these coherent components. This leads to a suppression of 
interferences between them at the single-electron level, 
thus causing its rapid relaxation in energy. As demonstrated by
experimental decoherence studies at $\nu=2$ 
through HOM interferometry\cite{Marguerite:2016-1} as 
well as by Mach-Zehnder interferometry\cite{Tewari:2016-1},
this is the dominant cause of electronic decoherence in
these
experiments so far. 

On the contrary, when the edge channel is not coupled 
to external dynamical degrees of freedom, 
many-body decoherence is not present and single-electron
decoherence only arises from the creation of electron/hole pairs within
the electronic fluid. This purely intrinsic process can be interpreted 
as the
spreading of electronic coherence associated with 
the injected single electron into higher order correlations. We expect
it to be less stringent than excitation emission into the external 
environment due to Pauli principle induced phase space limitations. 
The decoherence scenario is thus expected to be
significantly different and more favorable to decoherence control than
when the edge channel is capacitively coupled to other conductors. 

Inspired by this idea, we will therefore study electronic decoherence 
within an ideal $\nu=1$
quantum Hall edge channel. It is
solely influenced by the intrinsic properties of the edge channel, that
is the intrinsic and substrate material properties as well as 
its gating, thus giving us new insight
on the first question motivating the present work.

Cutting off the possibility to generate excitations within the
electromagnetic environment is also the basic idea behind passive 
decoherence
protection by sample design at $\nu=2$. The samples studied in
Refs.\cite{Altimiras:2010-2,Huynh:2012-1}
are based on blocking electronic relaxation and decoherence within one 
of
the two edge channels by closing the other one on itself. 

As known from previous studies\cite{Degio:2009-1,Ferraro:2014-1}, 
quantitatively studying the electronic decoherence scenarii
in these different situations requires an understanding of the effect of
effective screened Coulomb interactions on the electronic fluid. As we
shall recall now, in the linear response regime, it is completely 
encoded into the finite-frequency admittance matrix of the system.

\subsection{Interactions and edge-magnetoplasmon scattering}
\label{sec:models}

\subsubsection{General method}
\label{sec:models:introduction}

During their propagation, electronic excitations experience
screened Coulomb interactions within the conductor and
with charges located in nearby conductors. However, in a regime of
linear response for all conductors involved, interaction effects can be
described within the edge-magnetoplasmon scattering formalism, which
describes how the bosonic edge-magnetoplasmon modes are
altered within the interaction region. This is why the
bosonization framework provides the key for describing electronic 
coherence propagation along chiral edge channels.

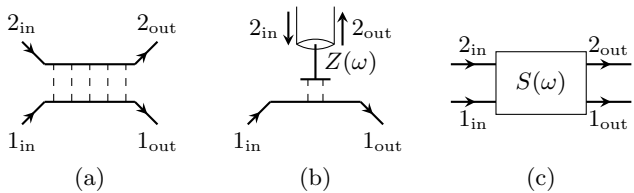
\begin{figure}
\begin{tikzpicture}
[
edge channel/.style={%
			thick
		},
		edge channel dir/.style={%
			thick,
			decoration={markings,mark=at position 0.55 with 
{\arrow{stealth}}},
			postaction={decorate}%
		},
		edge channel dir param/.style={%
			thick,
			decoration={markings,mark=at position #1 with {\arrow{stealth}}},
			postaction={decorate}%
		}%
	]

\def\lc{1.2}

\def\sep{0.5}
\begin{scope}[shift={(-3,0)}]
	\draw[thick](0,0) -- (\lc,0);
	\draw[edge channel dir] (-\lc/4,-\lc/4) -- (0,0);
	\node[below] () at (-\lc/4,-\lc/4) {$1_{\text{in}}$};
	\draw[edge channel dir] (\lc,0) -- ({\lc + \lc/4},-\lc/4);
	\node[below] () at ({\lc + \lc/4},-\lc/4) {$1_{\text{out}}$};
	\draw[thick] (0,\sep) -- (\lc,\sep);
	\draw[edge channel dir] (-\lc/4,\sep + \lc/4) -- (0,\sep);
	\node[above] () at (-\lc/4,\sep + \lc/4) {$2_{\text{in}}$};
	\draw[edge channel dir] (\lc,\sep) -- ({\lc + \lc/4},\sep + \lc/4);
	\node[above] () at ({\lc + \lc/4},\sep + \lc/4) {$2_{\text{out}}$};
	\foreach \i in {1,...,5}{
		\begin{scope}[shift={(-\lc/10+\lc/5*\i,0)}]
		\draw[dashed] (0,0) -- (0,\sep);
		\end{scope}
	}
	\node[below] at (\lc/2,-0.75) {(a)};
\end{scope}

\def\sep{0.3}
\def\size{0.5}
\begin{scope}[shift={(0,0)}]
	\draw[thick](0,0) -- (\lc,0);
	\draw[edge channel dir] (-\lc/4,-\lc/4) -- (0,0);
	\node[below] () at (-\lc/4,-\lc/4) {$1_{\text{in}}$};
	\draw[edge channel dir] (\lc,0) -- ({\lc + \lc/4},-\lc/4);
	\node[below] () at ({\lc + \lc/4},-\lc/4) {$1_{\text{out}}$};
	
	\draw[-stealth,thick] (\lc/5,4*\sep) -- (\lc/5,2.5*\sep) 
node[midway,left] {$2_{\text{in}}$};
	\draw[-stealth,thick] (4*\lc/5,2.5*\sep) -- (4*\lc/5,4*\sep) 
node[midway,right] {$2_{\text{out}}$};
	
	\draw[thick] (\lc/3,\sep) -- (2*\lc/3,\sep);

	\foreach \i in {1,...,2}{
		\begin{scope}[shift={(\lc/3-\lc/12+\lc/6*\i,0)}]
		\draw[dashed] (0,0) -- (0,\sep);
		\end{scope}
	}
	
	\draw  ({\lc/2+0.5*cos(120)},{4*\sep + \size + 0.5*sin(-120)}) --
	({\lc/2+0.5*cos(120)},{4*\sep + 0.5*sin(-120)}) 
arc(-120:-60:0.5)
-- ({\lc/2+0.5*cos(60)},{4*\sep + \size + 0.5*sin(-60)});
\draw ({\lc/2+0.5*cos(60)},{4*\sep + 0.5*sin(-60)}) 
arc(60:120:0.5);
	\draw[thick] (\lc/2,\sep) -- (\lc/2,{4*\sep+0.5*sin(-120)});
	
	\node[right] () at (\lc/2,1.7*\sep) {$Z(\omega)$};
	
		\node[below] at (\lc/2,-0.75) {(b)};
\end{scope}

\def\sep{0.5}
\begin{scope}[shift={(3,0)}]
	\draw[edge channel dir] (-\lc/2,0) -- (0,0)
	node[below,midway] {$1_{\text{in}}$};
	\draw[edge channel dir] (\lc,0) -- ({\lc + \lc/2},0)
	node[midway, below] {$1_{\text{out}}$};
	\draw[edge channel dir] (-\lc/2,\sep) -- (0,\sep)
	node[above,midway] {$2_{\text{in}}$};
	\draw[edge channel dir] (\lc,\sep) -- ({\lc + \lc/2},\sep)
	node[midway, above] {$2_{\text{out}}$};
	
	\draw (0,-\sep/3) rectangle (\lc,4*\sep/3);
	\node () at (\lc/2,\sep/2) {$S(\omega)$};

	\node[below] at (\lc/2,-0.75) {(c)};
\end{scope}

\end{tikzpicture}
	\caption{\label{fig:principe} (Color online) The edge-magnetoplasmon 
scattering
approach describes many situations, such as for example (a) two 
copropagating edge channels
capacitively coupled over a distance $l$, (b) a chiral edge channel
capacitively coupled to a linear external circuit described by a
frequency dependent impedance $Z(\omega)$. (c) Solving for the equation 
of motions leads to a frequency dependent scattering matrix $S(\omega)$
between the channel's edge-magnetoplasmon modes and the bosonic modes of
the other system.}
\end{figure}

More precisely, we consider a length $l$ region of a quantum Hall 
edge 
channel in which electrons experience intra-channel Coulomb 
interactions as well as Coulomb interactions with other edge channels 
(see Fig.~\ref{fig:principe}-(a)) or with
an external gate connected to an impedance (see 
Fig.~\ref{fig:principe}-(b)). For the edge channel under consideration, 
electronic degrees of freedom are described by the bosonic field 
$\phi(x,t)$ defined from the charge density by
\eqref{eq:bosonization:density}. Its equation of motion is given by
\begin{equation}
\label{eq:motion}
(\partial_t+v_F\partial_x)\phi(x,t)=\frac{e\sqrt{\pi}}{h}\,U(x,t)
\end{equation}
where $U(x,t)$ denotes the potential along the edge channel. Assuming 
we are in a linear screening regime within the edge channel as well as 
for the external elements capacitively coupled to it, the
potential $U(x,t)$ is linear in terms of both the bosonic fields 
associated with the other edge channels and bosonic dynamical 
variables describing other circuit elements.
In the case of a gate coupled to an external circuit, these would be 
the bosonic modes associated with the transmission line representation 
of the circuit's impedance. In the same way, the edge-magnetoplasmon 
modes of the current channel appear within source terms for the linear 
equations that describe bosonic modes for the other edge channels and 
circuit elements. 

The interaction region being of finite length, solving the full set of
equations of motion leads to an expression for the outgoing fields in 
terms of the incoming fields. 
Note that because the problem is time translation invariant, the 
solution can be expressed in terms of an elastic scattering matrix 
$S(\omega)$ linking the incoming and outgoing bosonic modes (see
Fig.~\ref{fig:principe} (c)).
In the present situation where all the incoming and outgoing channels, 
outside of the interaction region, correspond to non-interacting edge 
channels with the same Fermi velocity, energy is conserved. In terms of 
edge-magnetoplasmon scattering, this implies that the scattering matrix 
is unitary.
dire.

The edge-magnetoplasmon scattering matrix 
is directly related to the dimensionless 
finite-frequency admittance
$g_{\alpha,\beta}(\omega)=
R_KG_{\alpha,\beta}(\omega)$ ($R_K=h/e^2$ being
the quantum of resistance) defined as the ratio of the
derivative of total current coming into the sample through the edge
channel $\alpha$ with respect to the voltage applied to the reservoir
feeding the edge channel $\beta$. Such a relation had been derived in
the case of quantum wires \cite{Safi:1995-1,Safi:1995-2,Safi:1999-1}
which are non-chiral Luttinger liquids. In the present 
case of chiral quantum Hall edge
channel at integer filling fractions, it takes the following
form
\cite{Degio:2010-1}:
\begin{equation}
\label{eq:models:admittance}
g_{\alpha\beta}(\omega)=\delta_{\alpha,\beta}-S_{\alpha\beta}(\omega)\,.
\end{equation}
Relating edge-magnetoplasmon scattering to
response functions also puts some constraints on scattering amplitudes.

First of all, the dimensionless finite-frequency admittance 
$g(\omega)=1-S_{11}(\omega)$  
of the effective dipole formed by the interaction
region of the edge channel $1$ (lower part of 
Figs.~\ref{fig:principe}-(a-c)) 
and all grounded elements it is capacitively coupled to (upper part of 
Figs.~\ref{fig:principe}-(a-c)) is defined as:
\begin{equation}
g(\omega)=\left.
\frac{\partial \langle \widetilde{I}_1(\omega)\rangle}{\partial
\widetilde{V}_{1}(\omega)}\right|_{V_1=0}
\end{equation}
where $\tilde{I}_1(\omega)$ denotes the Fourier transform of the total
current $(\i_{1,\mathrm{in}}-i_{1,\mathrm{out}})(t)$ and $V_1(t)$
denotes the time dependent drive applied to the edge channel $1$ keeping
the reste at zero potential.
Being a physical response function, its analytic continuation to 
negative frequencies obeys the reality condition: 
$g(\omega)^*=g(-\omega)$. Consequentely, $t(\omega)=S_{11}(\omega)$ 
can be analytically
extended to negative frequencies by $t(-\omega)=t(\omega)^*$.

Next, the finite-frequency admittance $g(\omega)$ is the one of a
passive circuit. As such, it obeys the general property first proposed
by Cauer \cite{Cauer:1926} and 
then proven by Brune \cite{Brune:1931} of being positive real. With 
our
convention, this means that for $z=\sigma+i\omega$, $z\mapsto g(z)$ is
analytic in the half plane $\Re{(z)}<0$ and
\begin{subequations}
\label{eq:impedance-conditions}
\begin{align}
\Re{\left(g(z)\right)} &>0\quad \text{when}\ \sigma <0\\
\Im{\left(g(z)\right)} &= 0\quad \text{when}\ z\in \mathbb{R}^-
\end{align}
\end{subequations}
The analyticity condition ensures that the current response is causal 
and
the two other conditions express that, when driven by a time-dependent
voltage, the corresponding effective dipole dissipates energy and does
not produce it.
As we shall discuss, these conditions put some constraints on the 
low-frequency expansion of $t(\omega)$ and consequently on the 
effective 
interaction models that can be used. 

Finally, since the edge-magnetoplasmon scattering matrix depends on
the precise form of the electric potential within the wire $U(x,t)$,
analytical models are often approximative descriptions of the real
physics of the sample. However, Eq.~\eqref{eq:models:admittance}
suggests that 
edge-magnetoplasmon scattering amplitudes can be measured using
finite-frequency admittance measurements. This has indeed been
done in the case of the $\nu=2$ Quantum Hall edge channel system 
\cite{Bocquillon:2013-2}.

As will be discussed in Sec.~\ref{sec:decoherence}, the 
edge-magnetoplasmon scattering amplitudes are the key ingredients for
computing electronic decoherence \cite{Ferraro:2014-2}.
Before turning to this problem,
let us discuss several edge channel
models starting with the case of an ideal
$\nu=1$ edge channel with finite range intra-channel interactions. We
shall then consider the case of two interacting edge channels ($\nu=2$)
and discuss the case of specific geometries in which one of the edge
channels is closed.

\subsubsection{The $\nu=1$ case}
\label{sec:models:nu=1}

For a single edge channel with Coulomb intra-channel
interactions, the edge-magnetoplasmon scattering matrix
reduces to a frequency dependent transmission coefficient $t(\omega)$
which, in the absence of dissipation, satisfies 
$|t(\omega)|=1$.

Short-range effective screened Coulomb interactions correspond to a
renormalization of the edge-magnetoplasmon velocity and therefore to a 
linear dependence of the phase of $t(\omega)$ in 
$\omega$, $t(\omega)=\me^{\mi\omega\tau(l)}$
where $\tau(l)$ is the renormalized time of flight. By contrast, finite 
range interactions lead to a non-linear frequency dependence of 
the phase of $t(\omega)$. We shall write $t(\omega)=\me^{\mi \omega 
\tau(l,\omega)}$ where the time of flight now depends on $\omega$
through a frequency dependent velocity for the edge magnetoplasmons, 
$\tau(l,\omega)=l/v(\omega)$. Since $t(\omega)^*=t(-\omega)$,
$v(\omega)$ can be extended analytically to
negative frequencies by $v(-\omega)=v(\omega)$.

\begin{figure}
 \includegraphics[width=8cm]{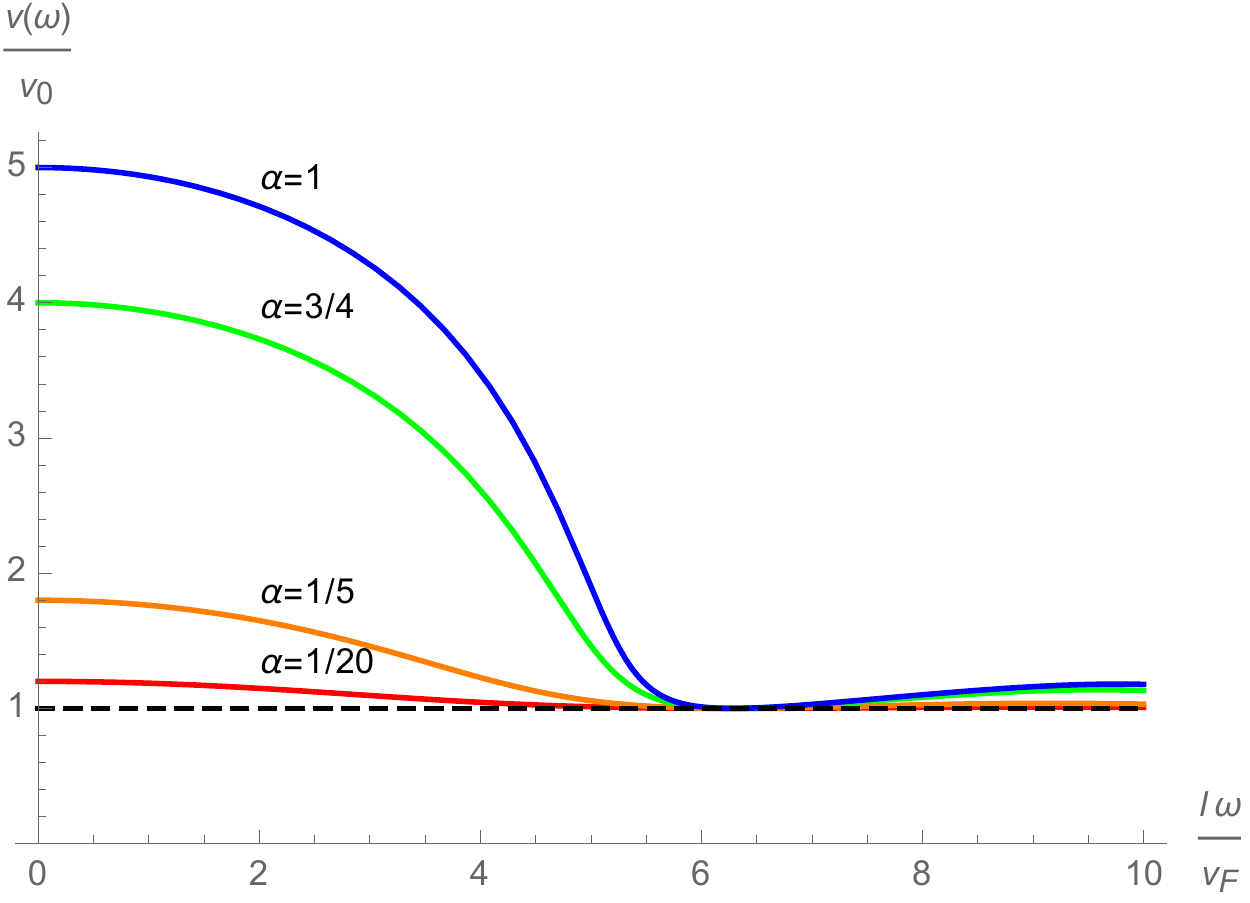}
	\caption{\label{fig:phenomenology:nu=1:long-range:velocity} (Color
	online) Velocity
 $v(\omega)/v_0$ corresponding to $\exp{(i\omega l/v(\omega))}$
 given by Eq.~\eqref{eq:nu=1:result} 
 in terms of $\omega l/v_F$ for $\alpha=1/20$ (graphene), $\alpha=1/5$,
 $\alpha=3/4$ (AsGa) and $\alpha=1$.
}
\end{figure}

A simple model of a $\nu=1$ edge channel with an
interaction region of length $l$, capacitance $C$ 
and bare Fermi velocity $v_F$
is presented in 
Appendix \ref{appendix:nu=1}. 
This model depends on a dimensionless coupling constant $\alpha =
(e^2/C)/(\hbar v_F/l)$ representing the ratio of the Coulomb energy for 
the interaction region to the associated kinetic energy. 
As expected, the edge-magnetoplasmon transmission amplitude 
$t(\omega)=e^{i\omega l/v(\omega)}$ exhibits a non-linear 
dependence of the phase:
\begin{equation}
\label{eq:nu=1:result}
 t(\omega) = \me^{\mi\omega l/v_F}
 \frac{1+A(\omega,l)\me^{-\mi\omega l/(2v_F)}}
 {1+A(\omega,l)\me^{\mi\omega l/(2v_F)}} 
 \end{equation}
where
 \begin{equation}
 \label{eq:nu=1:result:b}
 A(\omega,l)=4\alpha\,
 \sinc\left(\frac{\omega l}{2v_F}\right)\,.
\end{equation}
The edge-magnetoplasmon velocity $v(\omega)$ decreases from
$v_0=(1+4\alpha)v_F$ to its asymptotic value
$v_\infty=v_F$ showing some mild oscillations 
(see Fig.~\ref{fig:phenomenology:nu=1:long-range:velocity})
arising from
the sharp position dependence of the interaction potential at the
boundary of the interaction region. 

Realistic estimates for
the coupling constant $\alpha$ are given in Appendix
\ref{appendix:nu=1}. In AsGa, 
$\alpha\simeq 0.75$ for $v_F\simeq \SI{e5}{\meter/\second}$ thus
leading 
to a ratio $v_0/v_F=4$. By comparison, a similar estimate for
exfolliated graphene on a silicon oxyde surface \cite{Petkovic:2014-1}
leads to $\alpha\simeq 0.05$ assuming $v_F\simeq
\SI{e6}{\meter/\second}$, and thus to $v_0/v_F\simeq 1.2$. Provided it
has such a high Fermi velocity, this specific
form of graphene
may thus correspond to a weak coupling whereas
$\mathrm{AsGa}$ leads to strong coupling. 
A small coupling constant has drastic consequences on
electronic decoherence as will be discussed in 
Sec.~\ref{sec:decoherence_nu_1}. Therefore, studying single-electron
decoherence in the edge channels of graphene at $\nu=1$ may be a way
to test whether or not it is a weak or a strong coupling material.

We expect a more realistic model of intra-channel interactions to lead 
to a
qualitatively similar but smoother 
behavior 
of $v(\omega)$. 
Key features are the two different
asymptotic velocities $v_{0}$ and $v_{\infty}$ in the limits 
$\omega\rightarrow 0$ and $\omega\rightarrow +\infty$. 
The infrared velocity $v_0$ is the velocity of 
low energy edge-magnetoplasmon modes and should therefore be called the
plasmon velocity. Due to Coulomb interactions, it is expected to be
higher than the velocity of high-energy excitations which do
not experience interactions for a long time. Reasonable
phenomenological models for 
$v(\omega)$ should thus interpolate between $v_0$ and
$v_\infty$ with $v_0>v_\infty$.
However, as explained in appendix \ref{appendix:phenomenology}, the
relation between $t(\omega)$ and the finite-frequency admittance
combined to Eqs.~\eqref{eq:impedance-conditions} strongly constrains
the general form of the $t(\omega)$. It indeed rules out simple
phenomenological expressions for the
edge-magnetoplasmon velocity $v(\omega)$. Therefore,
we shall
discuss the ideal $\nu=1$ case using the long-range model presented in
Appendix~\ref{appendix:nu=1}.

\subsubsection{The $\nu=2$ case}
\label{sec:models:nu=2}

The $\nu=2$ edge channel system is the simplest and experimentally most
relevant case involving more than one channel. In this case, two
copropagating edge channels separated by approximately 
\SI{100}{\nano\meter} experience strong intra and inter-channel 
screened 
Coulomb interactions. Several models have been developed to describe 
this situation and are briefly reviewed here. 

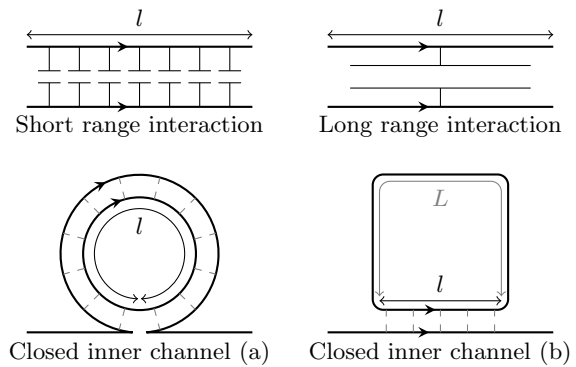
\begin{figure}
\begin{tikzpicture}
[
edge channel/.style={%
			thick
		},
		edge channel dir/.style={%
			thick,
			decoration={markings,mark=at position 0.45 with {\arrow{stealth}}},
			postaction={decorate}%
		},
		edge channel dir param/.style={%
			thick,
			decoration={markings,mark=at position #1 with {\arrow{stealth}}},
			postaction={decorate}%
		}%
	]

\def\lc{3}

\def\sep{0.8}
\def\sizecapa{0.32}
\def\lcapa{0.15}
\begin{scope}[shift={(-2,0)}]
	\draw[<->] (0,1.2*\sep) -- (\lc,1.2*\sep) node[midway, above] {$l$};
	\draw[edge channel dir] (0,0) -- (\lc,0);
	\draw[edge channel dir] (0,\sep) -- (\lc,\sep);
	\foreach \i in {1,...,7}{
		\begin{scope}[shift={(-0.1+0.4*\i,0)}]S
		\draw (0,0) -- (0,\sizecapa);
		\draw (0,\sep) -- (0,\sep-\sizecapa);
		\draw (-\lcapa,\sizecapa) -- (\lcapa,\sizecapa);
		\draw (-\lcapa,\sep-\sizecapa) -- (\lcapa,\sep-\sizecapa);
		\end{scope}
	}
	\node[below] at (\lc/2,0) {Short range interaction};
\end{scope}

\def\sep{0.8}
\def\sizecapa{0.25}
\def\lcapa{0.4*\lc}
\begin{scope}[shift={(2,0)}]
	\draw[<->] (0,1.2*\sep) -- (\lc,1.2*\sep) node[midway, above] {$l$};
	\draw[edge channel dir] (0,0) -- (\lc,0);
	\draw[edge channel dir] (0,\sep) -- (\lc,\sep);
	\begin{scope}[shift={(\lc/2,0)}]
		\draw (0,0) -- (0,\sizecapa);
		\draw (0,\sep) -- (0,\sep-\sizecapa);
		\draw (-\lcapa,\sizecapa) -- (\lcapa,\sizecapa);
		\draw (-\lcapa,\sep-\sizecapa) -- (\lcapa,\sep-\sizecapa);
	\end{scope}
	\node[below] at (\lc/2,0) {Long range interaction};
\end{scope}

\def\radius{0.25*\lc}
\def\sep{0.1*\lc}
\def\angle{5}
\def\anglebis{2}
\begin{scope}[shift={(-2,-3)}]
	\draw[edge channel dir] 
	(0,0) -- 
	({\lc/2-sin(\angle)*(\radius+\sep)},0) 
	arc(270-\angle:-90+\angle:\radius+\sep) 
	-- (\lc,0);
	\draw[edge channel dir] 
	(\lc/2,{(\radius+\sep)*cos(\angle)-\radius})
	arc(-90:-450:\radius);
	\draw[<->]
	({\lc/2-sin(\anglebis)*(\radius-0.5*\sep)},
	{(\radius+\sep)*cos(\angle)-(\radius-0.5*\sep)*cos(\anglebis)} )
	arc(270-\anglebis:-90+\anglebis:\radius-0.5*\sep)
	node[midway, below] {$l$};
	\begin{scope}[shift={(\lc/2,{(\radius+\sep)*cos(\angle)})}]
	 \foreach \i in {1,...,12}{
		\draw[dashed, gray] 
		(15+30*\i:\radius) -- (15+30*\i:\radius+\sep);
		}
	\end{scope}
	\node[below] at (\lc/2,0) {Closed inner channel (a)};
\end{scope}

\def\sep{0.1*\lc}
\def\sizeloop{0.6*\lc}
\begin{scope}[shift={(2,-3)}]
	\draw[edge channel dir] (0,0) -- (\lc,0);
	\draw[edge channel dir param=0.35, rounded corners] 
 (\lc/2-\sizeloop/2,\sep+\sizeloop) rectangle (\lc/2+\sizeloop/2,\sep);
 \draw[<->] 
 (\lc/2-0.9*\sizeloop/2,1.4*\sep) -- (\lc/2+0.9*\sizeloop/2, 1.4*\sep)
 node[midway, above]{$l$};
  \draw[rounded corners,gray, <->] 
 (\lc/2-0.9*\sizeloop/2,1.6*\sep) --
 (\lc/2-0.9*\sizeloop/2,\sep+0.95*\sizeloop) --
 (\lc/2+0.9*\sizeloop/2,\sep+0.95*\sizeloop)  node[midway, below]{$L$} --
 (\lc/2+0.9*\sizeloop/2, 1.6*\sep);
\foreach \i in {1,...,5}{
		\draw[dashed, gray] 
		({\lc/2-\sizeloop/2+\sizeloop/5*(\i-1/2)},0) -- 
		({\lc/2-\sizeloop/2+\sizeloop/5*(\i-1/2)},\sep);
		}
 	\node[below] at (\lc/2,0) {Closed inner channel (b)};
\end{scope}

\end{tikzpicture}
\caption{\label{fig:interactiontypes} Schematic 
view of the main types 
of interaction discussed at $\nu=2$. Short-range interaction 
corresponds 
to a capacitive coupling between charge densities at the same position 
in the two channels, and no coupling between different positions. Long 
range interaction describes a situation where the system behaves as one 
big capacitor. We are also interested in situations where the inner 
channel is closed on itself and interacts with the outer channel either 
along its whole length (a), or only on a small portion of the closed 
loop (b). In either of these cases, interactions can be short range or 
long range.}
\end{figure}

\paragraph{Co-propagating channels with short-range interaction}

In the presence of metallic side gates, Coulomb interactions are
screened and the charge density 
in one channel is capacitively coupled to the charge density at the 
same point in the other channel\cite{Levkivskyi:2008-1}. 
More precisely, charge density in 
channel $i$ at position $x$ and energy $\omega$ $\rho_i(x,\omega)$ is 
coupled to the local electrostatic potential $U$ through distributed 
capacitances: $\rho_i(x,\omega)=\mathcal{C}_{ij}U_j(x,\omega)$. 
This model, schematically depicted on 
Fig.~\ref{fig:interactiontypes}, is known to give a 
good description of interactions in experimental systems at small 
energies, a fact that has been directly probed in the frequency
\cite{Bocquillon:2013-2} and time \cite{Hashisaka:2017-1} domains and 
indirectly confirmed in Ref.\cite{Inoue:2013-1}. Within the 
interaction region, edge-magnetoplasmon eigenmodes are delocalized
over the two channels and propagate at different
velocities. This leads to
the following edge-magnetoplasmon scattering matrix\cite{Degio:2010-1}:
\begin{equation}
\label{eq:interactions:S-matrix:short-range}
 S(\omega)=\begin{pmatrix}
            p_{+}\me^{\mi \omega \tau_{+}} + 
            p_{-}\me^{\mi \omega \tau_{-}}
            & q \left(\me^{\mi \omega \tau_{-}}-
		      \me^{\mi \omega \tau_{+}} \right)
            \\
            q \left(\me^{\mi \omega \tau_{-}}-
		      \me^{\mi \omega \tau_{+}} \right)
	    & p_{+}\me^{\mi \omega \tau_{-}} + 
	      p_{-}\me^{\mi \omega \tau_{+}}
           \end{pmatrix}
\end{equation}
where
\begin{subequations}
\label{eq:interactions:p_pm}
\begin{align}
 p_{\pm}=\frac{1\pm\cos(\theta)}{2}, &\qquad q=\frac{\sin(\theta)}{2} \\
 \tau_{+}=\frac{l}{v_{+}}, & \qquad \tau_{-}=\frac{l}{v_{-}}.
\end{align}
\end{subequations}
In these equations, $\theta$ corresponds to the coupling strength, 
$v_{+}$ to the velocity of the slowest mode and $v_{-}$ to the one of 
the fastest mode. In the strong-coupling regime, 
$\theta=\pi/2$, the corresponding modes are a fast charge 
mode, which is symmetric across both channels and an antisymmetric slow
neutral mode\cite{Levkivskyi:2008-1}. 

\paragraph{Co-propagating channels with long-range interaction}

The second model for interacting co-propagating channels 
assume that local potentials $U$ are uniform on the whole length of 
the interaction region. The interaction region is a
capacitor (see Fig.~\ref{fig:interactiontypes}) and can be 
discussed in the 
spirit of the discrete element circuit models introduced by Büttiker
{\it et al} for quantum conductors and quantum Hall edge 
channels\cite{Pretre:1996-1,Christen:1996-1}.
This approach
leads to the following edge-magnetoplasmon
scattering matrix \cite{Grenier:2013-1}:
\begin{equation}
\label{eq:interactions:S-matrix:long-range}
 S(\omega)=\begin{pmatrix}
            p_{+}\mathcal{T}_{+}(\omega) + 
            p_{-}\mathcal{T}_{-}(\omega)
            & q \left(\mathcal{T}_{-}(\omega)-
		      \mathcal{T}_{+}(\omega) \right)
            \\
            q \left(\mathcal{T}_{-}(\omega)-
		      \mathcal{T}_{+}(\omega) \right)
	    & p_{+}\mathcal{T}_{-}(\omega) + 
	      p_{-}\mathcal{T}_{+}(\omega)
           \end{pmatrix}
\end{equation}
where $p_{\pm}$ and $q$ are given by Eq.~\eqref{eq:interactions:p_pm}
and other parameters are given in terms of the 
dimensionless parameter $x=\omega l/v_F$ by
\begin{equation}
 \mathcal{T}_{\pm}(\omega) =
	\frac
		{\me^{\mi x} - 1 + \mi \alpha_{\pm} x \me^{\mi x}}
		{\me^{\mi x} - 1 + \mi \alpha_{\pm} x }
\end{equation}
$\alpha_{\pm}$ being linked to the eigenvalues of the capacitance 
matrix $C_{\pm}$ by $\alpha_{\pm}=R_K C_{\pm} v_F / l$.

\subsubsection{The $\nu=2$ case with a loop}
\label{sec:models:nu=2:loops}

Fig.~\ref{fig:interactiontypes}(a) also depicts another situation 
that can be built with two copropagating edge channels, where the inner 
one is closed on itself over the length $l$ where interaction takes 
place\cite{Altimiras:2010-2}. In the geometry depicted 
on Fig.~\ref{fig:interactiontypes}(b), the same idea of a closed inner 
channel is used, but the copropagating distance over 
which interaction takes place is only a part of the total length of the 
loop. Such a geometry has been used for 
mitigating decoherence in electronic Mach-Zehnder interferometers
\cite{Huynh:2012-1}. Both geometries impose a periodicity condition on
the field for the inner channel: 
\begin{align}
 \phi_2(0,\omega)=\phi_2(l,\omega)\me^{\mi \omega \tau_L}
\end{align}
where $\tau_L=\frac{L}{v_+}$ is the time it takes for an excitation to 
cover the non-interacting length $L$ of the loop. The transmission
coefficient is then obtained in full generality as
\begin{align}
\label{eq:def_t_closedchannel}
 t(\omega)=
 S_{11}(\omega)
+\frac{S_{12}(\omega)S_{21}(\omega)}{\me^{-\mi\omega\tau_L}-S_{22}
(\omega)}\,.
\end{align}
As expected, in the absence of dissipation, we have a unitary $S$ 
matrix and this transmission coefficient has a modulus of $1$.  For 
short-range interaction, last equation specializes to
\begin{align}
  t(\omega)=
	-\me^{\mi \omega (\tau_+ + \tau_- - \tau_L)}
	\left(
		\frac{\me^{\mi \omega \tau_L} 
				 -p_{+}\me^{-\mi \omega \tau_+}
				 -p_{-}\me^{-\mi \omega \tau_-}
				 }
				 {\me^{-\mi \omega \tau_L}
					-p_{+}\me^{\mi \omega \tau_+}
					-p_{-}\me^{\mi \omega \tau_-}
				 }
	\right)
\end{align}
Of course, the special case (a) is recovered for $\tau_L=0$.

\section{Electronic decoherence}
\label{sec:decoherence}

Let us now explain how to obtain the outgoing electronic coherences
when a single-electron excitation is injected into the interaction
region. We will first review the main steps and the essentiel points of
the general methods developed for comparing the electronic decoherence
of Landau and Levitov quasi-particles \cite{Ferraro:2014-2}. Then, we
will discuss in details decoherence within a dissipationless single edge
channel and then in the $\nu=2$ edge channel system.

\subsection{General results}
\label{sec:decoherence:general}

In the bosonization framework, the interaction region is a frequency 
dependent
beam splitter for the edge-magnetoplasmon modes.
An incoming coherent state
for these modes is scattered exactly as a classical electromagnetic wave
on an optical beam splitter~\cite{Grenier:2013-1}. More precisely,
an incoming coherent edge magnetoplasmon of the
form $|\Lambda_1\rangle \otimes |\Lambda_2\rangle$ is transformed into
an outgoing state $|\Lambda'_1\rangle \otimes
|\Lambda'_2\rangle$ where for all $\omega >0$, 
$\Lambda'_\alpha(\omega)=
\sum_\beta S_{\alpha\beta}(\omega)\Lambda_\beta(\omega)$. 
Because single-electron states are described as
quantum superposition of coherent edge-magnetoplasmon states, an
exact description of the outgoing state after the interaction region can
be obtained. A single-electron state injected in edge channel 1
corresponds, with the notations given in 
appendix \ref{appendix:bosonization}, to
\begin{align}
	|\varphi_\text{e},F\rangle_1 &\otimes |F\rangle_2
	=
	\notag\\
	&\int_{-\infty}^{+\infty} \varphi_\text{e} (t)
	\frac{U^{\dagger}_{1}}{\sqrt{2\pi a}}
	\bigotimes_{\omega>0}\left(
	|\Lambda_{\omega}(t)\rangle_1
	\otimes
	|0_{\omega}\rangle_2 \right)\,
	\md t
\end{align}
and comes out of the interaction region as:
\begin{align}
\label{eq:decoherence:entangleoutgoingstate}
	\int \varphi_\text{e} (t)
	\frac{U^{\dagger}_{1}}{\sqrt{2\pi a}}
	\bigotimes_{\omega>0} \left(
		|t(\omega)\Lambda_{\omega}(t)\rangle_1
	\otimes
		|r(\omega)\Lambda_{\omega}(t)\rangle_2\right)\,
	\md t \,.
\end{align}
In this equation, we adopt the convention used in the remaining of this 
text that $S_{11}(\omega)=t(\omega)$ and $S_{21}(\omega)=r(\omega)$, 
other coefficients of $S$ being irrelevant as no injection is made in 
channel 2.
Tracing on the second edge channel degrees of freedom leads to the
reduced outgoing many-body density operator for the injection edge 
channel \cite{Degio:2009-1}:
\begin{align}
\label{eq:decoherence:first-step}
\rho_{1} =
 \int \varphi_{\text{e}}^{ }(t)\varphi_\text{e}^{*}(t') 
 \mathcal{D}_{\text{ext}}(t-t') 
 \psi^{\dagger}(t) |g(t)\rangle
 \langle g(t')| \psi(t')
 \md t \md t'
\end{align}
where $\mathcal{D}_{\text{ext}}(t-t')$ is the extrinsic decoherence 
coefficient  corresponding to the overlap of imprints left
in the environment by localized 
electrons injected at times $t$ and $t'$. It is given by
\cite{Degio:2009-1}~: 
\begin{equation}
 \mathcal{D}_{\text{ext}}(\tau) =
 \exp{
\left(\int_0^{+\infty}|r(\omega)|^2 (\me^{\mi \omega \tau} 
-1)\,\frac{\md\omega}{\omega}\right)
 }\,.
\end{equation}
The coherent edge-magnetoplasmon state $|g(t)\rangle$ in
Eq.~\eqref{eq:decoherence:first-step} corresponds 
to the cloud of electron/hole pairs
generated by Coulomb interactions when a localized electron
$\psi^\dagger(t)|F\rangle$ goes through the interaction region:
\begin{align}
 |g(t)\rangle =
	\bigotimes_{\omega>0}
	|(1-t(\omega))\Lambda_{\omega}(t)\rangle\,.
\end{align}
In the same way, in the $\nu=2$ case, the reduced density operator for
the inner edge
channel can be obtained by tracing out over the outer edge channel. This
leads to 
\begin{equation}
\rho_{2} = \int
\varphi_\text{e}(t)\varphi_\text{e}(t')\mathcal{D}_{\mathrm{inj}}(t-t')\
,
|\mathcal{E}_2(t)\rangle\langle\mathcal{E}_2(t')|\,\md t\md t'\,.
\end{equation}
where
\begin{equation}
|\mathcal{E}_2(t)\rangle =
\bigotimes_{\omega>0}|r(\omega)\Lambda_\omega(t)\rangle
\end{equation}
and the decoherence coefficient 
\begin{equation}
 \mathcal{D}_{\text{inj}}(\tau) =
  \exp{
	  \left(\int_0^{+\infty}|t(\omega)|^2(\me^{\mi \omega
	  \tau}-1)\,\frac{\md\omega}{\omega}\right)}
\end{equation}
is equal to the overlap of the outgoing states 
$|\mathcal{E}_1(t)\rangle$ 
of the injection edge channel
corresponding to two different injection times:
\begin{equation}
|\mathcal{E}_1(t)\rangle=\bigotimes_{\omega>0}
    |t(\omega)\Lambda_{\omega}(t)\rangle\,.
\end{equation}
This many-body description then gives access to all
electronic coherence functions after the interaction
region.

\subsection{Computing single-electron coherences}

Let us now turn to first order coherences in 
the 
outer and inner channels after interaction, denoted respectively by 
$\mathcal{G}^{(e)}_{\text{out},1}(t|t')$ and 
$\mathcal{G}^{(e)}_{\text{out},2}(t|t')$.

\subsubsection{Outer channel coherence}

When computing $\mathcal{G}^{(e)}_{\text{out},1}(t|t')$, the final 
results appear as a sum of two terms. The first one corresponds to 
a modification of  the Fermi sea which, under the right condition, 
can be seen as the contribution of
electron-hole pairs generated by Coulomb interaction
vacuum state (namely the Fermi sea). This one is called the
\emph{modified vacuum}. Under the same condition, the second 
contribution comes from the incoming excitation elastically scattered
or after interaction induced relaxation. This one is called the
\emph{wavepacket} contribution. These two contributions
can be written as\cite{Ferraro:2014-1} 
\begin{subequations}
\label{eq:decoherence:result}
	\begin{align}
	\label{eq:decoherence:result:MV}
    \mathcal{G}^{(e)}_{\text{MV},1}(t|t')&=
    \int \varphi_\text{e}^{ }(t_{+}) \varphi_\text{e}^{*}(t_{-})
    \mathcal{D}(t,t',t_{+},t_{-})
    \\
    &
    \langle \psi^{\dagger}(t') \psi(t) \rangle_F
    \langle \psi(t_{-}) \psi^{\dagger}(t_{+}) \rangle_F
    \, \md t_+ \md t_- 
    \notag
    \\
	\label{eq:decoherence:result:WP}
    \mathcal{G}^{(e)}_{\text{WP},1}(t|t')&=
    \int \varphi_\text{e}^{ }(t_{+}) \varphi_\text{e}^{*}(t_{-})
    \mathcal{D}(t,t',t_{+},t_{-})
    \\
    &\langle \psi(t) \psi^{\dagger}(t_{+}) \rangle_F
    \langle \psi(t_{-}) \psi^{\dagger}(t') \rangle_F
    \, \md t_+ \md t_-
    \notag
	\end{align}
\end{subequations}
where
\begin{align}
\label{eq:decoherence:gamma:definition}
  \mathcal{D}&(t,t',t_{+},t_{-}) =
  \notag \\
  &\gamma_+ (t_+ - t') \gamma_- (t_+ - t)
	\gamma^*_+ (t_- -t) \gamma^*_- (t_- -t')
\end{align}
is the effective single particle decoherence coefficient
which takes into account both the 
action of environmental degrees of freedom and of electron-hole pairs 
cloud created in the injection channel. It is determined by the two
functions
\begin{align}
\gamma_{\pm}(t)=
	\exp \left(
					\pm \int_0^{\infty} \frac{\md \omega}{\omega}
					(1-t(\omega))(\me^{\mi\omega t}-1)
			 \right) 
\end{align}
Explicit expressions for the two contributions
\eqref{eq:decoherence:result:MV} and
\eqref{eq:decoherence:result:WP} are given in Ref.\cite[Supplementary
Material]{Ferraro:2014-2} 
and form the
starting point of the numerical evaluation of the outgoing electronic
coherence in the frequency domain (see 
Sec.~\ref{sec:decoherence:numerics}). 

An important quantity is the
elastic scattering amplitude $\mathcal{Z}(\omega)$ 
for an incoming single-electron excitation 
at energy $\hbar\omega>0$ which determines the inelastic
scattering probability
$\sigma_{\mathrm{in}}(\omega)=1-|\mathcal{Z}(\omega)|^2$. Its 
expression is
given by
\begin{equation}
\mathcal{Z}(\omega)= 1-\int_0^\omega B_-(\omega')\,\md\omega'
\end{equation}
where $B_-$ is defined as the regular part of the Fourier transform of
$\gamma_-$ and therefore satisfies the integral equation
\begin{equation}
\omega B_-(\omega)=t(\omega)-1+\int_0^\omega
B_-(\omega')(t(\omega-\omega')-1)\,\md\omega'\,
\end{equation}
with initial condition $B_-(0^+)=-t'(\omega=0^+)$.

\subsubsection{Inner channel coherence}

Using the reduced density matrix $\rho_2$ for the inner channel, 
any coherence function we are interested in can be computed. The main 
result is 
strikingly simple: $\mathcal{G}^{(e)}_{\text{out},2}(t|t')$ is of the 
same 
exact form as $\mathcal{G}^{(e)}_{\text{MV},1}(t|t')$ if we replace the 
function $t(\omega)$ in the decoherence coefficient with $1+r(\omega)$. 
The 
fact that there is no wavepacket term emphasizes that no 
electron has been injected into the inner channel: only a cloud of 
electron/hole
pairs is created.

\subsubsection{Numerical method}
\label{sec:decoherence:numerics}

As shown in Ref.~\citep[Supplementary Material]{Ferraro:2014-2}, the
numerical evaluation consists in evaluating multiple integrals of 
factors.
The implementation is quite straightforward, even though the main
difficulty comes from the number of nested integrals (four for each
point of the electronic coherence). For this, we use a frequency
representation of the coherence. We discretize the input coherence on a
grid using two directions, $\omega$ and $\delta \omega$. $\omega$ is the
conjugate of $t - t'$ and thus encodes the
frequency dependence in the Wigner function. $\delta \omega$ is the
conjugate of $(t+t')/2$, and thus gives access to time dependance in the
Wigner function. When there are $n$ points in the
input coherence in each direction $\omega$ and $\delta\omega$, a naive
implementation would require an $\mathcal{O}(n^6)$ computation time.
However, by
exploiting the structure of the expressions, we have been able to
lower the total complexity to $\mathcal{O}(m\times n^4)$ where $n$
denotes the number of points in the direction $\omega$ and $m$ the
number of points in the direction $\delta\omega$. This structures allows
us to decouple the two directions and, as such, we can have a better
numerical evaluation by lowering the discretization step in the
direction $\omega$, without touching to the direction $\delta \omega$,
as long as we have enough information about the time evolution of the
Wigner function. With these refinements and using the OpenMP parallel
framework, a post-interaction coherence is computed within five to ten
minutes on a 64~cores computer.

Exactly as in our previous work\cite[Supplementary
Material]{Ferraro:2014-2}, discretization steps are chosen by looking at
errors. The trace of the excess single-electron coherence is the total
charge injected and should not change. If this already very sensitive
indicator is not enough, we compute the average outgoing electric
current from the outgoing excess single-electron coherence and compare
it to its value obtained by applying edge-magnetoplasmon scattering to 
the
incoming average current. All graphs presented in the following
exhibit errors smaller than $5\%$ for those tests.

\subsection{Decoherence at $\nu=1$}
\label{sec:decoherence_nu_1}

Let us first discuss electronic decoherence by using a crude
physical picture 
for a single edge channel in which we have a low-frequency
($\omega\lesssim \omega_c$) edge-magnetoplasmon velocity
$v_0$ greater than the high frequency ($\omega\gtrsim \omega_c$) 
velocity $v_\infty$. This is an oversimplification of the model 
presented in Sec.~\ref{sec:models:nu=1} but it presents the key feature 
of having
distinct high and low energy edge-magnetoplasmon velocities.

Since for $\omega \gtrsim \omega_c$, edge
magnetoplasmons travel at the velocity $v_\infty$, decoherence only
arises from the effective edge-magnetoplasmon scattering phase 
$\tilde{t}(\omega)=t(\omega)\me^{-\mi\omega \tau_\infty}$ which is
roughly $1$ 
for $\omega \gtrsim \omega_c$ and $\me^{-\mi\omega\Delta\tau}$ for
$\omega\lesssim
\omega_c$, where $\Delta\tau=\tau_\infty-\tau_0$ denotes the difference 
of
time of flights between high and low energy edge magnetoplasmons.
As interactions have an effective 
bandwidth $\sim \omega_c$,
creation of electron/hole pair excitations happens close to the Fermi 
level (within one to a few $\omega_c$). Consequently, for electronic 
excitations injected at a much 
higher energy, the corresponding low energy edge-magnetoplasmon modes
can be viewed as an effective distinct environment for the high-energy
electronic excitations\cite{Degio:2009-1}. 

At lower energies, electronic decoherence also arises from the
$\omega$-dependence of the edge-magnetoplasmon velocities but, at low
enough frequency, a perturbative approach in $\omega R_KC_\mu$ can be
used. As we shall see, this leads to an expression of the inelastic 
scattering probability in terms of the effective description of the 
interaction region as a discrete element circuit, going beyond 
the series addition of the electrochemical capacitance $C_\mu$ and the 
relaxation resistance $R_K/2$.

In the following, we shall first explore these high and low energy
limiting regimes of electronic decoherence and then discuss
the full physical 
picture of electronic decoherence and relaxation within a single
isolated edge channel.

\subsubsection{High energy decoherence and relaxation}
\label{sec:decoherence_nu_1:high-energy}

For a single-electron excitation injected at high energy,
the contribution to electronic
coherence $\varphi_\text{e}^{}(t)\,\varphi_\text{e}^{*}(t')$ picks up 
an effective
decoherence coefficient\cite{Degio:2009-1} $\mathcal{D}(t-t')$:
\begin{equation}
\label{eq:decoherence:high-energy:effective-decoherence}
\Delta\mathcal{G}^{(e)}_{\text{WP}}(t|t')\simeq
\varphi^{}_e(t)\,\varphi^{*}_e(t')\,\mathcal{D}(t-t')
\end{equation}
which, at $\nu=1$, is equal to the overlap
$\langle g(t')|g(t)\rangle$ of
the electron/hole pair clouds generated by Coulomb interactions:
\begin{equation}
\label{eq:decoherence:high-energy:decoherence-coefficient}
\mathcal{D}(\tau)=
	\exp{\left(
				\int_0^{+\infty}
				|1-\widetilde{t}(\omega)|^2
				\left(\me^{\mi\omega\tau}-1\right)
				\frac{\md\omega}{\omega}
			\right)}\,.
\end{equation}
This description is analogous to the one used in the weak-coupling
description of dynamical Coulomb blockade across a tunnel junction
\cite{Ingold:1992-1}. 
The relaxation kernel
\begin{equation}
\widetilde{\mathcal{D}}(\omega')
=	\int_{-\infty}^{+\infty}
	\me^{-\mi\omega\tau}\mathcal{D}(\tau)\,\md\tau\,.
\end{equation}
can then be decomposed into an elastic and an inelastic
part: $\widetilde{\mathcal{D}}(\omega')=2\pi
(Z_\infty\delta(\omega')+d(\omega'))$ where
\begin{equation}
\label{eq:Z-infinity}
Z_\infty=
	\exp{\left(
				-\int_0^{+\infty} |1-\tilde{t}(\omega)|^2
				\frac{\md\omega}{\omega}
			\right)}
\end{equation}
is nothing but the high-energy limit of the elastic scattering
probability $|\mathcal{Z}(\omega)|^2$. 
The
inelastic
part $d(\omega)$ describes electronic relaxation: it represents the
probability that the electron has lost energy $\omega$. It is 
determined by the integral equation
\begin{subequations}
\label{eq:decoherence:nu=1:relaxation-tail}
\begin{align}
\omega\, d(\omega)&= |1-\tilde{t}(\omega)|^2\nonumber \\
&+\int_0^\omega |1-\tilde{t}(\omega')|^2d(\omega-\omega')\,\md\omega'\,.
\end{align}
\end{subequations}
which can readily be solved on a computer using the inital condition
that
$d(\omega\rightarrow 0^+)\to
\lim_{\omega\rightarrow 
0^+}\left(|1-\tilde{t}(\omega)|^2/\omega\right)$.
It can 
also be expressed as a formal series
corresponding to the various processes involving the emission of an
increasing number of pairs of electron/hole excitations, exactly the 
same structure than in the
dynamical Coulomb blockade theory \cite{Ingold:1992-1}.
With these notations, the elastic part of the outgoing Wigner 
function is well separated from the
inelastic part:
\begin{subequations}
\label{eq:decoherence:nu=1:high-energy-relaxation}
\begin{align}
\Delta \mathcal{W}^{(e)}_{\text{WP}}(t,\omega)&=Z_\infty 
\mathcal{W}_{\varphi_\text{e}}(t,\omega)
\\
&+
\int_0^{\omega}d(\omega')
\mathcal{W}_{\varphi_\text{e}}(t,\omega+\omega')\,\md\omega'
\end{align}
\end{subequations}
where $\mathcal{W}_{\varphi_\text{e}}(t,\omega)$ denotes the Wigner function 
associated
to the incoming wavepacket $\varphi_\text{e}$.
The incoming electron loses energy through electron/hole pair creation
within a few $\hbar\omega_c$ of the Fermi sea. As shown in Appendix 
\ref{appendix:dissipation}, in the present regime, one can show that for
high-energy electrons, the amount of energy dissipated though
electron/hole pair creations is small compared to their injection
energy, thus providing us with an a posteriori validation of our
approach.

The low energy electron/hole pairs will then propagate along at the low
energy edge-magnetoplasmon velocity. In a first approximation, 
the physical picture for the
decoherence and relaxation of single-electron excitations injected at
high energy thus involves the incoming electron and its relaxation tail
(described by Eq.~\eqref{eq:decoherence:nu=1:high-energy-relaxation})
propagating at the high-energy velocity $v_\infty$ and the corresponding
low energy electron/hole pairs propagating at the low energy 
edge-magnetoplasmon velocity $v_0$. This simple picture justifies 
interpreting $v_\infty$ as the velocity of hot electrons whereas $v_0$ 
is
viewed as a plasmon velocity.

\subsubsection{Low energy decoherence and relaxation}
\label{sec:decoherence_nu_1:low-energy}

At low frequency,
the effective dipole associated with
the interaction region does not respond to a dc bias and can thus be 
described in terms of a frequency
dependent admittance $G(\omega)$ 
in series with a capacitor
$C_\mu$ (see
Fig.~\ref{fig:effective-circuit}). As explained in Appendix
\ref{appendix:circuit}, the corresponding transmission 
coeffficient
$t(\omega)=1-g(\omega)$ has modulus one if and only if
$\Re{(1/G(\omega))}=R_K/2$ meaning that the circuit involves the 
relaxation
resistance $R_q=R_K/2$ in series with a purely reactive impedance. The
simplest model for this pure reactance consists of 
an $LC$ circuit depicted on the left panel of 
Fig.~\ref{fig:effective-circuit}. The $RC$-time $\tau_0=R_KC_\mu$ of the
circuit corresponds to the time of flight of low energy
edge magnetoplasmons across the interaction region. Deviations from this
behavior will lead to single-electron decoherence. 

At low energy, a
perturbative approach detailed in Appendix 
\ref{appendix:perturbation-theory} leads to its descrition
in terms of the discrete element circuit parameters $\tau_0$, $L$ and
$C$.
The inelastic scattering probability 
across the interaction region is then given by 
\begin{align}
\label{eq:decoherence:inelastic-development}
\sigma_{\text{in}}^{(\text{pert})}(\omega)&= 
	\frac{11\alpha_3^2}{180}  \left(\omega\tau_0\right)^6
+	\frac{5\alpha_3\alpha_5}{42}  \left(\omega\tau_0\right)^8
+	\mathcal{O}\left((\omega\tau_0)^9\right)\,.
\end{align}
where
the inductance $L$ is directly related to the $\alpha_3$ coefficient
and the capacitance $C$ only contributes to the next
order:
\begin{subequations}
\label{eq:decoherence:circuit-parameters}
\begin{align}
\tau_0 &= R_KC_\mu \\
\alpha_3&=\frac{L/R_K}{R_KC_\mu}-\frac{1}{12}\\
\alpha_5&=\frac{1}{80}-\frac{1}{4}\frac{L/R_K}{R_KC_\mu}+
\left(\frac{L/R_K}{R_KC_\mu}\right)^2\left(1+\frac{C}{C_\mu}\right)
\end{align}
\end{subequations}
This connects the inelastic scattering probability for an incoming
electron to the low-frequency discrete element circuit description 
for the interaction region.

A complementary understanding can be obtained by
relating the finite-frequency admittance to the edge magnetoplasmon's
effective velocity $v(\omega)$ within the interaction region using 
$t(\omega)=\exp{(i\omega l/v(\omega))}=1-g(\omega)$ (see Appendix
\ref{appendix:perturbation-theory}).
The effective circuit of Fig.~\ref{fig:effective-circuit}
corresponds to
a low-frequency expansion of $v(\omega)$ 
of the form:
\begin{subequations}
\label{eq:decoherence:plasmon-velocity}
\begin{align}
\frac{v(\omega)}{v_0}&=1+
\left(\frac{1}{12}-\frac{L/R_K}{R_KC_\mu}\right)(R_KC_\mu
\omega)^2\\
&-\left[
\frac{C}{C_\mu}\left(\frac{L/R_K}{R_KC_\mu}\right)^2
-\frac{1}{12}\frac{L/R_K}{R_KC_\mu}
+\frac{1}{180}\right]\left(R_KC_\mu\omega\right)^4
\label{eq:decoherence:plasmon-velocity:b}\\
&+\mathcal{O}\left((\omega R_KC_\mu)^6\right)
\end{align}
\end{subequations}
where $R_KC_\mu$ is the low-frequency time of flight $l/v_0$.
This expansion directly connects the discrete circuit element 
parameters 
$L$ and
$C$ to the low-frequency behavior of $v(\omega)$. 
The value $L=C_\mu R_K^2/12$ corresponds to a frequency dependency
$v(\omega)=v_0+\mathcal{O}\left((R_KC_\mu\omega)^4\right)$.
For $0\leq L<C_\mu R_K^2/12$, the velocity of edge magnetoplasmons 
starts first to increase quadratically at low-frequency, 
whereas $v(\omega)$ directly starts decreasing for $L>C_\mu R_K^2/12$. 
Note that a higher inductance contributes to a stronger slow-down
of the edge magnetoplasmons
with increasing frequency, as expected for an inductive effect. 
The order $4$ term given by 
Eq.~\eqref{eq:decoherence:plasmon-velocity:b} 
describes the behavior of
the plasmon velocity beyond this first order and contributes to
its decrease with increasing frequency. 

Coming back to the electronic inelastic scattering probability 
given by Eq.~\eqref{eq:decoherence:inelastic-development}, 
the case where $L=R_K^2C_\mu/12$ minimizes
its growth: the first non zero term is at order $(\omega\tau_0)^{10}$. 
This reflects the fact that for $L=R_K^2C_\mu/12$, the distorsion of a 
percussional current pulse is minimal at low-frequencies. 

When $\alpha_3\neq 0$, the first and second non trivial orders in 
$\omega\tau_0$ compete as soon as $\alpha_3\alpha_5<0$, which means that
they also compete in the expansion of the edge-magnetoplasmon time
of flight as a function of frequency. This is the case when using the
phenomenological form for the edge-magnetoplasmon velocity given by 
Eq.~\eqref{eq:phenomenology:nu=1:velocity}.

\begin{figure}
\begin{tikzpicture}
[
edge channel/.style={%
			thick
		},
		edge channel dir/.style={%
			thick,
			decoration={markings,mark=at position 0.55 with 
{\arrow{stealth}}},
			postaction={decorate}%
		},
		edge channel dir param/.style={%
			thick,
			decoration={markings,mark=at position #1 with {\arrow{stealth}}},
			postaction={decorate}%
		}%
	]

\begin{scope}[shift={(-3,0)}]
\def\lc{2}
\def\l{0.75}
	\draw[fill=gray!50!white] (-\l,-\l/2) rectangle (0,\l/2);
	\draw[fill=gray!50!white] (\lc+\l,-\l/2) rectangle (\lc,\l/2);
	\draw[edge channel dir] (0,0) -- (\lc,0);
	\draw[thick] (-\l/2,-\l/2) -- (-\l/2,-\l) -- (\lc+\l/2,-\l) -- 
(\lc+\l/2,-\l/2);
	\draw[thick] (\lc/2,-\l) -- (\lc/2,-3*\l);
	\begin{scope}[shift={(\lc/2,-2*\l)}]
		\draw[fill=white] (0,0) circle (\l/2);
		\draw[samples=200,domain=-pi:pi,thick] 
		plot({\x/12},{1/8*sin(\x*180/pi)});
		\node[right] () at (\l/2,0) {$V(t)$};
	\end{scope}
	\begin{scope}[shift={(\lc/2,-3*\l)}]
		\draw[thick] (-\l/2,0) -- (\l/2,0);
		\foreach \i in {1,...,4}{
			\draw[thick] (-\l/2-\l/10+\i*\l/4,0) -- 
({-\l/2-\l/10+(\i-1)*\l/4},-\l/4);
			}
	\end{scope}
\end{scope}

\begin{scope}[shift={(1.5,1)}]
 \def\lcapa{0.5}
 \def\sepcapa{0.2}
 \def\lind{0.7}
 \def\lr{0.6}
 \def\sr{0.15}
 \def\lc{1.2}
 \def\l{0.5}
 
 	\begin{scope}[shift={(\lc/2,\l/2)}]
		\draw[thick] (-\l/2,0) -- (\l/2,0);
		\foreach \i in {1,...,4}{
			\draw[thick] (-\l/2-\l/10+\i*\l/4,0) -- 
({-\l/2-\l/10+(\i+1)*\l/4},\l/4);
			}
  \end{scope}
 \draw[thick] (\lc/2,0) -- (\lc/2,\l/2);
 \draw[thick] (0,-\l/2) -- (0,0) -- (\lc,0) -- (\lc,-\l);
 \draw[thick] (\lc-\lcapa/2,-\l) -- (\lc+\lcapa/2,-\l);
 \draw[thick] (\lc-\lcapa/2,-\l-\sepcapa) -- 
(\lc+\lcapa/2,-\l-\sepcapa);
\node[right] () at (\lc+\lcapa/2,-\l-\sepcapa/2) {$C$};
 \draw[thick,decoration={aspect=0.25,segment length = 
4,coil},decorate] (0,-\l/2) -- (0,-\l/2-\lind);
\node[left] () at (-\lcapa/2,-\l/2-\lind/2) {$L$};
 \draw[thick] (\lc,-\l-\sepcapa) -- (\lc,-\l-\lind) -- (0,-\l-\lind) -- 
(0,-\l/2-\lind);
 \draw[thick] (\lc/2,-\l-\lind) -- (\lc/2,-2*\l-\lind-\lr);
 \draw[thick,fill=white] (\lc/2-\sr,-3*\l/2-\lind) -- 
(\lc/2-\sr,-3*\l/2-\lind-\lr) -- (\lc/2+\sr,-3*\l/2-\lind-\lr) -- 
(\lc/2+\sr,-3*\l/2-\lind) -- cycle;
\node[right] () at (\lc/2+\lcapa/2,-3*\l/2-\lind-\lr/2) {$R_{q}$};
\draw[thick] (\lc/2-\lcapa/2,-2*\l-\lind-\lr) -- 
(\lc/2+\lcapa/2,-2*\l-\lind-\lr);
 \draw[thick] (\lc/2-\lcapa/2,-2*\l-\lind-\lr-\sepcapa) -- 
(\lc/2+\lcapa/2,-2*\l-\lind-\lr-\sepcapa);
\node[right] () at (\lc/2+\lcapa/2,-2*\l-\lind-\lr-\sepcapa/2) 
{$C_{\mu}$};
 \draw[thick] (\lc/2,-2*\l-\lind-\lr-\sepcapa) -- 
 (\lc/2,-4*\l-\lind-\lr-\sepcapa) ;
 	\begin{scope}[shift={(\lc/2,-3*\l-\lind-\lr-\sepcapa)}]
		\draw[fill=white] (0,0) circle (\l/2);
		\draw[samples=200,domain=-pi:pi,thick] 
		plot({\x/16},{1/12*sin(\x*180/pi)});
		\node[right] () at (\l/2,0) {$V(t)$};
	\end{scope}
	\begin{scope}[shift={(\lc/2,-4*\l-\lind-\lr-\sepcapa)}]
		\draw[thick] (-\l/2,0) -- (\l/2,0);
		\foreach \i in {1,...,4}{
			\draw[thick] (-\l/2-\l/10+\i*\l/4,0) -- 
({-\l/2-\l/10+(\i-1)*\l/4},-\l/4);
			}
	\end{scope}
\end{scope}

\end{tikzpicture}
	\caption{\label{fig:effective-circuit} (Color online) Left panel: 
effective dipole
associated with the interaction region. Right panel: equivalent 
effective
$ZC$-circuit at low-frequency. The resistive part of $Z(\omega)$ is the
relaxation resistance $R_q=R_K/2$ and its imaginary part comes from an 
$LC$
circuit.}
\end{figure}
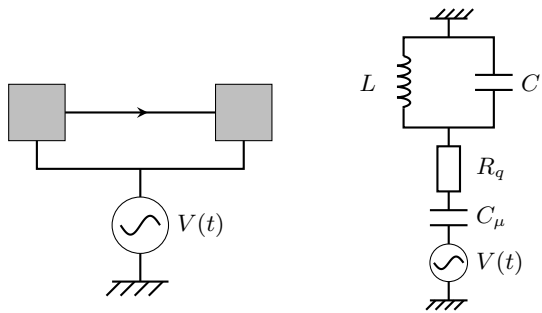

\subsubsection{Numerical results at $\nu=1$}
\label{sec:decpherence_nu_1:numerics}

Let us now illustrate these discussions by using an $\omega$-dependent
unit modulus transmission $t(\omega)$ given by the long-range
interaction model of Sec.~\ref{sec:models:nu=1}.
The corresponding $R_KC_\mu$ time is $l/v_0$ and expressions for the
inductance $L$ and capacitance $C$ of the discrete element circuit 
are given by Eq.~\eqref{eq:circuits:nu=1:parameters}.  
We will discuss both the case of a strong-coupling material 
($\alpha=0.75$) and of a weak-coupling material
($\alpha\simeq 0.05$).

\begin{figure}
\includegraphics{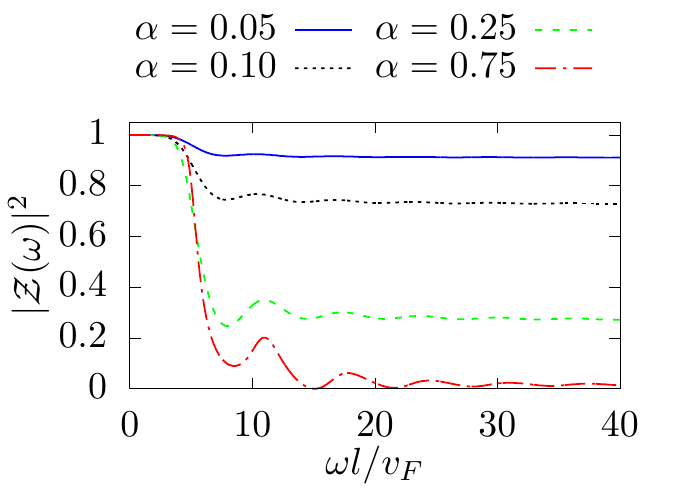}
\caption{\label{fig:decoherence:nu=1:result:elastic-probability} 
(Color online)
Elastic scattering probability for a single-electron excitation 
as a function of $\omega l/v_F$ for the long-range interaction model 
given by Eq.~\eqref{eq:appendix:nu=1:result}, for different values of 
the coupling constant $\alpha$. }
\end{figure}

Fig.~\ref{fig:decoherence:nu=1:result:elastic-probability} presents 
the elastic scattering probability $|\mathcal{Z}(\omega)|^2$
as a function of $\omega R_KC_\mu=\omega l/v_0$ for these two values 
of the coupling constant
as well as for intermediate values $\alpha=1/10$ and
$\alpha=1/4$. 
At strong coupling, 
the low energy almost flat plateau close to unity is followed by a very 
strong 
decay of $|\mathcal{Z}(\omega)|^2$ when $\omega R_KC_\mu\gtrsim 2\pi$ 
towards a very 
small value. The low-coupling case also leads to a decay of the elastic 
scattering probability when $\omega R_KC_\mu\gtrsim 2\pi$ but towards a 
higher value, $Z_{\infty}\simeq 0.9$. 

\begin{figure}
\includegraphics[width=7cm]{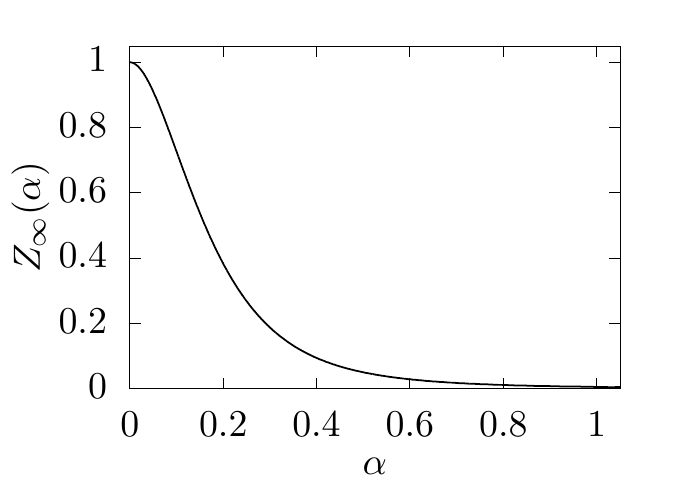}
\caption{\label{fig:Z-infinity}
(Color online) Asymptotic elastic scattering probability for high 
energy electrons $Z_\infty(\alpha)$ given by Eq.
\eqref{eq:Z-infinity} as a function of the coupling constant $\alpha$ 
for the
model introduced in Sec. \ref{sec:models:nu=1}.
The inset shows the relaxation tail $d(\omega)$ defined by Eq.
\eqref{eq:decoherence:nu=1:relaxation-tail} which gives the probabilituy
distribution for energy loss $\hbar\omega$ by an incoming very high
energy electron as a function of $\omega l/v_F$ for the same values as 
Fig.~\ref{fig:decoherence:nu=1:result:elastic-probability}, with the 
same color code.
}
\end{figure}

Fig \ref{fig:Z-infinity} depicts the asymptotic value $Z_\infty$ of 
the elastic scattering probability $|\mathcal{Z}(\omega)|^2$ at high
energy as a function of the coupling constant $\alpha$ in the model of
Sec. \ref{sec:models:nu=1}. Note that this is also the asymptotic 
elastic
scattering probability for a finite energy single electron excitation
in the limit $l\gg v_0/\omega_0$. 
We clearly see the difference between weak
and strong coupling on electronic decoherence of high energy 
excitations: for $\alpha=0.05$, $Z_\infty\simeq 0.91$ whereas for
$\alpha=0.75$, $Z_\infty\simeq 0.015$.

\begin{figure}
\includegraphics{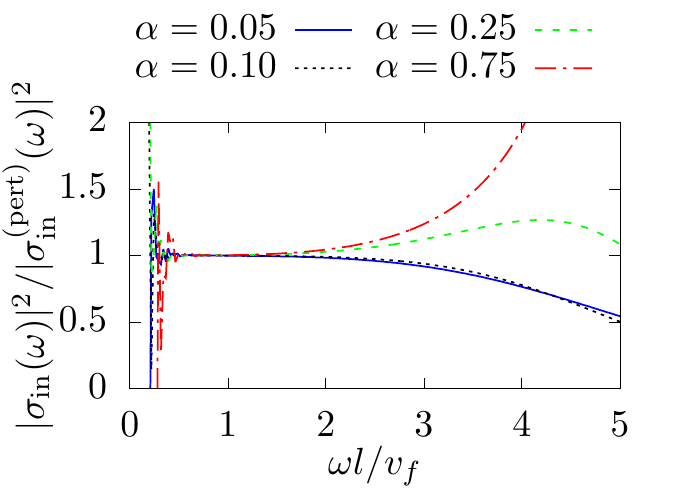}
\caption{\label{fig:decoherence:nu=1:result:perturbative}
(Color online) Ratio of inelastic scattering probabilities for the full 
model to its 
perturbative circuit expansion 
\eqref{eq:decoherence:inelastic-development}
at low energy. Numerical errors at small $\omega l/v_F$ are due to the
rapid decay of the
dominant
$(\omega l/v_F)^6$ asymptotic behavior of the inelastic scattering 
probability
at very low energies.
}
\end{figure}

Fig. \ref{fig:decoherence:nu=1:result:perturbative} 
depicts the ratio
of the full inelastic
scattering probability to the perturbative expression as a function of
$\omega$. It shows that
the perturbative result
is only valid at low energies, that is significantly 
before the drop of the elastic scattering 
probability, when the inelastic scattering probability
is still very close to unity. 
Understanding the full behavior of the elastic scattering
probability indeed requires a full non perturbative approach even at 
weak
coupling because, at higher injection energies, multiple low energy
electron/hole pair emissions coexist
with the emission of single electron/hole pair of higher energy. 
Properly
accounting for all these processes requires the full knowledge of
the frequency dependance of $g(\omega)$ for which the simplest discrete
element circuit descriptions are not sufficient.

\begin{figure}
\includegraphics{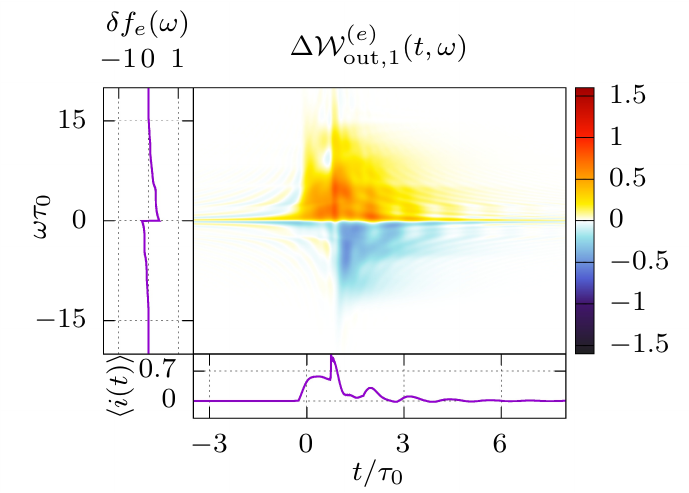}
\includegraphics{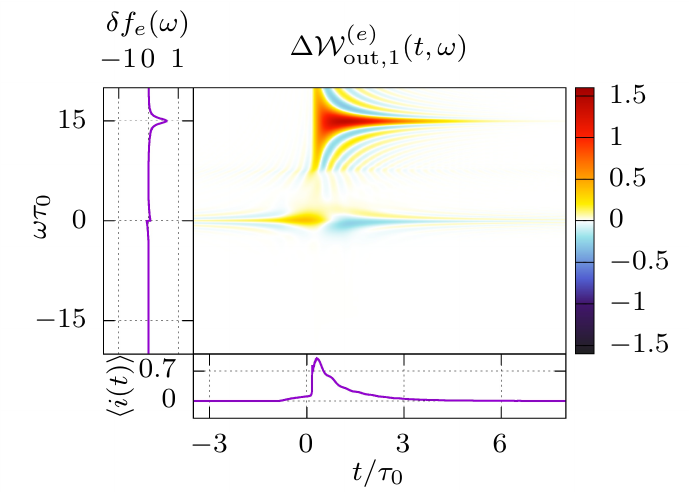}
\caption{\label{fig:decoherence:nu=1:result:high-energy} 
(Color online) Wigner distribution function of an incoming wavepacket 
injected at
energy $\omega_{0}R_K C_\mu=15$. 
Top panel: outgoing single-electron
coherence for $\alpha=0.75$.
Bottom panel: outgoing
single-electron coherence for $\alpha=0.05$.
$t=0$ corresponds to the expected time of reception for a free 
propagation at the low energy velocity $v_0$.
}
\end{figure}

Fig.~\ref{fig:decoherence:nu=1:result:high-energy} presents the 
electronic
decoherence of an incoming wavepacket injected at energy 
$\omega_{0} R_K C_\mu=15$. In the weak-coupling case, 
we clearly see the separation in energy between the elastically 
scattered electronic excitation together with its relaxation tail at 
high energy 
and the resulting electron/hole pairs close to the Fermi level. 
This is expected
since 
the elastic scattering
probability is quite high at the injection energy. 
The temporal separation which is a result of the difference between the
hot-electron velocity $v_\infty$ and the plasmon velocity $v_0$ is also
clearly visible on the average electric current $\langle i(t)\rangle$: 
the sharp
rise of the current corresponds to the arrival of the elastically
scattered quasi-particle and
$t=0$ corresponds to propagation at the fastest velocity $v_0$.  

By contrast, in the strong-coupling case, electronic decoherence is
much stronger. 
The relaxation tail of the incoming excitation is visible 
as a sharp rise of the current which arrives 
later than the beginning of the neutral electron/hole pair cloud. As 
expected the difference
between the plasmon and high-energy electron velocities is also more 
important than
in the weak-coupling case.

\begin{figure}
	\includegraphics{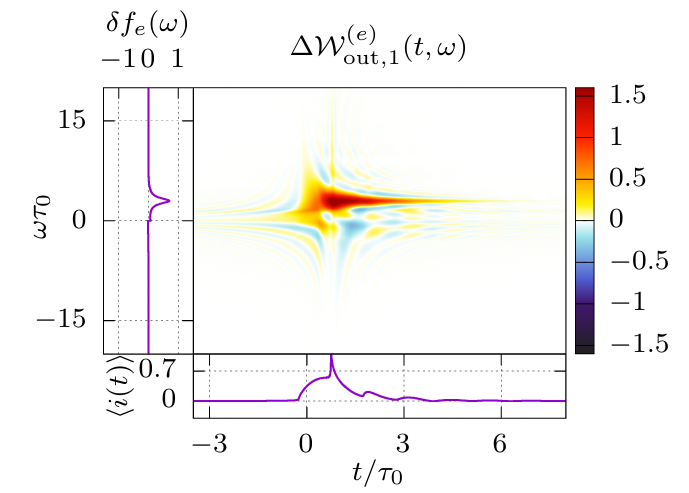}
	\includegraphics{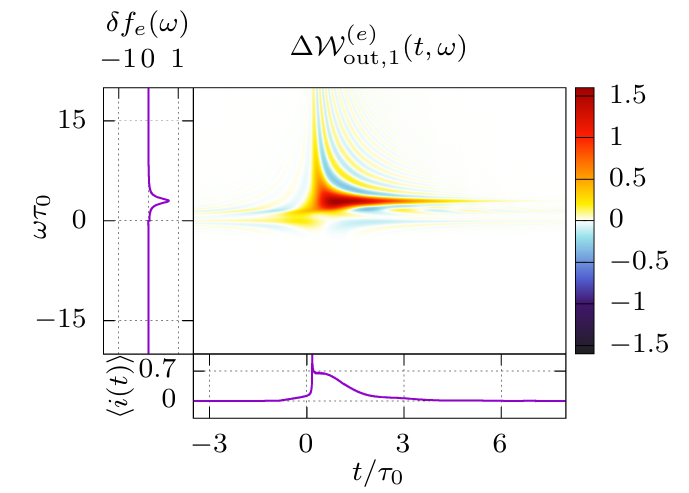}
\caption{\label{fig:decoherence:nu=1:result:low-energy} 
(Color online) Wigner distribution function of an incoming wavepacket 
injected at
energy $\omega_{\mathrm{e}}R_KC_\mu=3$. 
Top panel: outgoing single-electron
coherence for $\alpha=0.75$.
Bottom panel: outgoing 
single-electron coherence for $\alpha=0.05$.
$t=0$ is the expected time of reception for a free propagation
at the low energy velocity $v_0$.
}
\end{figure}

These results can be compared to the ones 
depicted on Fig.~\ref{fig:decoherence:nu=1:result:low-energy} which 
presents
the electronic decoherence of an incoming wavepacket injected at
$\omega_{0}R_K C_\mu\simeq 3$, an energy lower than the 
previously discussed threshold. 
Most of its spectral
weight is below the threshold.
The Landau quasi particle 
propagates without experiencing much decoherence in both cases. 
We also see that it propagates at the low energy edge-magnetoplasmon
velocity $v_0$. 
As expected, the incoming excitation seems less altered 
at weak coupling ($\alpha=0.05$) than at strong coupling 
($\alpha=0.75$).

\begin{figure}
 \includegraphics{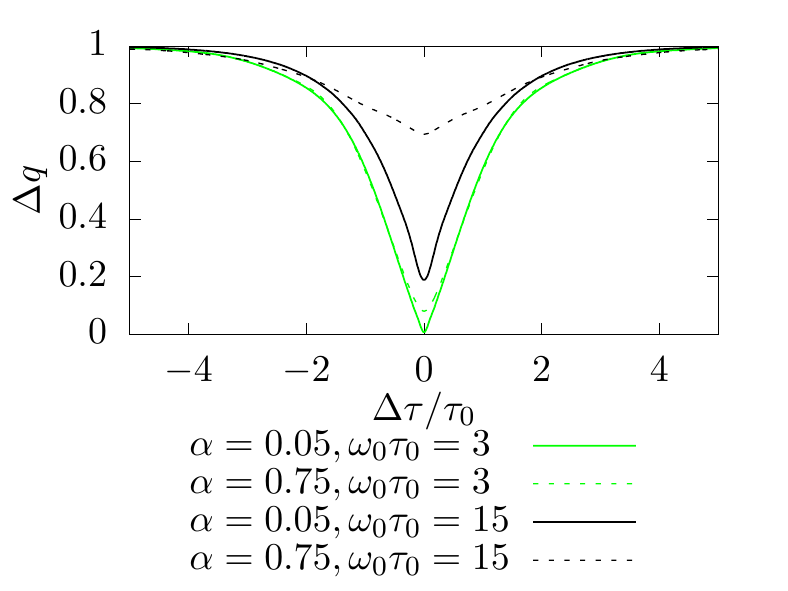}
 \caption{\label{fig:HOM:resultsnu1}
 (Color online) Theoretical results of an Hong-Ou-Mandel interferometry 
experiment obtained from the Wigner functions displayed in 
Figs.~\ref{fig:decoherence:nu=1:result:high-energy} and 
\ref{fig:decoherence:nu=1:result:low-energy}. 
As expected from the Wigner functions themselves, low 
energy excitations ($\omega_0\tau_0=3$) present a high contrast HOM 
dip. The results for high-energy exitations ($\omega_0\tau_0=15$) are 
clearly different between a weak coupling ($\alpha=0.05$) and a strong
coupling ($\alpha=0.75$) material, thus providing a clear 
signature of the protection against decoherence offered by weak coupling
materials.}
\end{figure}

The main tool to test robustness to decoherence that can be used in 
electron quantum optics is an Hong Ou Mandel 
experiment\cite{Bocquillon:2013-1,Olkhovskaya:2008-1}. It is then 
natural to think that strong and weak coupling regimes would lead to 
quantitatively different results in such experiments. In order to
answer this question, we have computed the HOM signal, which is the
excess HOM normalized noise obtained as the overlap of the incoming 
Wigner functions \cite{Ferraro:2013-1}, in both cases. Results are 
shown on Fig.~\ref{fig:HOM:resultsnu1} for both injection energies and 
both coupling values. As was discussed when looking at the Wigner 
functions, theses curves confirm that weak coupling materials would lead 
to a stronger protection against decoherence.

\subsubsection{Commenting on $\mathrm{AsGa}$ vs
graphene.}

As discussed above, exfolliated graphene on a silicon oxyde surface 
may correspond to a weak coupling
value of $\alpha$ and thus to much lower electronic decoherence. 
Moreover, 
provided velocities in graphene are much higher than in 
$\mathrm{AsGa}$, the
crossover energy between the low and high energy regimes should be much 
higher
for fixed device dimensions. For example, a $l=\SI{20}{\micro\meter}$ 
propagation
distance corresponds to $\omega/2\pi=v_F/l\simeq
\SI{500}{\giga\hertz}$ for $v_F=\SI{e6}{\meter\per\second}$ and to 
$\SI{50}{\giga\hertz}$ for $v_F=\SI{e5}{\meter\per\second}$.

The single-electron source based on the mesoscopic capacitor that has
been developed in $\mathrm{AsGa}$ generates electronic excitations at 
an energy comparable to this crossover scale. With our estimated
parameters, strong electronic decoherence is expected for a propagation 
above $\SI{30}{\micro\meter}$ when injecting at an energy of the order 
of $\SI{40}{\micro\electronvolt}$\footnote{These figures correspond to 
the ideal 
$\nu=1$ case which is not the case that has been experimentally 
studied. In the
experiments, extrinsic decoherence indubed by the second edge channel 
leads to
much shorter coherent propagation distance for such energy resolved 
excitations.}. 
Although no single-electron source has been developed yet for
graphene in the quantum Hall regime, 
the ratio of estimated high-energy velocities in the two materials 
suggests
a propagation 
distance of the order of $\SI{200}{\micro\meter}$ in a $\nu=1$ ideal 
channel before any significant step in the inelastic scattering
probability
manifests itself in graphene. 
Moreover, as discussed in the previous section,
even for such long propagation distances,
electronic decoherence would be much lower in a weak coupling material
compared to the case of a strong coupling material (see Fig.
\ref{fig:Z-infinity}). 

Of course, this discussion has been made within the framework of our model
for electronic propagation within an ideal $\nu=1$ edge channel.
In practice,
it is known that edge magnetoplasmons propagating along quantum
Hall edge channels experience
dissipation\cite{Volkov:1988-1,Kumada:2011-1,Bocquillon:2013-2,Petkovic:2014-1,Kumada:2014-1}. 
This is one of the possible causes for missing energy in 
electronic relaxation experiments\cite{Degio:2010-4}.
Investigating edge-magnetoplasmon dissipation effects on 
single-electron decoherence is certainly very important but this would
go beyond the scope of the present paper. Nevertheless,
we think that the main
point stressed in the present paragraph, that is the effect of the Fermi
velocity difference on the coupling constant and on the length to time
scale conversion may
lead to important differences between strong and weak-coupling materials
concerning single-electron decoherence. 
As suggested by Fig. \ref{fig:HOM:resultsnu1}, HOM experiments may
offer clear discriminating signatures of weak versus strong coupling
materials but this would require the experimental development of single
electron sources for Landau quasi-particles injection in graphene
quantum Hall edge channels.

On the experimental side, a Mach-Zehnder interferometer has recently
been demonstrated with encapsulated monolayer graphene sheet embedded within
hexagonal boron nitride\cite{Wei:2017-1}. The beam splitters 
exploit same-spin intervalley
scattering at a pn junction and the interferometer's geometry is
controlled by Coulomb exchange interactions. Surprisingly, a contrast of
$90\ \%$ has been observed at low bias 
in a parameter regime where one arms consists
of one carrier edge channel and the other or two and for an arm length of
$\SI{1.2}{\micro\meter}$. Such a high contrast remains up to a bias
voltage larger than $\SI{200}{\micro\volt}$. 
Although decoherence mechanisms have not been yet studied in great
detail for this device, we think that such a surprisingly high contrast as well as our
discussion of coupling constant and high energy velocity effects
call for intensive studies of single-electron decoherence in a material such
as graphene. 

\subsection{Decoherence at $\nu=2$}
\label{sec:decoherence_nu_2}

Let us now turn to the $\nu=2$ case, which has already been studied in
relation with experiments\cite{Ferraro:2014-1,Wahl:2013-1}. In the
present case, we shall briefly recall the results obtained using the
dispersionless model for edge-magnetoplasmon scattering between two
strongly coupled copropagating edge channels (short-range interactions
in Sec.~\ref{sec:models:nu=2})
before discussing the influence of the finite range of interactions
in an Hong-Ou-Mandel experiment.

\subsubsection{Short-range interactions}
\label{sec:decoherence_nu_2:short-range}

\begin{figure}
 \includegraphics{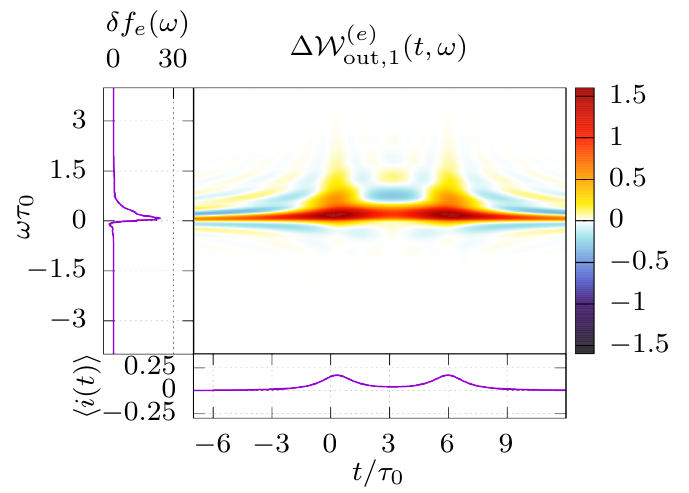}
 \includegraphics{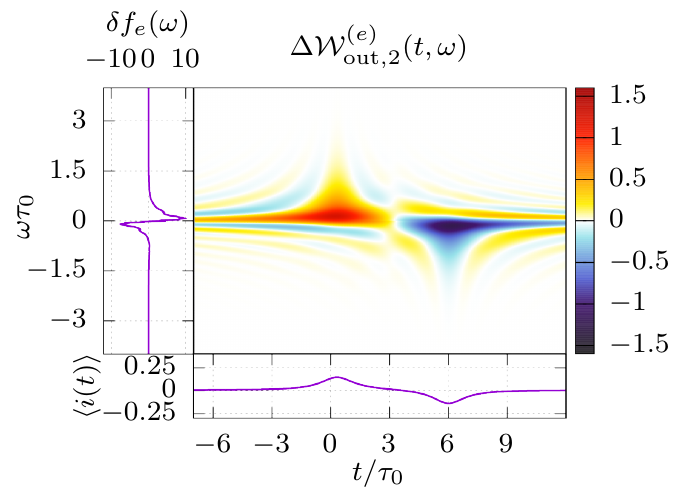}
\caption{(Color online) Wigner function for the outer (top) and inner (bottom) 
channel, for a Levitov excitation of width $\tau_0$. We use 
short-range interaction with parameters $\theta=\pi/2$, 
$\tau_+=6\tau_0$ and $\tau_-=\tau_+/20$. Since we inject a coherent 
state of plasmons, it fractionalizes into half-excitations and exhibits 
the behaviour of spin-charge separation, with the apparition of a fast 
symmetric mode over the two channels mode and a slow antisymmetric one. 
}
 \label{fig:inner_channel_lev}
\end{figure}

\begin{figure}
\includegraphics{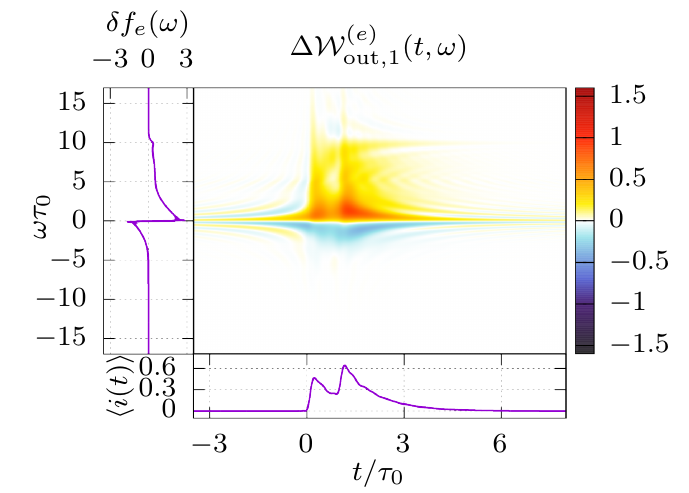}
\includegraphics{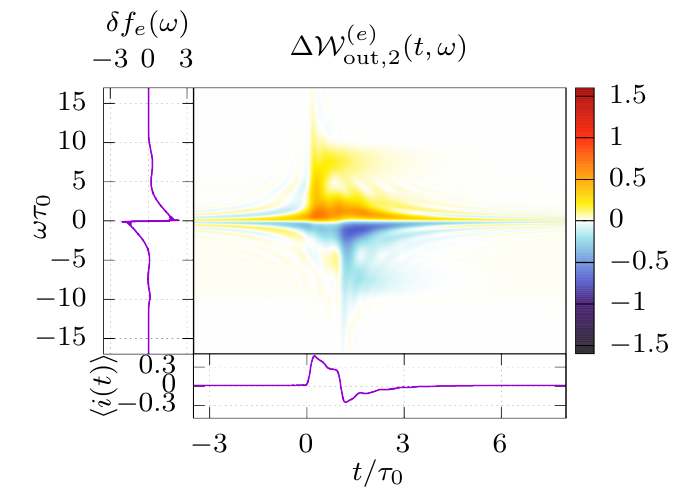}
\caption{(Color online) Wigner function for the outer (top) and inner (bottom) channel 
for a Landau excitation with parameters $\omega_0 \tau_0 =10$. 
Interaction parameters are $\theta=\pi/2$, $\tau_+=\tau_0$ and 
$\tau_-=\tau_+/20$. In that case, the incoming state is a superposition 
of coherent plasmonic states. Interactions lead to the destruction of 
coherences between those states, and the end result is therefore a 
statistical mixture of coherent plasmonic states, whose energy content 
is no more resolved around $\omega_0$. In the time domain, since all 
bosonic states exhibit spin-charge separation when they pass through 
the interaction region, we recover once again this type of separation 
for the electric current.}
 \label{fig:inner_channel_landau}
\end{figure}

Numerical results for both outer and inner channel coherences in the 
specific case of short-range interaction at strong coupling are 
presented on Fig.~\ref{fig:inner_channel_lev} for the Leviton source 
and on Fig.~\ref{fig:inner_channel_landau} for an energy-resolved 
excitation. Two distinct behaviours can be seen on these results. In the 
case of the Leviton source, the emitted state is a coherent state of 
plasmons created by the application of a classical voltage drive to an 
ohmic contact. Its evolution is dominated by fractionalization: we
observe
a simple separation of the incoming 
packet into two modes, one symmetric over the two channels and 
the other antisymmetric (see Fig.~\ref{fig:inner_channel_lev}). 
In the outer channel, we recover exactly a fractionalization of the 
incoming excitation into two Levitons with charges $-e/2$, as was 
predicted in various theoretical works\cite{Berg:2009-1,Grenier:2013-1} 
and demonstrated experimentally\cite{Bocquillon:2013-2,Kamata:2014-1,
Freulon:2015-1}.

As recalled in Sec.~\ref{sec:decoherence-physics}, a Landau type excitation 
illustrates a different scenario:
before fractionalization takes place, many-body decoherence 
leads to 
a fast energy relaxation with a strong decay of the 
weight around the injection energy, as can be seen on the upper panel 
of Fig.~\ref{fig:inner_channel_landau}.
This theoretical scenario and the corresponding quantitative 
predictions\cite{Ferraro:2014-1,Wahl:2013-1} have recently been 
confirmed by experiments \cite{Marguerite:2016-1}. The lower panel of 
Fig.~\ref{fig:inner_channel_landau} shows the electronic coherence in 
the inner channel. Although most excitations are created close to the 
Fermi level, we also see excitations created around the injection 
energy (for electrons) and close to the opposite (for the holes), which 
are the inner channel equivalent of the elastically scattered part 
still present in the outer channel.

\subsubsection{Long-range interactions}
\label{sec:decoherence_nu_2:long-range}

\begin{figure}
 \centering
\includegraphics{%
	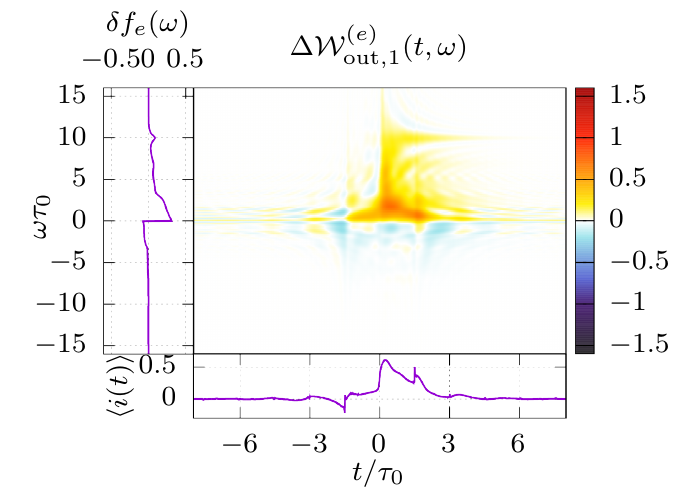}
\caption{(Color online) Wigner function of the outer channel for a Landau excitation 
with $\omega_0\tau_0=10$ going out of a long-range interaction zone in 
the strong interaction regime ($\theta=\pi/2$), with parameters 
$l/v_F=1.5\tau_0$, $\alpha_+=1/2$ and $\alpha_-=1/59$. The finite-frequency admittance of this interaction 
region has the same low energy limit than a short-range interaction 
region with parameters $\tau_+=\tau_0/2$, $\tau_-=\tau_+/20$.
Differences between the long and short-range cases are the apparition of
excitations at earlier times, three main 
peaks in the current instead of two, and a more complex 
pattern at low 
energies. }
 \label{fig:wigner:longrangenu2}
\end{figure}

At $\nu=2$, 
a long-range interaction model can be studied (see
Sec.~\ref{sec:models:nu=2}) and 
may be experimentally relevant at higher energies 
\cite{Bocquillon:2013-2}. The outgoing Wigner functions for excitations 
crossing a long-range interaction region are shown on 
Fig.~\ref{fig:wigner:longrangenu2}. Several 
qualitative differences with the short-range case can be observed. 
First, we see 
non-vanishing coherence and current at negative times, the reference 
being given by the time taken for a free excitation to cross this 
interaction region. This is due to the long-range characteristics of 
interactions: as soon as the incoming excitation enters the interaction 
region, it influences 
the whole 
interaction region and the contribution of the resulting low energy
electron/hole pairs can be seen near its ends. This means that a 
first current peak should begin at a time $\tau=l/v_F$ before 
the arrival of the ``real'' excitation as can be seen on the figure. 
Speaking of current, the bottom panel shows
that the outgoing current has three main peaks, compared to the two 
obtained in a short-range setting.

\begin{figure}
 \centering
 \includegraphics{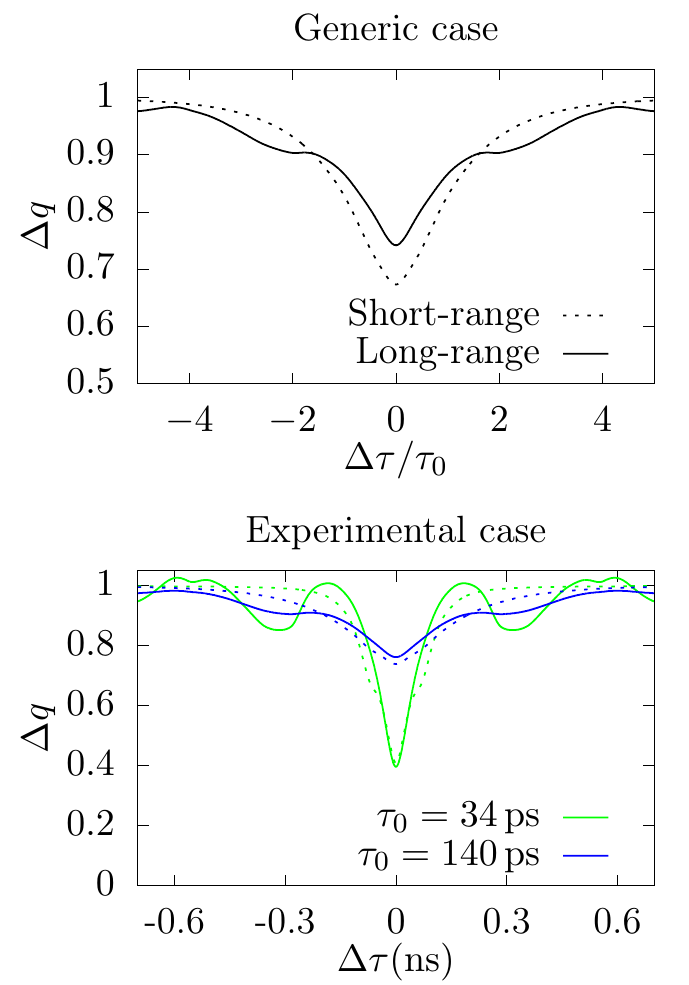}
\caption{(Color online) Top panel: Predicted results of an Hong-Ou-Mandel 
experiment after an interaction region in the short- and long-range 
cases, at $\nu=2$. The interaction parameters are the ones given in the 
caption of Fig.~\ref{fig:wigner:longrangenu2} and corresponds to 
interaction region of the same lengths and with the same low energy 
behaviour in terms of velocities. The main difference between the
predictions of the two models are the depth of HOM dip at 
$\Delta\tau=0$ and the secondary dips at $\Delta\tau= \pm1.5\tau_0$ in 
the long-range case which are due to low energy side excitations seen 
on Fig.~\ref{fig:wigner:longrangenu2}. The wider time spreading of the 
outgoing coherence also leads to a wider HOM dip.
Bottom panel: Plot of the HOM curves for the long (full lines) and 
short range (dotted lines) models with parameters corresponding to
the experiment \cite{Marguerite:2016-1}.
}
 \label{fig:HOM:resultsnu2}
\end{figure}

It is then natural to ask wether or not these differences can
be detected by an HOM experiment. To answer this question, the top 
panel of
Fig.~\ref{fig:HOM:resultsnu2} displays our prediction for both 
the short and long-range interaction models assuming interaction regions of 
the same length and the same incoming excitations. As seen from this
figure, these two interaction models lead to qualitatively different HOM 
curves: the long-range one shows a wider dip, as expected of the 
wider time spreading of the outgoing excitation and 
more ``secondary dips'' than the short-range model. 
This last feature can be traced back to 
the three main peaks in the outgoing Wigner function
computed using
the long-range model
compared to the two peaks of short-range interactions.

To comment on the experimental state of the art\cite{Marguerite:2016-1,Freulon:2015-1}, 
we have plotted on the
bottom panel of Fig.~\ref{fig:HOM:resultsnu2} the HOM predictions for
parameters corresponding to the recently published experimental results
in Ref.\cite{Marguerite:2016-1} Unfortunately the 
side lobes that would enable
us to differentiate between the two interaction models occur for a
time shift comparable or greater than $\SI{300}{\pico\second}$. 
However, 
probing time shifts larger than
$\SI{200}{\pico\second}$ brings us to values too close to the
half-period of the drive which is typically $\SI{1}{\nano\second}$.
In such situations, it is not possible anymore to forget about the
excitation emitted in the other half period: we cannot rely
on a single-electron
decoherence computation for a quantitative theory/experiment comparison.
Probing such large time shifts while comparing to our present
theoretical predictions would therefore require lowering the drive frequency $f$
thus deteriorating the signal to noise ratio of the low-frequency
current noise measurements.

In our opinion, this calls for complementary
investigations and/or experimental developments 
in order to determine which interaction model for the $\nu=2$ edge channel
system would
be the best at
reproducing the full HOM curves in detail. By contrast, samples
specifically designed for blocking relaxation processes are
likely to give much stronger experimental signatures as will be
discussed in the forthcoming section.

\section{Decoherence control}
\label{sec:decoherence-control}

In this section, we will consider passive decoherence control by sample
design in the $\nu=2$ edge channel system. The idea is to combine the
efficient screening of the edge channel used to propagate the injected
electronic excitation to the blocking of energy transfers by closing the
other edge channel. In a first experiment\cite{Altimiras:2010-2}, electronic relaxation in 
the outer edge channel has been partially blocked by letting the outer 
channel propagate along a closed inner
edge channel as depicted on Fig.~\ref{fig:interactiontypes}(a). 
In a more recent Mach-Zehnder interferometry experiment,
electronic decoherence has been partially blocked by bordering the
propagating edge channel by closed loops \cite{Huynh:2012-1} as depicted
on Fig.~\ref{fig:interactiontypes}(b).

We shall now discuss electronic decoherence within both types of
samples. We will first discuss what happens to Levitons
by looking at edge-magnetoplasmon scattering in the time domain.
Understanding this scattering in the frequency domain will then enable
us to discuss electronic decoherence of a Landau excitation injected at
various energies in Sec.~\ref{sec:decoherence-control:decoherence}. 
Finally, a realistic possible
sample design for probing the blocking of single-electron decoherence
with HOM interferometry will be discussed in 
Sec.~\ref{sec:decoherence-control:sample}.

\subsection{Magnetoplasmon scattering}
\label{sec:decoherence-control:scattering:time}

\subsubsection{Time domain}

Let us start by analyzing what happens to a percussionnal voltage pulse 
$V(t)=V_0\delta (t-t_0)$ sent across
this type of interaction zone. 
The outgoing voltage pulse can be obtained from the inverse Fourier 
transform of $t(\omega)$.
Using equation \eqref{eq:def_t_closedchannel}, we can rewrite the 
transmission coefficient in the generic case as
\begin{subequations}
\begin{align}
 t(\omega)
	&= S_{11}(\omega) \label{eq:closedloop:direct}
	\\
  &+\me^{\mi \omega \tau_L} S_{12}(\omega)S_{21}(\omega)
		\sum_{n=0}^{\infty}\me^{\mi n \omega\tau_L}S_{22}(\omega)^n
		\label{eq:closedloop:turn}
\end{align}
\end{subequations}
This expression has a clear physical meaning. Indeed, all excitations 
recovered in channel 1 after the interaction region of size $l$ 
correspond to one of the following paths: term
\eqref{eq:closedloop:direct} correspond to incoming excitations directly 
crossing
the region in channel 1 wheareas
terms \eqref{eq:closedloop:turn} corresponds to incoming excitations
generating
excitations in channel 2 ($S_{21}$) which go 
round the closed loop and create excitations back in 
channel 1 ($S_{12}$). This can either happen after one lap round the loop 
($\me^{\mi\omega\tau_L}$) or after $n+1$ laps, in which case we need 
to 
take into account the fact that excitations in channel 2 crossed the 
interaction region in the second channel $n$ times ($S_{22}^n$) and made $n$ 
more laps ($\me^{\mi n \omega \tau_L}$).

In the case of short-range interactions, $t(\omega)$ can be rewritten as
a sum of 
complex exponentials
\begin{align}
 t(\omega)
	&= p_+ \me^{\mi \omega \tau_+} + p_- \me^{\mi \omega \tau_-} \\
	&+ \sum_{n=0}^{\infty} \sum_{k=0}^{n+2} w_{n,k} 
	\me^{\mi \omega
			\left( (n+1)\tau_L+ k\tau_+ +(n+2-k)\tau_- \right)
			}
	\notag
\end{align}
where the weights $w_{n,k}$ are given by\footnote{In this equation, we 
adopt 
the convention that $\binom{n}{k}=0$ if $k>n$ or $k<0$.}
\begin{align}
 w_{n,k} 
 = q^2 
	\left[
		\binom{n}{k} p_{+}^{n-k}p_{-}^{k}
	+ \binom{n}{k-2} p_{+}^{n+2-k} p_{-}^{k-2}
	\right.
	\notag \\
	\left.
	-2\binom{n}{k-1} p_{+}^{n+1-k} p_{-}^{k-1}
	\right]
\end{align}
This equation shows that the outgoing voltage for 
a localized excitation of charge $-e$ created at time $t_0$ corresponds 
to the generation of a percussional current pulse with charge $-ep_{+}$ 
emitted at time $t_0+\tau_{+}$, another one with charge $-ep_{-}$ at 
time $t_0+\tau_{-}$, and an infinity of others at times $t_0 + 
(n+1)\tau_L + k\tau_+ + (n+2-k)\tau_-$ with charges $-ew_{n,k}$. Total 
current is conserved, since $p_{+}+p_{-}=1$ and $\forall n, 
\sum_{k=0}^{n+2}w_{n,k} = 0$.

\begin{figure}
 \begin{center}
	 \includegraphics{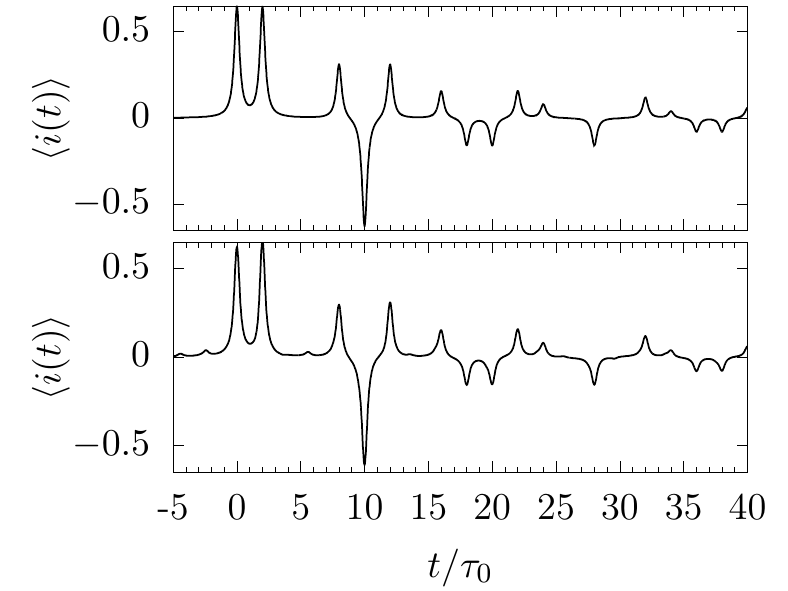}
	 \caption{Outgoing current for an incoming leviton excitation of 
width $\tau_0/4$ after an interaction with a closed loop. Parameters 
$\tau_{-}=\tau_0$, $\tau_{+}=3\tau_0$ and $\tau_{L}=7\tau_0$. Such 
parameters, while not experimentally reasonable, allow a good 
visualization of the physical properties of this current. Indeed, we 
see first the two peaks corresponding to standard fractionalization 
when crossing the interaction region, followed by a series of three 
peaks corresponding to excitations having crossed two times the 
interaction region and going round the loop once (first corresponds to 
two crossings in the symmetric mode, then one antisymmetric and 
one symmetric, third one is two crossings in antisymmetric mode), and 
so on. Top panel: as given by the analytical computation presented in 
this section. Bottom panel: as recovered when integrating the 
numerically obtained outgoing Wigner function over all energies.}
	\label{fig:leviton_current_closechannel}
 \end{center}
\end{figure}

For the Leviton source, with the exact same reasoning, the outgoing 
state is a sum of time-shifted Leviton excitations with suitable 
charges. Fig.~\ref{fig:leviton_current_closechannel} shows the outgoing 
current for this type of environment computed in two different ways. 
The top panel of Fig.~\ref{fig:leviton_current_closechannel} 
corresponds to an analytical computation of the expected current in the 
way we just exposed. The bottom panel 
is obtained from our numerical code used to compute the outgoing
single-electron coherence, by integrating
the resulting excess Wigner distribution function over the energy.
The very good  agreement between the two results illustrates the 
validity of our numerical approach.

\subsubsection{Frequency domain}

Let us now turn to the transmission 
coefficient as a function of energy. As stated before, since
$|t(\omega)|^2=1$, this system behaves as an 
effective $\nu=1$ system but it has a much richer texture than 
the model presented in Sec.~\ref{sec:models:nu=1}.

\begin{figure}
 \begin{center}
	 \includegraphics{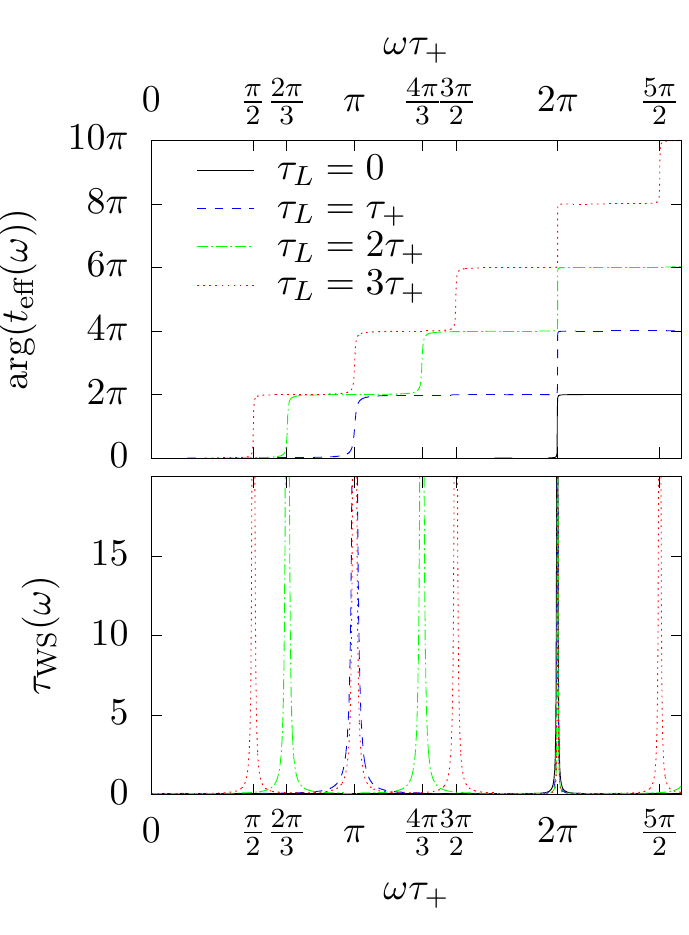}
	 \caption{(Color online) Phase of the transmission coefficient (top panel) and the 
associated dwelling time $\tau_{\text{WS}}(\omega)$(lower panel) 
for a short-range  interaction with weak coupling ($\theta=\pi/10$) and 
parameters $\tau_-=\tau_+/20$, for 4 different 
geometries for the loop. We see that the phase jumps each time 
$\omega(\tau_+ + \tau_L)\simeq 2n\pi$, with a stronger jump when 
$\omega\tau_+ = 2\pi$. These jumps are the signature of a quasi bound 
state (scattering resonance) at corresponding energy inside the loop.}
	\label{fig:transmission_lowcoupling}
 \end{center}
\end{figure}

First of all, let us consider short-range interactions at 
weak coupling. The closed inner channel can be seen as a 
Fabry-Pérot interferometer with low transparency on one side and totally
reflecting on the other part. The interaction region can then be viewed
as a cavity which is connected to a transmission line.
As in optics, the phase of its reflexion
coefficient, which is here the edge-magnetoplasmon transmission
$t(\omega)$, exhibits sharp resonances. They can
arise from quasi-bound scattering states within the interaction
region seen as a cavity, which appear as peaks in the Wigner-Smith time 
delay
\begin{equation}
\tau_{\text{WS}}(\omega)=\frac{1}{2\pi \mi}
\frac{\md \log{(t(\omega))}}{\md \omega}
\end{equation}
which represents a dwelling time within the cavity.
These resonances are sharply visible in the weak-coupling regime
presented on Fig.~\ref{fig:transmission_lowcoupling}. The top panel depicts the 
phase of $t_{\text{eff}}(\omega)=\me^{-\mi \omega \tau_-}t(\omega)$, 
and displays strong jumps of $2\pi$ every time 
$\omega(\tau_+ + \tau_L)\simeq 2n\pi$. These jumps lead to strong
resonances in the Wigner Smith time delay as seen on the lower panel.

\begin{figure}
 \begin{center}
  \includegraphics{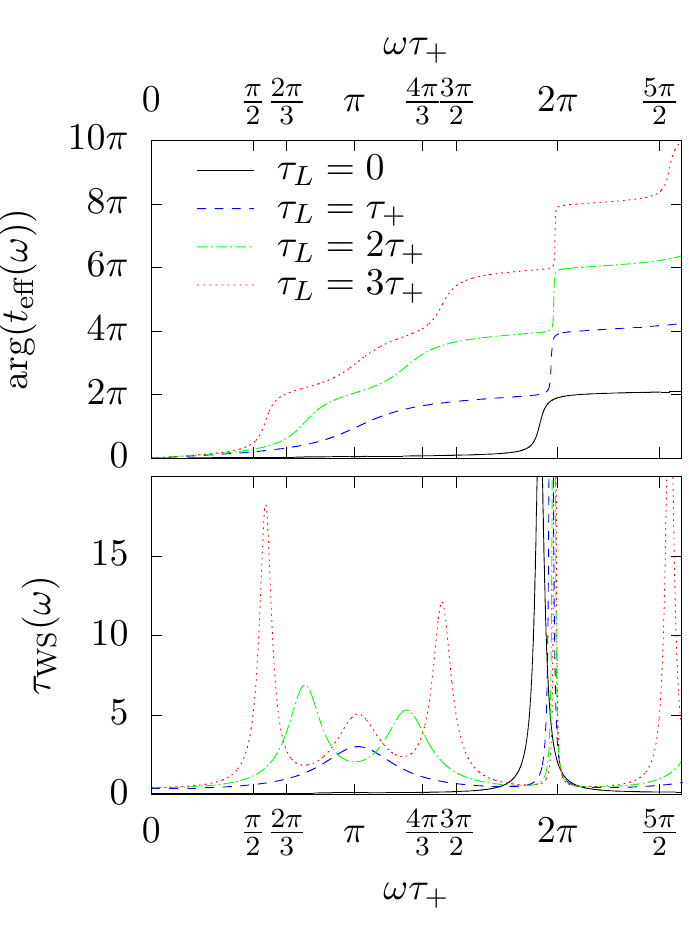}
	 \caption{(Color online) Phase of the transmission coefficient (top panel) and associated 
dwelling time in the closed inner channel (lower panel) for 
a short-range interaction with strong coupling 
($\theta=\pi/2$) and the same 4 
different geometries for the loop as the low-coupling case. We see that 
the phase does not go from one plateau to another, but still exhibits 
jumps at values close to the ones seen before, the jump at 
$\omega\tau_+=2\pi$ being once again the strongest.  The 
corresponding quasi bound states inside the loop are therefore 
broadened in energy.}
	\label{fig:transmission_highcoupling}
 \end{center}
\end{figure}

Let us now turn to the strong-coupling case ($\theta=\pi/2$). As is 
expected from the comparison with a Fabry-Pérot interferometer with 
higher transparency, the quasi bound states inside the loop are 
broadened in energy, as can be seen on Fig.~\ref{fig:transmission_highcoupling}. 

\subsection{Electronic decoherence}
\label{sec:decoherence-control:decoherence}

\begin{figure}
	\includegraphics{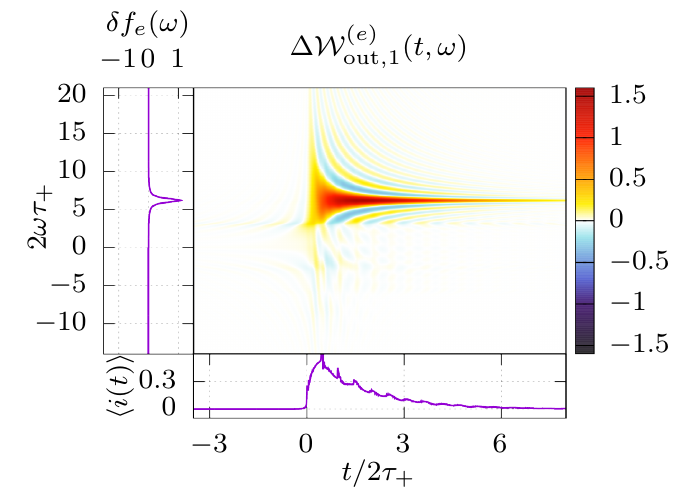}
	\includegraphics{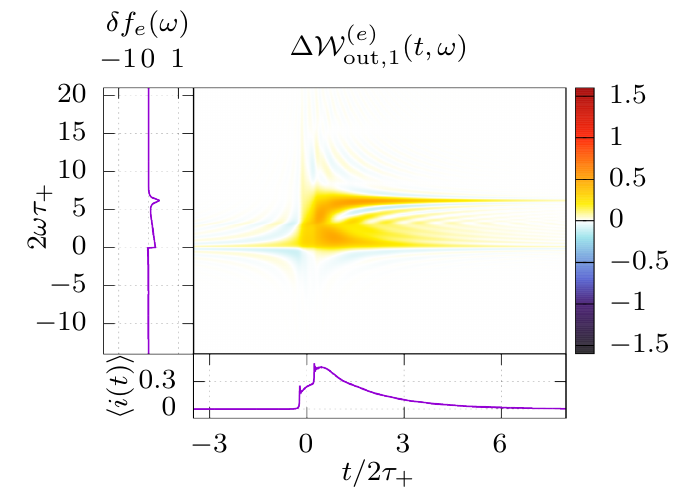}
\caption{(Color online) Outgoing 
Wigner function for an incoming Landau excitation of 
duration
$\tau_0=0.8\tau_+$. Interaction parameters are
$\theta=\pi/2$ and $\tau_-=\tau_+/10$. Top panel: 
short-range interaction with a closed environment of type (a) 
($\tau_L=0$). Bottom panel: copropagation along an open channel on the 
same distance with same interaction parameters. For both graphs, the 
incoming excitation is at an energy 
$\omega_0\tau_+=\pi$ below the energy resonances
of the loop. 
When interacting with a closed channel (upper 
panel), relaxation is highly suppressed compared to copropagation along 
an open channel (lower panel). Because the injection energy is below
closed channel resonances, the outgoing occupation number remains close 
to the incoming one. Electron/hole pair creation is responsible of the 
spikes that appear on the average electric current which are 
characteristic of the closed channel geometry. 
}
 \label{fig:wigner_strongcoupling_below}
\end{figure}

\begin{figure}
	\includegraphics{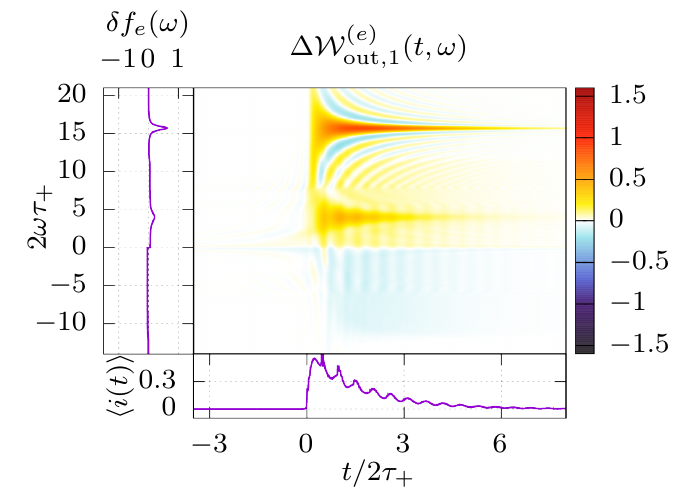}
	\includegraphics{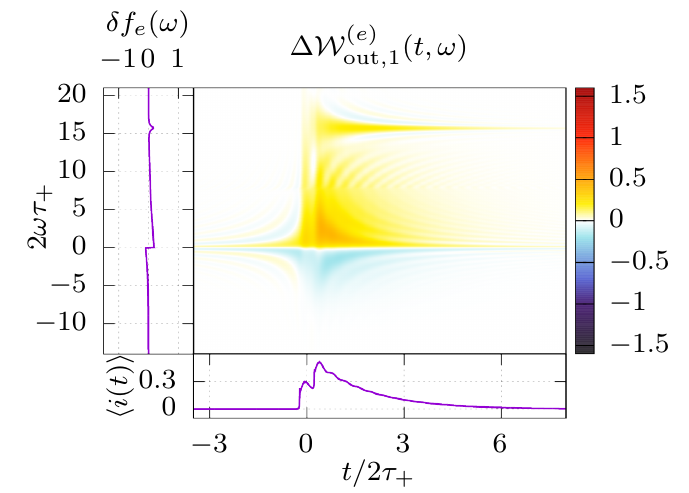}
	\caption{(Color online) Same as Fig.~\ref{fig:wigner_strongcoupling_below}, 
but for an incoming excitation above the resonance energy,
$\omega_0\tau_+=5\pi/2$.
Energy relaxation involves the emission of electron/hole pairs 
at the resonance energy, leading to a second peak in the energy 
distribution.}
 \label{fig:wigner_strongcoupling_above}
\end{figure}

We now discuss electronic relaxation and decoherence of Landau
excitations at strong coupling in the closed channel geometry depicted
on Fig.~\ref{fig:interactiontypes}(a) ($\tau_L=0$).
Numerical results for the Wigner function of an electron emitted below 
the energy of the first resonance of the closed resonator 
and one emitted between the first and the 
second resonances are shown on the upper panels of
Figs.~\ref{fig:wigner_strongcoupling_below} and 
\ref{fig:wigner_strongcoupling_above}. These results are compared, on the 
bottom panel of each figure, to a situation where the interaction 
region is of the same length but the inner channel is not closed onto 
itself. The geometry with a closed channel exhibit much less electronic 
decoherence in comparison with the open channel geometry.

In the first 
situation depicted on Fig.~\ref{fig:wigner_strongcoupling_below},
electron/hole pair generation is inhibited because the electronic energy
is off resonance with the cavity and therefore, relaxation is blocked.
As a result, no decoherence happens and the excitation leaves the 
interaction region pretty much unchanged.

\begin{figure}
	\includegraphics{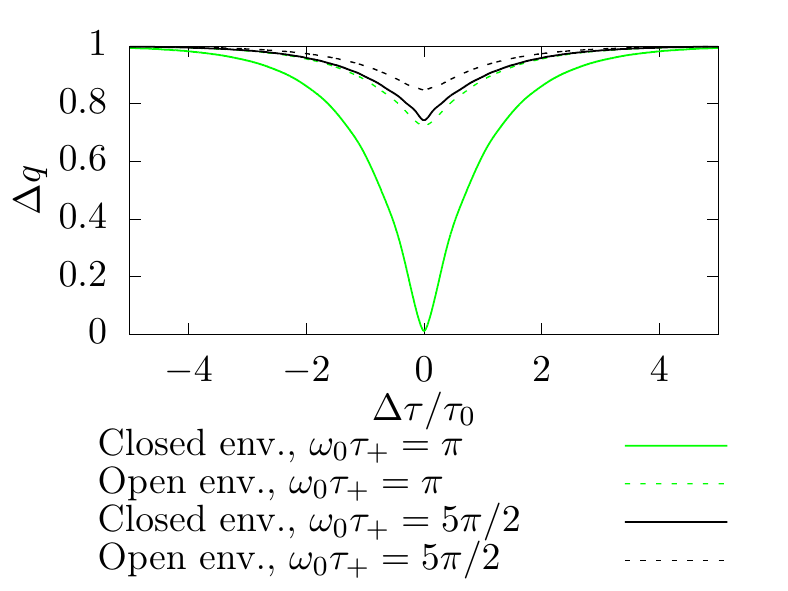}
 \caption{\label{fig:HOM_closechannel_strongcoupling}
(Color online) Results of an Hong-Ou-Mandel experiment for the 4 Wigner 
distributions presented in this section. The bigger depth of the HOM 
dip for the loop environment proves that closing the environment on 
itself provides a net advantage compared to the open case. 
Specifically, in the case where the excitation is emitted below the 
first level in the loop ($\omega\tau_+=\pi$), we see a dip going nearly 
all the way down to $0$, which denotes a quasi-complete protection from 
decoherence.}
\end{figure}

When the Landau
excitation is injected above the first resonance
(see upper panel of Fig.~\ref{fig:wigner_strongcoupling_above}),
it relaxes by emitting electron/hole pairs precisely at the
energy given by the first resonance. This relaxation leads to a peak in
the electronic distribution at the final energy of the electron, which 
is its injection energy minus the resonance energy.
The characteristic features of the interaction-generated
electron/hole pair cloud are the temporal oscillations of $\Delta
\mathcal{W}^{(e)}(t,\omega)$ for $\omega$ below the peak associated with the
relaxed electron. 
HOM
interferometry can then be used to check whether or not we are protected
against decoherence.
As shown on Fig.~\ref{fig:HOM_closechannel_strongcoupling}, 
the HOM dip for wavepackets propagating along a closed inner channel 
should be bigger that their opened counterpart, going even down close 
to zero for an excitation emitted below the first resonance.

\subsection{A realistic sample proposal}
\label{sec:decoherence-control:sample}

In this section, we discuss a possible geometry in which Landau excitations such 
as the one emitted by a single-electron source\cite{Feve:2007-1} would
be protected 
against decoherence. 

One may naively think that loops smaller than the size of 
dots used to emit the excitation would be needed, which seems unreasonable 
experimentally. Luckily, previous experimental studies\cite{Talyanskii:1997-1} have shown 
that the speed of electronic excitations in top-gated regions of the 
2DEG are smaller than the ``free'' velocity, a fact that can be 
checked using available experimental data on the energies of the 
quantum dot. The energy $\hbar\omega_0$ of 
Landau particles emitted by 
the dot used in Ref. \cite{Marguerite:2016-1} is around 
\SI{60}{\micro\electronvolt}, 
the size of the dot being \SI{2}{\micro\meter}, leading to a relevant 
velocity in gated region of the 2DEG $v^{\text{gate}}\sim
\SI{5.8e4}{\meter\per\second}$. 
The dwelling time of excitations in the dot is 
$\tau_0\simeq\SI{100}{\pico\second}$, leading to a typical width in 
energy of about $1/10$th of the injection energy. Consequently, 
a safe limit for blocking decoherence would be to have a loop 
such that $\omega_0 (\tau_+ + \tau_L) < 3\pi/2$. The edge-magnetoplasmon
modes populated within the incoming electronic excitations have
their energies
below the resonance, 
even when considering the resonance width.

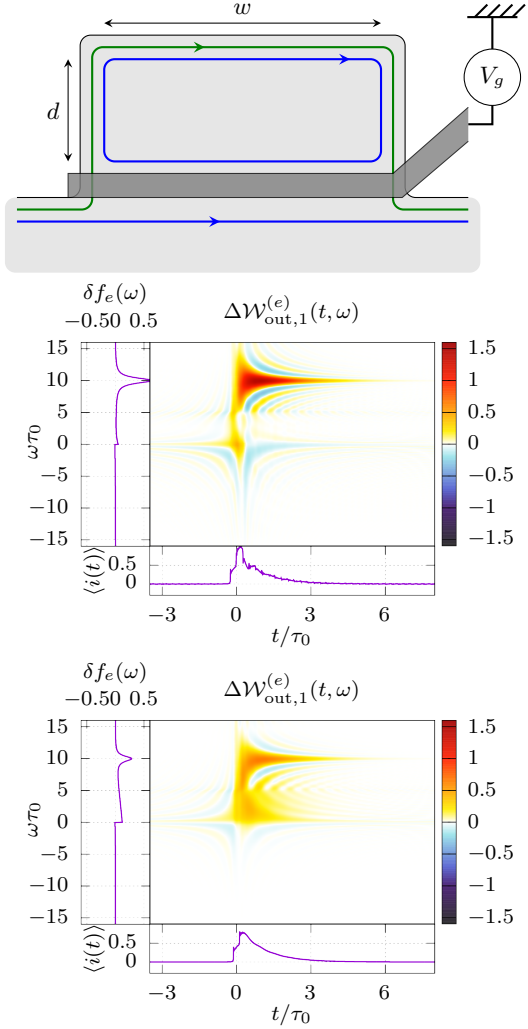
\begin{figure}
 \centering
 \begin{tikzpicture}
[
edge channel/.style={%
			thick
		},
		edge channel dir/.style={%
			rounded corners,
			thick,
			decoration={markings,mark=at position 0.45 with {\arrow{stealth}}},
			postaction={decorate}%
		},
		edge channel dir param/.style={%
			thick,
			decoration={markings,mark=at position #1 with {\arrow{stealth}}},
			postaction={decorate}%
		}%
	]

\def\l{2}
\def\xsource{0}
\def\xloop{0.5*\l}
\def\hloop{\l}
\def\wloop{2*\l}
\def\xhom{3*\l}
\def\wgate{0.16*\l}
\def\sep{0.08*\l}

\fill[rounded corners, gray!20!white]
(\xsource-\sep,-0.5*\l) -- (\xsource-\sep,0) --(\xloop-\sep,0) 
-- (\xloop-\sep,\hloop+\sep) -- ++(\wloop+2*\sep,0) 
--++(0,-\hloop-\sep) -- (\xhom+\sep,0) -- (\xhom+\sep,-0.5*\l) -- cycle;
\draw[rounded corners] (\xsource,0) --(\xloop-\sep,0) 
-- (\xloop-\sep,\hloop+\sep) -- ++(\wloop+2*\sep,0) 
--++(0,-\hloop-\sep) -- (\xhom,0);

\draw[edge channel dir, green!50!black] (\xsource,-\sep) -- 
(\xloop,-\sep) -- 
(\xloop,\hloop) -- ++ (\wloop,0) -- ++(0,-\hloop-\sep)
-- (\xhom,-\sep);

\draw[edge channel dir, blue] (\xsource,-2*\sep) -- (\xhom,-2*\sep);

\draw[edge channel dir, blue] (\xloop+\sep,\wgate+\sep) -- 
(\xloop+\sep,\hloop-\sep) -- 
(\xloop+\wloop-\sep,\hloop-\sep) -- 
(\xloop+\wloop-\sep,\wgate+\sep) -- cycle;

\draw[black,fill=gray, opacity=0.8] 
(\xhom,0.6*\l) -- (\xloop+\wloop,\wgate) -- (\xloop-\wgate,\wgate) -- 
(\xloop-\wgate,0) -- (\xloop+\wloop+\wgate*0.38,0) -- 
(\xhom,0.6*\l-1.41*\wgate);


\draw[thick] (\xhom,0.6*\l-0.705*\wgate) --++ (\wgate,0) -- 
(\xhom+\wgate,1.2*\l);
\node[draw, circle, fill=white] () at (\xhom+\wgate,0.8*\l) {$V_g$};

	\begin{scope}[shift={(\xhom+\wgate,1.2*\l)}]
	\draw[thick] (-\l/6,0) -- (\l/6,0);
	\foreach \i in {1,...,4}{
		\draw[thick] (-\l/6-\l/30+\i*\l/12,0) 
-- ({-\l/6-\l/30+(\i+1)*\l/12},\l/12);
		}
	\end{scope}

\draw[stealth-stealth] (\xloop-2*\sep,\wgate+\sep) 
--(\xloop-2*\sep,\hloop-\sep) node[midway,left] {$d$};
\draw[stealth-stealth] (\xloop+\sep,\hloop+2*\sep) 
--(\xloop+\wloop-\sep,\hloop+2*\sep) node[midway,above] {$w$};

\end{tikzpicture}
	\includegraphics{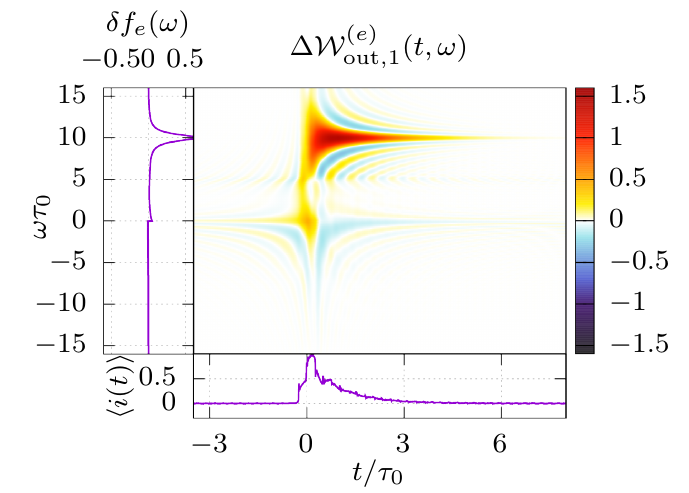}
 \includegraphics{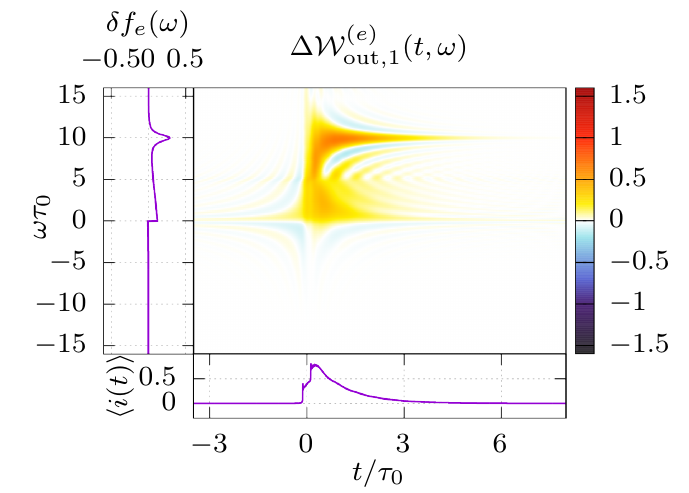}
	\caption{(Color online) Top panel: a possible experimental design 
for testing decoherence control on a Landau excitation. Here, 
the 2DEG (in light gray) defines a cavity delimited by a top 
gate shifting the electron density so that only the outer edge 
channel can pass through. This creates a region with a closed inner 
channel. The single-electron source as well as the QPC of the HOM probe 
should be located close to the loop.
The loop corresponds to 
$\tau_+=(w+2d)/v^{\text{chem.}}_{+}$ and $\tau_L=w/v^{\text{gate}}_{F}$, 
where $v^{\text{chem.}}_{+}$ denotes the speed of the slow mode in 
chemically defined edge channels, whereas $v^{\text{gate}}_{F}$ is the 
Fermi velocity in an edge channel propagating along a metallic gate.
Middle panel: outgoing Wigner function when
$w=\SI{1.5}{\micro\meter},h=\SI{0.5}{\micro\meter}$ for an incoming 
excitation with parameters $\omega_0\tau_0=10$ and 
$\tau_0=\SI{100}{\pico\second}$. The velocities are 
$v^{\text{chem.}}_{+}=\SI{1e5}{\meter\per\second}$ 
and $v^{\text{gate}}_{F}=\SI{5.8e4}{\meter\per\second}$. 
Bottom panel: Wigner function in the case where the gate closing the 
loop is used to either let both channels through or none (times of 
flight are equivalent in those two cases). Decoherence would be far 
more important in such cases where the inner channel is not closed on 
itself.}
 \label{fig:sampledesign}
\end{figure}

A sample design with a loop of total size 
\SI{4}{\micro\meter} is sketched on Fig.~\ref{fig:sampledesign}. We predict
protection against decoherence for the single-electron excitations we are 
interested in. Of course, by tuning the dot parameters for emitting 
excitations at lower energies, decoherence protection would still be 
possible even with two times larger loops 
(see Appendix 
\ref{appendix:experimentaldecohencecontrol}, Fig.~\ref{fig:appendix:biglooplowenergy}).
The design presented here would allow a test of 
decoherence protection for single-electron excitations emitted by the 
mesoscopic capacitor driven by square pulses. Electronic decoherence and 
relaxation of energy resolved single-electron excitations being 
stronger than for an out of equilibrium distribution generated by a 
biased QPC, such an experiment would provide a stronger test of the 
potential of sample design for decoherence protection. 

Finally, as was presented on Fig.~\ref{fig:wigner_strongcoupling_above}, larger loops with
$2\pi\le\omega_0(\tau_+ + \tau_L)\le4\pi$ give access to the physics 
of electronic excitations accompanied by a single plasmon around an
energy given by the first scattering resonance of the loop. 
This allows the probing of new hybrid quantum single-electron and 
single-plasmon excitations and calls for new protocol measurements to 
fully characterize these excitations.

\section{Conclusion}

To conclude, we have addressed the question of decoherence control 
for single-electron excitation propagating
within chiral edge channels. This work is focused on purely passive
decoherence control through 
the properties of the material itself and sample design.

To get an insight on the influence of the material, we have discussed electronic
decoherence within an ideal single chiral channel. Using a
semi-realistic model for long-range interactions, we have found that a
high bare Fermi velocity may be significantly more promising for limiting
decoherence because it leads to a lower coupling constant, a point that
has indeed been
overlooked, and because it
amplifies the distance covered within a given time. We have found that the
conjugation of these two effects could lead to a drastic decrease of electronic
decoherence over distances of $10$ to $\SI{100}{\micro\meter}$ as long as
dissipative effects could be neglected.
We think that this calls for more thorough experimental studies 
to explore the potential of different materials for electron quantum
optics. Moreover,
our analysis once
again stresses the importance of performing electronic decoherence
experiments in setups where finite-frequency a.c. transport could also
be measured. 

We have also shown that passive decoherence protection through sample design
could be tested for excitations emitted by the mesoscopic
capacitor in the single-electron regime using an HOM experiment. We have
proposed a
realistic design for demonstrating this effect. Moreover, our study suggests
that such sample could be used
for emitting single edge magnetoplasmons thus opening the way to 
hybrid electron-and-photon quantum optics.

\appendix
\section{Bosonization}
\label{appendix:bosonization}

Bosonization provides a description of a 1D chiral relativistic gaz of
fermions in terms of bosonic degrees of freedom corresponding to charge
density waves.
At a given chemical potential $\mu$, the excess charge density $n(x,t)$ 
is the normal ordered product $\normord{(\psi^\dagger\psi)}(x)$ with 
respect to the corresponding Fermi energy. It is expressed in terms of a 
quantum bosonic field $\phi$
\begin{equation}
\label{eq:bosonization:density}
\normord{(\psi^\dagger\psi)}(x,t)=\frac{1}{\sqrt{\pi}}
\,(\partial_{x}\phi)(x,t)
\end{equation}
whose mode decomposition can be written in terms of creation 
$b^\dagger(\omega)$ and destruction operators $b(\omega)$ called edge-magnetoplasmon modes
\begin{equation}
\label{eq:bosonization:bosonic-field}
\phi(x,t)=\frac{-\mi}{\sqrt{4\pi}}\int_{0}^{+\infty}
\left(b(\omega)\me^{\mi\omega
(x/v_F-t)}-\mathrm{h.c.}\right)\,\frac{\md\omega}{\sqrt{\omega}}\,.
\end{equation}
where $v_F$ denotes the Fermi velocity of fermionic excitations in this 
chiral channel. The edge-magnetoplasmon modes can be expressed in terms 
of the fermionic mode operators $c(\omega)$ and $c^\dagger(\omega)$ 
defined by
\begin{equation}
\psi(x,t)=
\int_{-\infty}^{+\infty}c(\omega)\,\me^{\mi\omega(x/v_{F}-t)}\,
\frac{\md\omega}{\sqrt{2\pi v_F}}\,
\end{equation}
through 
\begin{equation}
b^{\dagger}(\omega)
	=\frac{1}{\sqrt{\omega}}\int_{-\infty}^{+\infty}
	c^{\dagger}(\omega+\omega')c(\omega')\,\md\omega'\,.
\end{equation}
This immediately shows that $b^{\dagger}(\omega)$ creates a coherent
superposition of electron/hole pairs with energy $\hbar\omega$. Using
Eq.~\eqref{eq:bosonization:density}, the finite-frequency modes of the
excess electronic current $i(x,t)=-e v_F n(x,t)$ are directly 
proportional to the edge-magnetoplasmon modes, 
$i(\omega>0)=-e\sqrt{\omega}b(\omega)$. The electronic operator can be 
expressed in terms of these bosonic modes through
\begin{equation}
\label{eq:bosonization:fermions}
	\psi(x,t)=\frac{\mathcal{U}}{\sqrt{2\pi a}} \,
	\exp{\left(\mi\sqrt{4\pi}\phi(x,t)\right)}
\end{equation}
where $a$ is an ultraviolet cutoff that gives the length scale below 
which bosonization is not valid and $\mathcal{U}$ (resp. 
$\mathcal{U}^\dagger$) is the ladder operator suppressing (resp. adding) 
one electron from the reference vacuum.

The fermionic operator $\psi^\dagger(x,t)$ thus performs two things: it
shifts the vacuum state to add one electronic charge $-e$ to it and 
then it acts as a displacement operator on the edge-magnetoplasmon 
modes with parameter $\Lambda_\omega(x,t)=
\me^{-\mi\omega (x/v_F-t)}/\sqrt{\omega}$:
\begin{equation}
D[\Lambda(x,t)]
= \exp{
		\left(
			\int_0^{+\infty} 
			\!
			\left(
				\Lambda_\omega(x,t)b^\dagger(\omega)
				-\text{h.c.}
			\right)
			\md \omega
		\right)
	}\,.
\end{equation}
As discussed in
Ref.~\cite{Grenier:2013-1}, a classical time-dependent voltage drive 
$V(t)$ generates an edge-magnetoplasmon coherent state with parameter
$\Lambda_\omega[V(t)]=-e\widetilde{V}(\omega)/h\sqrt{\omega}$. The
coherent state of parameter $\Lambda_{\omega}(x,t)$ thus corresponds to 
the single-electron state generated by a voltage pulse 
$V(t)=-(h/e)\delta(t-x/v_F)$ generating a percussional current pulse 
carrying a single-electron charge.

\section{A long-range model for $\nu=1$}
\label{appendix:nu=1}

In this section, we derive an exact expression for the edge-magnetoplasmon transmission 
coefficient in the $\nu=1$ case using a simple model of Coulomb
interaction based on discrete elements in the spirit of Büttiker's
treatment of high frequency quantum transport \cite{Pretre:1996-1}.
Electrons within the interaction region see the electric
potential $U(x,t)$ given by a capacitive coupling inside a finite 
length 
region 
of size $l$:
\begin{align}
 U(x,t) = 
	\begin{cases}
	 0 & \text{if } x\notin \left[-\frac{l}{2},\frac{l}{2}\right] \\
	 \frac{1}{C}\int_{-\frac{l}{2}}^{\frac{l}{2}} n(y,t)\md y & 
\text{else.}
	\end{cases}
\end{align}
where the excess density of charges $n$ is itself linked to the bosonic 
field $\phi$ through equation \eqref{eq:bosonization:density}. 
Eq.~\eqref{eq:motion} can be recasted as 
a closed equation on $\phi$ expressed in the 
frequency domain as
\begin{equation}
 \left(-\mi\omega + v_F \partial_x\right) \phi(x,\omega) 
 = \frac{e^2}{hC}
	\left( 
		\phi\left(-\frac{l}{2},\omega\right) - 
		\phi\left(\frac{l}{2},\omega\right) 
	\right)\,.
\end{equation}
Expressing $\phi(x,\omega)$ as $\me^{\mi\omega x/v_F} 
\varphi_{\omega}(x)$ leads to
\begin{align}
 \partial_x \varphi_{\omega}(x) 
 = &
 \frac{e^2}{v_F h C} \me^{-\mi\omega x/v_F}
 \\
  &\left(
		\me^{-\mi \omega l/(2 v_F) }
		\varphi_{\omega}\left(-\frac{l}{2}\right)
		-\me^{\mi \omega l/(2 v_F) }
		\varphi_{\omega}\left(\frac{l}{2}\right)
	\right)
	\notag
\end{align}
which can be integrated over the whole interaction region to give us 
a relation between $\varphi_{\omega}\left(-\frac{l}{2}\right)$ and 
$\varphi_{\omega}\left(\frac{l}{2}\right)$. Finally, the solution reads
\begin{equation}
 \phi\left(\frac{l}{2},\omega\right) =
 t(\omega)\phi\left(-\frac{l}{2},\omega\right)
\end{equation}
where
\begin{subequations}
\label{eq:appendix:nu=1:result}
\begin{align}
 t(\omega) &= \me^{\mi\omega l/v_F}
 \frac{1+A(\omega,l)\me^{-\mi\omega l/(2v_F)}}
 {1+A(\omega,l)\me^{\mi\omega l/(2v_F)}} \\
 A(\omega,l)&=
 \frac{4e^2/C}{hv_F/l}\,
 \sinc\left(\frac{\omega l}{2v_F}\right)\,
\end{align}
\end{subequations}
in which we recognize the kinetic energy scale $hv_F/l$ as well as the
dimensionless ratio $\alpha=e^2l/Chv_F$ of the electrostatic energy
$e^2/C$ to this kinetic energy scale, which quantifies the strength of
Coulomb interactions in this system. Note that, at least for
sufficiently long edge channels, this coupling constant
does not depend on the length $l$ since $C$ also scales as $l$.

As expected, the transmission coefficient $t(\omega)$ is of modulus 1
because no energy can be lost in a $\nu=1$ setup without any
dynamical environment. The quantity of interest is therefore the phase 
of $t(\omega)$.

In the limit where Coulomb interaction effects can be neglected 
($\alpha\to 0$), $t(\omega)=\me^{\mi\omega l/v_F}$ showing that the 
bare 
Fermi velocity is recovered. The opposite limit of ultrastrong Coulomb 
interactions ($\alpha\to\infty$) leads to $t(\omega)=1$, that is an 
infinite edge-magnetoplasmon velocity. However, at fixed coupling 
$\alpha$, the edge-magnetoplasmon velocity tends to $v_\infty=v_F$ when 
$\omega l/v_F\gg 1$. At low frequency, we 
find that the time of flight of edge magnetoplasmons is renormalized 
thus
leading to an increased renormalized plasmon velocity  
\begin{equation}
\frac{v_0}{v_\infty}=1+\frac{4e^2/C}{hv_\infty/l}\,.
\end{equation}
compared to the velocity at high frequency which is
the bare Fermi velocity $v_F$.

To estimate an order of magnitude of this ratio, 
let us remind that $C$ being the
capacitance of the interaction region that is roughly similar to a 1D
wire,
$C\simeq 2\pi
\varepsilon_0\varepsilon_r l$ up to a geometrical factor for
large $l$, that is when boundary effects are small. Consequently,
$\alpha$ does not depend on $l$ but behaves as \citep{Grenier:2013-1}:
\begin{equation}
\alpha \simeq \frac{
\alpha_{\text{qed}}}{\pi\varepsilon_r}\times \frac{c}{v_F}\times
\left(\text{Geometrical\ Factor}\right)
\end{equation}
where $\alpha_{\text{qed}}$ denotes the fine-structure constant,
$\varepsilon_r$ the relative permittivity of the material and $v_F$
the bare Fermi velocity. 

For $\mathrm{AsGa}$, one usually estimates $v_F\simeq\SI{e5}{\meter/\second}$ and
$\varepsilon_r\simeq 10$ thus leading to 
\begin{equation}
\alpha\simeq 0.75\times (\text{Geometrical\ Factor})
\end{equation}
Assuming a geometrical factor of order $1$, this gives a velocity
for low-energy magnetoplasmons of the order of $v_0\sim
\SI{4e5}{\meter/\second}$ which is compatible to what is
observed in $\nu=2$ edge channel systems \cite{Kamata:2010-1}. 
Let us remind
that the edge-magnetoplasmon velocity depends on the details of
the electric potential seen by electrons near the edge of the 2DEG
and therefore of the conception of the sample. 
This is precisely used in the above reference to modulate it by
polarising gates.

In the case of graphene, a common estimation for 
the Fermi velocity is of the order of $v_F\simeq 
\SI{1e6}{\meter/\second}$ and $\varepsilon_r\simeq 14$
\cite{Petkovic:2013-1,Petkovic:2014-1} thus leading to
\begin{equation}
\alpha\simeq 0.054
\end{equation}
when using a geometrical factor equal to unity. The coupling constant is
much lower and therefore $v_0/v_F\simeq 1.2$. Let us stress that, as far
as we know, no direct measurement of $v_F$ in quantum Hall edge channels
of graphene have been performed but if this commonly discussed value is
confirmed, this would put graphene in a totally different coupling range
than AsGa.

For intermediate values of the coupling constant $\alpha$, as shown on Fig.~\ref{fig:phenomenology:nu=1:long-range:velocity}, the edge-magnetoplasmon 
velocity deduced from $t(\omega)$ presents a decay from $v_0$ to a
regime with small oscillations above the asymptotic
value of $v_F$. 

Expanding the phase of $t(\omega)$ in powers of $\omega\tau_0$ leads to
\begin{subequations}
\begin{align}
\phi(\omega) &= \omega\tau_0+\frac{\alpha}{3}(\omega\tau_0)^3\nonumber \\
&+ \frac{8\alpha}{90}\left(\alpha^2+2\alpha-1/8\right)(\omega\tau_0)^5 +\mathcal{O}
\left((\omega\tau_0)^7\right)
\end{align}
\end{subequations}
which, as explained Appendix \ref{appendix:perturbation-theory}, gives us the low energy
expansion of the inelastic scattering probability.

\section{Energy dissipation}
\label{appendix:dissipation}

Let us discuss energy dissipation through the creation of electron/hole
pairs in the $\nu=1$ case. To begin with, this discussion makes sense 
when there is a clear separation in energy between the injected
electron after relaxation and the electron/hole excitations generated by
Coulomb interaction (see Sec.~\ref{sec:decoherence_nu_1:high-energy}). 
In the following
discussion, we shall thus assume that the spectral weight of the
incoming electron as well as of the contribution
$\mathcal{G}^{(e)}_{\text{WP,1}}$ to the outgoing coherence are well
above the vicinity of the Fermi level. 

The incoming average energy comes from the injected
electron and is equal to
\begin{equation}
E_{\text{in}}=
	\hbar\int_0^{+\infty}
	\left|\tilde{\varphi}_\text{e}(\omega)\right|^2\omega\,\frac{\md\omega}{2\pi
	v_F}\,.
\end{equation}
using the convention 
\begin{equation}
\tilde{\varphi}_\text{e}(\omega)=v_F\int_{-\infty}^{+\infty}
\varphi_{\text{e}}(-v_Ft)\,\me^{\mi
\omega t}\,\md t
\end{equation}
for defining the electronic wavepacket in the frequency domain from the
original wavefunctiuon $\varphi_{\text{e}}$ in the spatial domain.

The outgoing average energy then consists of two parts: the energy carried
by the injected electron which has flewn across the interaction
region either elastically or inelastically, and the energy of electron/hole
excitations created by its passing through. The first contribution is
\begin{subequations}
\begin{align}
E_{\text{out}}^{(\text{e})} 
&= Z_\infty E_{\text{in}} \\
&+ \hbar\int_{(\mathbb{R}^+)^2}\left|\varphi_\text{e}(\omega)\right|^2
	 (\omega-\omega')d(\omega')\,\md\omega'\frac{\md\omega}{2\pi v_F}
\end{align}
\end{subequations}
The first line corresponds to elastic scattering and the second line
to inelastic processes in which the electron has fallen down 
from $\hbar\omega$ to $\hbar(\omega-\omega')$. 
There, the integrals are extended to $+\infty$ safely because of our
working hypothesis: the relaxation tail is well above the Fermi level.
We then use that 
$\int_0^{+\infty}d(\omega')\,\md\omega'=1-Z_\infty$ and the 
normalization condition of the wavepacket to rewrite this as
\begin{equation}
E_\text{out}^{(\text{e})}=E_{\text{in}}-\hbar\int_0^{+\infty}
\omega'd(\omega')\, \md\omega'
\end{equation}
Energy conservation, which is true on average, shows that
the dissipated energy in electron/hole pair creation is equal to
\begin{equation}
E_{\text{out}}^{(\text{diss})}=\hbar\int_0^{+\infty}\omega 
d(\omega)\,\md\omega
\end{equation}
Recognizing that $\int_0^{+\infty} \omega d(\omega)\md\omega$ 
corresponds
to the derivative of the decoherence coefficient $\mathcal{D}(\tau)$
when $\tau\rightarrow 0^+$ leads to
\begin{equation}
\label{eq:ENERGY}
E_{\text{out}}^{(\text{diss})}=
\hbar\int_0^{+\infty}\left|1-\tilde{t}(\omega)\right|^2
\,\md\omega\,.
\end{equation}
Using the transmission coefficient given by Eq.~\eqref{eq:appendix:nu=1:result}, the
dissipated energy is given by 
\begin{equation}
\label{eq:dissipated-energy:nu=1}
E_{\text{out}}^{(\text{diss})}=\frac{h v_F}{\pi\,l}
\int_0^{+\infty}\frac{64\alpha^2\sin^4(u)\,\md u}{(u+2\alpha\sin(2u))^2+
16\alpha^2\sin^2(u)}\,.
\end{equation}
which converges both in the UV and the IR. 

\begin{figure}
\includegraphics[width=8cm]{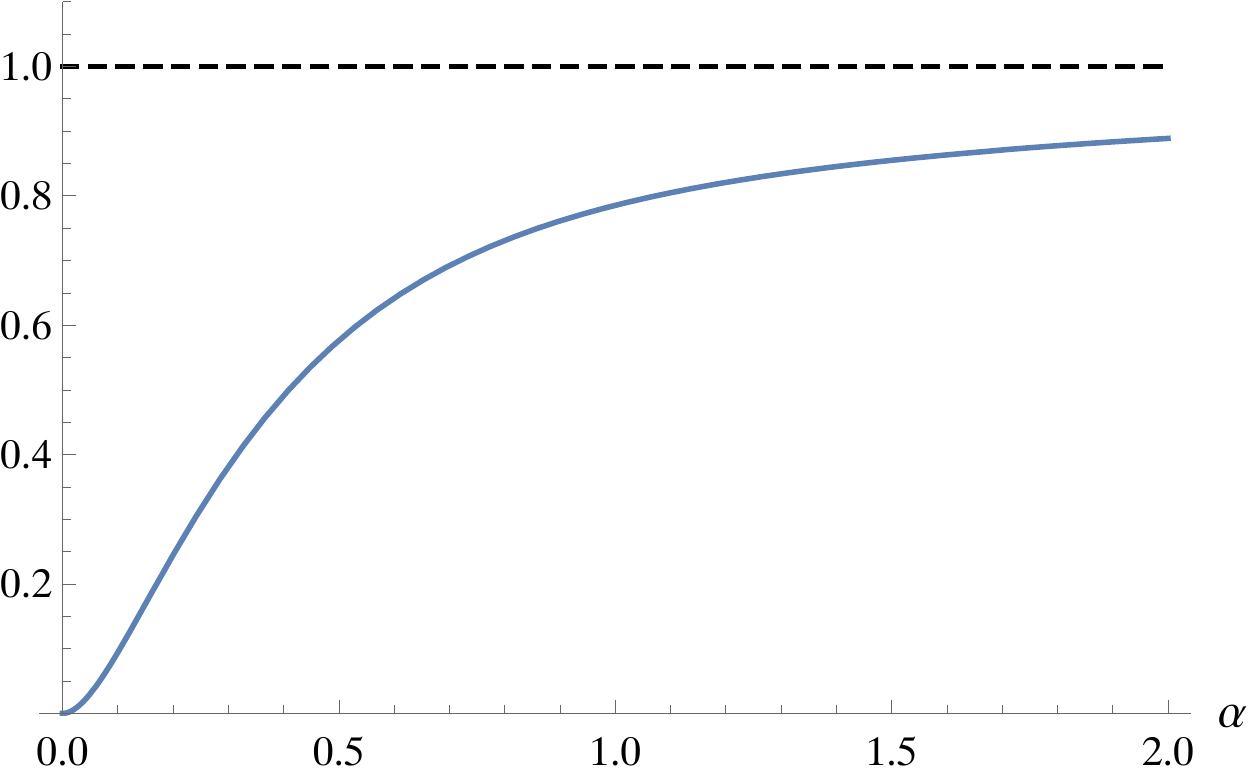}
\caption{\label{fig:energy-dissipation}
Dependence on the coupling constant $\alpha$ of 
$E_{\text{out}}^{(\text{diss})}(\alpha,hv_F/l)$ in units of
$hv_0/l$ where 
$E_{\text{out}}^{(\text{diss})}(\alpha,hv_F/l)$ denotes
the average energy
dissipated by a hot electron  given by 
Eq.~\eqref{eq:dissipated-energy:nu=1} corresponding to the model discussed
in Appendix \ref{appendix:nu=1}. 
}
\end{figure}

Fig.~\ref{fig:energy-dissipation} presents the 
numerical evaluation of the dissipated energy in units of 
$hv_0/l$, where $v_0=(1+4\alpha)v_F$ is the low energy edge-magnetoplasmon velocity in this model.
We observe that it saturates to $1$ at large coupling.
The finiteness of the dissipated energy validates a posteriori
that the high-energy description
of electronic decoherence is valid as long as 
the average energy of the incoming excitation is large compared
to $\alpha hv_0/l$.

As a final check, one can rederive Eq.~\eqref{eq:ENERGY} 
by considering the reduced
density operator for the low energy electron/hole pair excitations.
When assuming that even after relaxation, the wavepacket remains
well separated from the Fermi sea, one can assume that $\langle
\psi(t_-)\psi^\dagger(t_+)\rangle_F\simeq v_F^{-1}\delta(t_+-t_-)$ in
\eqref{eq:decoherence:result:MV} 
and therefore $\mathcal{G}^{(e)}_{\text{MV,1}}(t|t')$ can be
approximated by an expression
which corresponds to the statistical mixture of states $|g(t)\rangle$
ponderated by $|\varphi_\text{e}(t)|^2$. This naturally comes from the physical
image of the incident electron emerging from the interaction in a
quantum superposition of the
coherent electron/hole pair clouds $|g(t)\rangle$ attached to the
electron being at position
$v_Ft$. Two different positions $v_Ft$ and $v_Ft'$ of the electron being
perfectly distinguishable, what comes out is the statistical mixture
of coherent electron/home pair clouds for the low energy
edge-magnetoplasmon modes. Computing the average energy 
stored in this statistical mixture precisely
leads to \eqref{eq:ENERGY} since all the states $|g(t)\rangle$ carry the
same average energy.

\section{Discrete element circuit description}
\label{appendix:circuit}

In this appendix, we discuss the circuit synthesis for the edge-magnetoplasmon transmission amplitude in the case of an ideal
$\nu=1$ edge channel and we obtain
its first non trivial Cauer form. We then connect the discrete circuit
element parameters to the parameters of the model presented in 
Appendix \ref{appendix:nu=1}.

\subsection{Circuit synthesis for an ideal $\nu=1$ edge channel}

Using the relation $t(\omega)=1-R_KG(\omega)$ where $G(\omega)$ is the
finite-frequency admittance of the discrete element circuit of Fig.~\ref{fig:effective-circuit}, 
the transmission amplitude $t(\omega)$ can be expressed in terms of the
impedance $Z(\omega)$ as
\begin{equation}
\label{eq:effective-circuit:transmission}
t(\omega)=\frac{1+\omega C_\mu\Im{(Z(\omega))}+i\omega
C_\mu(R_K-\Re{(Z(\omega))})}{1+\omega C_\mu\Im{(Z(\omega))}-i\omega
C_\mu\Re{(Z(\omega))}}\,.
\end{equation}
Consequently, 
$t(\omega)$ is a pure
phase if and only if $\Re{(Z(\omega))}=R_K/2$ at all 
frequencies.
We can then write
\begin{equation}
t(\omega)
=\frac{1+\mi\alpha(\omega)}{1-\mi\alpha(\omega)}=
\exp{\left(2i\arctan{(\alpha)}\right)}
\end{equation}
where 
\begin{equation}
\label{eq:effective-circuit:alpha}
\alpha(\omega) = \frac{\omega R_KC_\mu}{2}\,
\frac{1}{1+\omega C_\mu \Im{(Z(\omega))}}\,.
\end{equation}
With our conventions, the reactance $\Im{(Z(\omega))}$ is a strictly
decreasing function of $\omega$ \cite{Foster:1924-1}. Since, by
definition, the
electrochemical capacitance $C_\mu$ contains the low-frequency
divergence of the ZC circuit, it is expected to be regular at low
frequency, starting with a zero at $\omega=0$ and then alternating
poles and zeroes. A suitable low-frequency
expansion of $t(\omega)$ can then be obtained using a Cauer form of
circuit synthesis which leads to a continuous fraction expansion of the
finite-frequency admittance.

The simplest case corresponds to the circuit depicted on the right panel
of Fig.~\ref{fig:effective-circuit}. It leads to 
\begin{equation}
\label{eq:effective-circuit:alpha:LC}
\alpha(\omega)=\frac{\omega R_KC_\mu}{2}\,
\frac{1-\omega^2LC}{1-\omega^2L(C+C_\mu)}
\end{equation}
Expanding $2\arctan{(\alpha(\omega))}$ in powers of $\omega R_KC_\mu$
then leads to the low-frequency finite-frequency admittance
up to order $(\omega R_KC_\mu)^6$:
\begin{subequations}
\label{eq:circuits:admittance-expansion}
\begin{align}
g(\omega)&=-i\omega R_KC_\mu + \frac{1}{2}(\omega R_KC_\mu)^2 \\
&-i\left[\frac{L/R_K}{R_KC_\mu}-\frac{1}{4}\right](\omega R_KC_\mu)^3 \\
&+\left[\frac{L/R_K}{R_KC_\mu}-\frac{1}{8}\right](\omega R_KC_\mu)^4\\
&-i\left[\left(1+\frac{C}{C_\mu}\right)
\left[\frac{L/R_K}{R_KC_\mu}\right]^2+\frac{1}{16}-\frac{3}{4}\frac{L/R_K}{R_KC_\mu}\right]
(\omega R_KC_\mu)^5
\end{align}
\end{subequations}
which then leads
to Eqs.~\eqref{eq:decoherence:circuit-parameters}. 

Being described by two parameters ($L$ and $C$) besides
$C_\mu$ and $R_q=R_K/2$, this circuit provides an expansion of $\phi(\omega)$
up to order $(\omega R_KC_\mu)^5$. 
In order to capture the low-frequency behavior of $\phi(\omega)$ to the
next non trivial orders ($7$ and $9$), 
we need to go one step further in the Cauer form of the
circuit. This would correspond to adding another $LC$ impedance in series
with the capacitor $C$. This process can then be iterated to reconstruct
the full $\omega$ dependence of $\Im{(Z(\omega))}$.

\subsection{Extracting the discrete element parameters}
\label{appendix:perturbation-theory:velocities}

Let us now derive the discrete element circuit parameters for the 
interaction model at $\nu=1$ considered in 
Appendix \ref{appendix:nu=1}. Expanding the 
admittance at low frequency and identifying this expansion with
\eqref{eq:circuits:admittance-expansion} leads to
\begin{subequations}
	\label{eq:circuits:nu=1:parameters}
\begin{align}
R_KC_\mu &= \tau_0=l/v_0\\
\frac{L/R_K}{R_K C_\mu} &=\frac{1+4\alpha}{12}\\
\frac{C}{C_\mu} &=
\frac{1+4\alpha}{5}
\end{align}
\end{subequations}
The inductance $L$ as well as the capacitance $C$ increase when
increasing the effective Coulomb interaction strength. 
This is expected since increasing Coulomb interactions tend to increase
the velocity ratio $v_0/v_\infty$. 
In this model
the ratio of $L/C$ to $R_K^2$ remains constant and equal to $5/12$.
Note that for $L=R_K^2C_\mu/12$ and $C=C_\mu/5$,
$\alpha_3=\alpha_5=0$: the first non trivial contribution in
$\phi(\omega)$ appears at order $(\omega R_KC_\mu)^7$.

\section{Phenomenological models for plasmon velocity}
\label{appendix:phenomenology}

Let us discuss problems that arise for 
some phenomenological expressions
for the edge magnetoplasmons in the ideal $\nu=1$ case. 

We first consider the phenomenological expression 
\begin{equation}
\label{eq:phenomenology:nu=1:velocity}
\frac{v(\omega)}{v_0}=
\frac{1+\frac{v_\infty}{v_0}(\omega/\omega_c)^2}{1+(\omega/\omega_c)^2}
\end{equation}
which 
interpolates between $v_0$ at low frequency and $v_\infty$ at high
frequency, the crossover scale being $\omega_c$. 
We shall
denote by $\tau_0=l/v_0$. 
The finite-frequency admittance 
only depends on the dimensionless variable $\omega\tau_0$ and parameters
$0<v_\infty/v_0\leq 1$ and
$\omega_c\tau_0>0$.
Compared to the long-range interaction model detailed in Appendix
\ref{appendix:nu=1}, this phenomenological expression
avoids oscillations in the edge-magnetoplasmon
velocity and it depends on one more parameter than just $l/v_0$ and
the coupling constant. However, as we will see now, 
is it not physically acceptable!

A first hint of a problem comes from the low energy expansion using a
discrete element circuit description that reproduces the same
$t(\omega)$ dependance uo to order $5$. 
Then, under this condition, the electrochemical capacitance
$C_\mu$, the inductance $L$ and the capacitance $C$ of the first ladder
in the Cauer expansion are given by:
\begin{subequations}
\begin{align}
R_KC_\mu &= \tau_0\\
\frac{L/R_K}{R_K C_\mu} &=\frac{1}{12}+
\left(1-\frac{v_\infty}{v_0}\right)\,\frac{1}{(\omega_c\tau_0)^2}\\
\frac{C}{C_\mu} &=
	\frac{\frac{1}{720}+
	\left(1-\frac{v_\infty}{v_0}\right)\left[\frac{1}{60(\omega_c\tau_0)^2}-\frac{1}{(\omega_c\tau_0)^4}\right]
	}{
\left[\frac{1-\frac{v_\infty}{v_0}}{(\omega_c\tau_0)^2}+\frac{1}{12}\right]^2}
\end{align}
\end{subequations}
As expected, the eigenfrequency $1/\sqrt{LC_\mu}$ corresponds, up
to renormalization, to $\omega_c$. 
Since $v_\infty\leq v_0$ these expressions give a physical value for the
inductance $L$ but $C/C_\mu$ sometimes
becomes negative!
This is a serious hint that Eq.~\eqref{eq:phenomenology:nu=1:velocity} is not a
physically meaningful $\omega$-dependance for the edge-magnetoplasmon
velocity. This can be seen by considering the analytical continuation of
$\Re(1-t(\omega))$ to the complex plane $s=\sigma+\mi \omega$ which must
be positive for $\sigma<0$: it exhibits singularities (and thus negativities) on the negative
real axis ($\sigma <0$ and $\omega=0$).

In the same way, a phenomenological edge-magnetoplasmon velocity with a
sharper high-energy stabilization towards $v_\infty$ such
as\cite{Neuenhahn:2008-1,Neuenhahn:2009-1}
\begin{equation}
\frac{v(\omega)}{v_\infty}=1+\frac{v_0-v_\infty}{v_\infty}\me^{-(\omega\tau_c)^2}\,.
\end{equation}
is not physical within our framework because 
the analytical continuation of
$\Re{(1-\me^{\mi \omega l/v(\omega)})}$ also presents singularities in
the half plane $\sigma+\mi \omega$ for $\sigma <0$.

\section{Low energy perturbative expansion}
\label{appendix:perturbation-theory}

Here, we consider low energy excitations that have
almost all their spectral weight below $\omega_c$. The relevant base
velocity is $v_0$ and therefore, we define the effective
transmission amplitude as
$\tilde{t}(\omega)=t(\omega)\,\me^{-\mi\omega\tau_0}$ so that the 
deviation
from $\tilde{t}(\omega)=1$ for $0<\omega\lesssim\omega_c$ is 
small. 

Assuming that
$\tilde{t}(\omega)=1$ for $\omega\lesssim \omega_c$,
the electronic excitation experiences no decoherence for the part 
which is located below $\omega_c$: it simply moves at the 
plasmon velocity $v_0$. This is consistent with the high-energy picture discussed in 
Sec.~\ref{sec:decoherence_nu_1:high-energy}: although a high-energy
electronic excitation moves forward at the velocity $v_\infty$ together
with its relaxation tail, the electron/hole pairs created close
to the Fermi level move at the plasmon velocity $v_0$. 
The idea is thus to perform a perturbative expansion in terms 
of $\omega\tau_0$ of the rescaled edge-magnetoplasmon transmission 
coefficient 
$\tilde{t}(\omega)$. 

At low frequency, the edge-magnetoplasmon transmission coefficient is of the form
\begin{equation}
\tilde{t}(\omega)=\exp{\left(\mi\sum_{k\geq 2}
\alpha_k(\omega\tau_0)^k\right)}
\end{equation}
where $\tau_0$ is a typical time scale of the problem and $\alpha_k$
dimensionless couplings. Note that only odd powers of $\omega\tau_0$ 
need to be considered because, as discussed in 
Sec.~\ref{sec:models:introduction},
$t(\omega)^*=t(-\omega)$.

To obtain the inelastic scattering probability 
$\sigma_{\text{in}}(\omega)=1-|\mathcal{Z}(\omega)|^2$, we shall expand
perturbatively in $\omega\tau_0$ the elastic scattering amplitude
\begin{equation}
\mathcal{Z}(\omega)=1+\int_0^{\omega}B_-(\omega')\,\md\omega'
\end{equation}
where
\begin{equation}
 B_-(\omega) =
	\sum_{n=1}^{\infty} \frac{1}{n!}
		\left(\frac{t(\omega)-1}{\omega}\right)^{\ast n} (\omega)
\end{equation}
is expanded as a series of convolution powers $(\cdot)^{\ast n}$.
Denoting $P(\omega)= \frac{t(\omega)-1}{\omega}$, we have
\begin{subequations}
 \begin{align}
 \frac{P(\omega)}{\tau_0}
				&=\mi \alpha_3 \left(\omega \tau_0 \right)^2
					 +\mi \alpha_5 \left(\omega \tau_0 \right)^4
					 -\frac{\alpha_3 ^2}{2}
						\left(\omega \tau_0\right)^5
						\notag \\
					 &+\mi \alpha_7 \left(\omega \tau_0 \right)^6
					 -\alpha_3\alpha_5
						\left(\omega\tau_0\right)^7
						\notag \\
					&+\mathcal{O}\left(\left(\omega\tau_0\right)^8\right)
	\\
	\frac{P^{\ast 2}(\omega)}{\tau_0}
		&= -\frac{\alpha_3^2}{30}\left(\omega \tau_0\right)^5
			-\frac{2\alpha_3 \alpha_5}{105} \left(\omega\tau_0\right)^7
		\notag\\	
		&+\mathcal{O}\left(\left(\omega\tau_0\right)^8\right)
	\\
	\frac{P^{\ast 3}(\omega)}{\tau_0}
		&= \mathcal{O}\left(\left(\omega\tau_0\right)^8\right)
 \end{align}
\end{subequations}
Consequently, the expansion of $\mathcal{Z}(\omega)$ up to order $(\omega\tau_0)^8$
only involves the 2nd convolution power of $P$. This corresponds to two
edge-magnetoplasmon emission processes. Processes with higher
multi-plasmon emission will only contribute to higher powers in
$\mathcal{Z}(\omega)$'s expansion. Limiting ourselves to this order
leads to:
\begin{align}
 \mathcal{Z}(\omega) &=
	1
	+ \frac{\mi\alpha_3}{3} \left(\omega\tau_0\right)^3
	+ \frac{\mi\alpha_5}{5} \left(\omega \tau_0\right)^5
	\notag \\
		&- \frac{31}{360}\alpha_3^2
		\left(\omega \tau_0\right)^6
	+ \frac{\mi\alpha_7}{7} \left(\omega \tau_0\right)^7
	\notag \\
&-\frac{106}{105}\alpha_3\alpha_5\left(\omega \tau_0\right)^8	
+\mathcal{O}\left(\left(\omega\tau_0\right)^9\right)
\end{align}
which gives the final result for the inelastic scattering probability 
$\sigma_{\text{in}}(\omega)=1-|\mathcal{Z}(\omega)|^2$:
\begin{align}
\sigma_{\text{in}}(\omega) &=
 \frac{11\alpha_3^2}{180} \left(\omega \tau_0\right)^6
 +\frac{5 \alpha_3\alpha_5}{42} \left(\omega \tau_0\right)^8
 	\notag \\
	&+\mathcal{O}\left(\left(\omega\tau_0\right)^{9}\right)
\end{align}
thus recovering Eq.~\eqref{eq:decoherence:inelastic-development}. 
Note that keeping only the first convolution power in the expansion would lead to
\begin{align}
\sigma_{\text{in}}^{(1)}(\omega)&=
\frac{\alpha_3^2}{18} \left(\omega \tau_0\right)^6
 - \frac{7\alpha_3\alpha_5}{60} \left(\omega \tau_0\right)^8
     \notag \\
	     &+\mathcal{O}\left(\left(\omega\tau_0\right)^{9}\right)
\end{align}
which is the inelastic scattering probability arising from single edge-magnetoplasmon emission.

\section{More experimentally relevant Wigner functions}
\label{appendix:experimentaldecohencecontrol}

In this appendix, we show some more Wigner functions for loops built as 
in Fig.~\ref{fig:sampledesign} of different sizes, and excitations of 
different energies. Velocities parameters are the same as in the main 
text, and Landau excitations have a typical time 
$\tau_0=\SI{100}{\pico\second}$. All other parameters are shown below 
the corresponding Wigner functions. On all figures, the top panel shows a 
closed loop, whereas middle panel shows the case where both edge 
channels would stay outside of the loop and experience standard 
interaction along a length $w$. The bottom panel then displays the expected 
results of an HOM experiment for both cases. Using these figures, we 
can gain a more quantitative understanding of how changing the loop 
size or the injection energy impacts the experimentally accessible 
quantities.

\begin{figure}
	\includegraphics{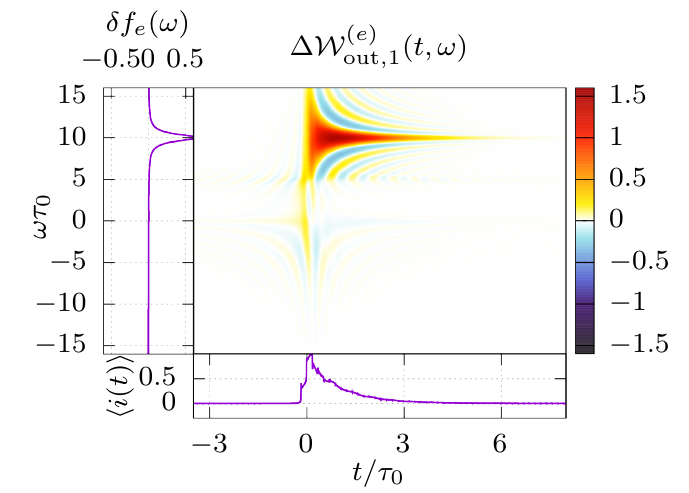}
	\includegraphics{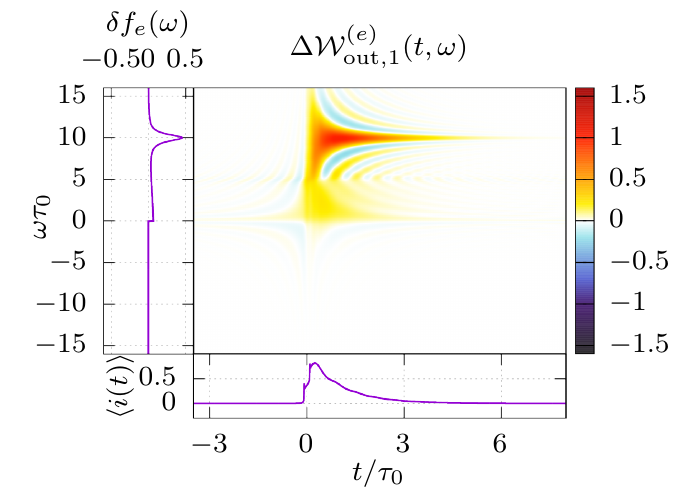}
	\includegraphics{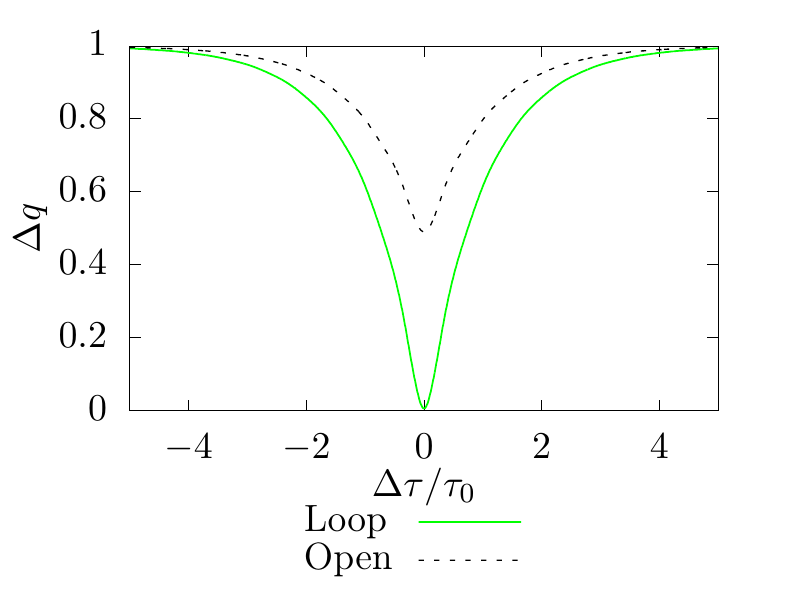}
	\caption{(Color online) Parameters are $w=\SI{1}{\micro\meter}$, 
$d=\SI{0.4}{\micro\meter}$, $\omega_0\tau_0=10$. The energy of the 
particle is $\SI{60}{\micro\electronvolt}$, the resonance energy is at 
$\SI{120}{\micro\electronvolt}$.}
 \label{fig:appendix:smallloop}
\end{figure}

\begin{figure}
	\includegraphics{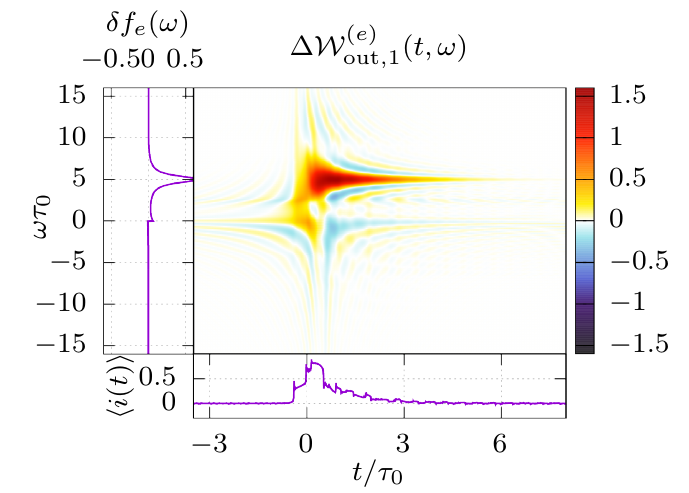}
	\includegraphics{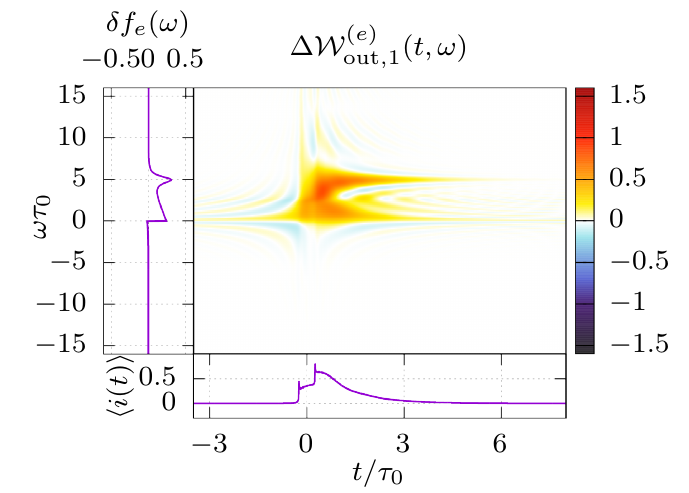}
	\includegraphics{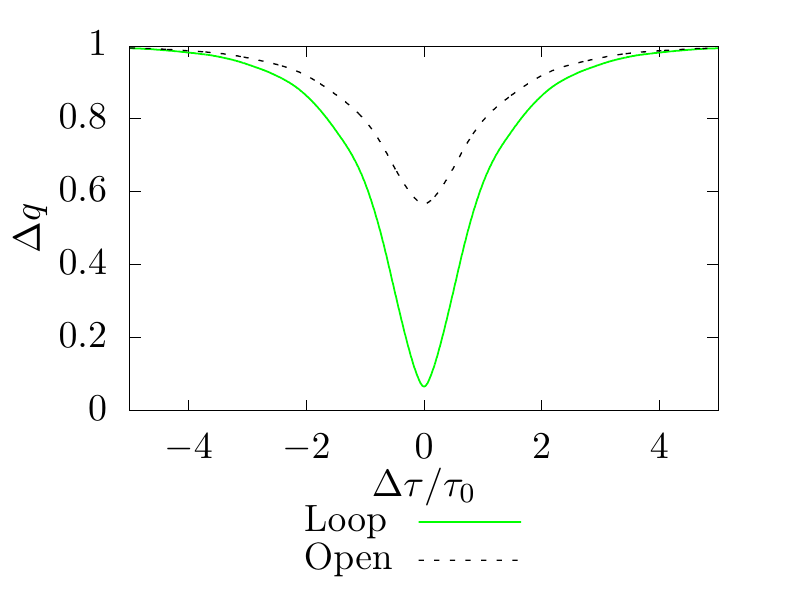}
	\caption{(Color online) Parameters are $w=\SI{3}{\micro\meter}$, 
$d=\SI{0.5}{\micro\meter}$, $\omega_0\tau_0=5$. The energy of the 
particle is at $\SI{30}{\micro\electronvolt}$, the resonance energy at 
$\SI{45}{\micro\electronvolt}$. We see that sending a smaller energy 
excitation allows for larger loops.}
 \label{fig:appendix:biglooplowenergy}
\end{figure}

\begin{figure}
	\includegraphics{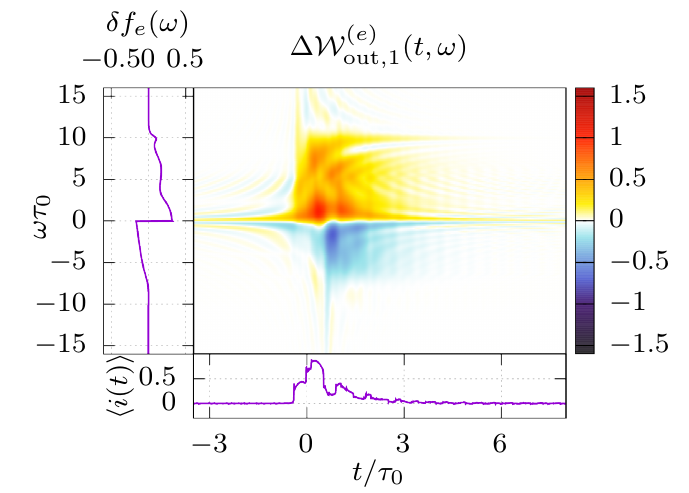}
	\includegraphics{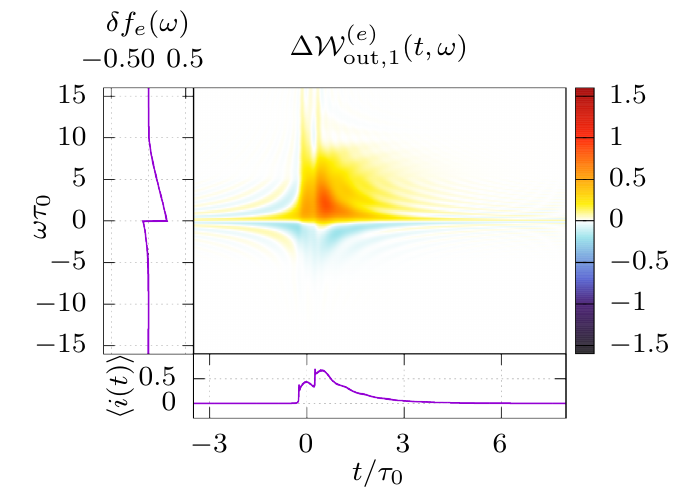}
	\includegraphics{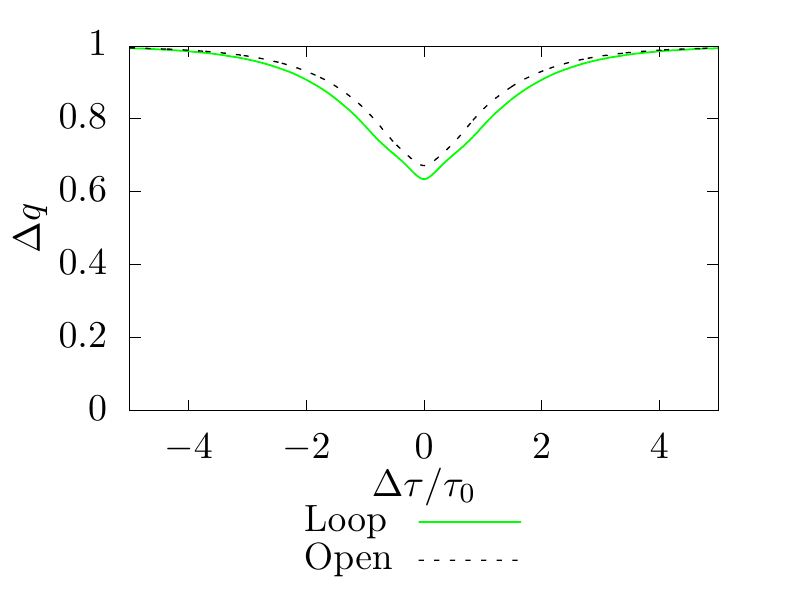}
	\caption{(Color online) Parameters are $w=\SI{3}{\micro\meter}$, 
$d=\SI{0.5}{\micro\meter}$, $\omega_0\tau_0=10$. The energy of the 
particle is at $\SI{60}{\micro\electronvolt}$, the resonance energy at 
$\SI{45}{\micro\electronvolt}$. In that case, the loops does not allow 
protection from decoherence, and a plasmon state is emitted along with 
the electron.}
 \label{fig:appendix:bigloophighenergy}
\end{figure}

\acknowledgments{We thank C. Bauerle, J.M.~Berroir, E. Bocquillon, 
V. Freulon,
F.D. Parmentier, P. Roche and B.~Pla\c{c}ais for useful discussions.
This work is supported by the ANR grant ''1shot reloaded''
(ANR-14-CE32-0017) and ERC Consolidator grant ''EQuO'' (No. 648236).
}


\begin{thebibliography}{74}
\expandafter\ifx\csname natexlab\endcsname\relax\def\natexlab#1{#1}\fi
\expandafter\ifx\csname bibnamefont\endcsname\relax
  \def\bibnamefont#1{#1}\fi
\expandafter\ifx\csname bibfnamefont\endcsname\relax
  \def\bibfnamefont#1{#1}\fi
\expandafter\ifx\csname citenamefont\endcsname\relax
  \def\citenamefont#1{#1}\fi
\expandafter\ifx\csname url\endcsname\relax
  \def\url#1{\texttt{#1}}\fi
\expandafter\ifx\csname urlprefix\endcsname\relax\def\urlprefix{URL }\fi
\providecommand{\bibinfo}[2]{#2}
\providecommand{\eprint}[2][]{\url{#2}}

\bibitem[{\citenamefont{Bauerle et~al.}(2018)\citenamefont{Bauerle, Glattli,
  Meunier, Portier, Roche, Roulleau, Takada, and Waintal}}]{Bauerle:2018-1}
\bibinfo{author}{\bibfnamefont{C.}~\bibnamefont{Bauerle}},
  \bibinfo{author}{\bibfnamefont{D.C.}~\bibnamefont{Glattli}},
  \bibinfo{author}{\bibfnamefont{T.}~\bibnamefont{Meunier}},
  \bibinfo{author}{\bibfnamefont{F.}~\bibnamefont{Portier}},
  \bibinfo{author}{\bibfnamefont{P.}~\bibnamefont{Roche}},
  \bibinfo{author}{\bibfnamefont{P.}~\bibnamefont{Roulleau}},
  \bibinfo{author}{\bibfnamefont{S.}~\bibnamefont{Takada}}, \bibnamefont{and}
  \bibinfo{author}{\bibfnamefont{X.}~\bibnamefont{Waintal}}
  (\bibinfo{year}{2018}), \bibinfo{note}{arXiv:1801.07497}.

\bibitem[{\citenamefont{Splettstoesser and Haug}(2017)}]{Splettstoesser:2017-1}
\bibinfo{editor}{\bibfnamefont{J.}~\bibnamefont{Splettstoesser}}
  \bibnamefont{and} \bibinfo{editor}{\bibfnamefont{R.}~\bibnamefont{Haug}},
  eds., \emph{\bibinfo{title}{Single-Electron control in Solid State Devices}},
  vol. \bibinfo{volume}{254} (\bibinfo{year}{2017}).

\bibitem[{\citenamefont{Bocquillon et~al.}(2014)\citenamefont{Bocquillon,
  Freulon, Parmentier, Berroir, Pla{\c c}ais, Wahl, Rech, Jonckheere, Martin,
  Grenier et~al.}}]{Bocquillon:2014-1}
\bibinfo{author}{\bibfnamefont{E.}~\bibnamefont{Bocquillon}},
  \bibinfo{author}{\bibfnamefont{V.}~\bibnamefont{Freulon}},
  \bibinfo{author}{\bibfnamefont{F.}~\bibnamefont{Parmentier}},
  \bibinfo{author}{\bibfnamefont{J.}~\bibnamefont{Berroir}},
  \bibinfo{author}{\bibfnamefont{B.}~\bibnamefont{Pla{\c c}ais}},
  \bibinfo{author}{\bibfnamefont{C.}~\bibnamefont{Wahl}},
  \bibinfo{author}{\bibfnamefont{J.}~\bibnamefont{Rech}},
  \bibinfo{author}{\bibfnamefont{T.}~\bibnamefont{Jonckheere}},
  \bibinfo{author}{\bibfnamefont{T.}~\bibnamefont{Martin}},
  \bibinfo{author}{\bibfnamefont{C.}~\bibnamefont{Grenier}},
  \bibnamefont{et~al.}, \bibinfo{journal}{Ann. Phys. (Berlin)}
  \textbf{\bibinfo{volume}{526}}, \bibinfo{pages}{1} (\bibinfo{year}{2014}).

\bibitem[{\citenamefont{Bertoni et~al.}(2000)\citenamefont{Bertoni, Bordone,
  Brunetti, Jacoboni, and Reggiani}}]{Bertoni:2000-1}
\bibinfo{author}{\bibfnamefont{A.}~\bibnamefont{Bertoni}},
  \bibinfo{author}{\bibfnamefont{P.}~\bibnamefont{Bordone}},
  \bibinfo{author}{\bibfnamefont{R.}~\bibnamefont{Brunetti}},
  \bibinfo{author}{\bibfnamefont{C.}~\bibnamefont{Jacoboni}}, \bibnamefont{and}
  \bibinfo{author}{\bibfnamefont{S.}~\bibnamefont{Reggiani}},
  \bibinfo{journal}{Phys. Rev. Lett.} \textbf{\bibinfo{volume}{84}},
  \bibinfo{pages}{5912} (\bibinfo{year}{2000}).

\bibitem[{\citenamefont{Ionicioiu et~al.}(2001)\citenamefont{Ionicioiu,
  Amaratunga, and Udrea}}]{Ionicioiu:2001-1}
\bibinfo{author}{\bibfnamefont{R.}~\bibnamefont{Ionicioiu}},
  \bibinfo{author}{\bibfnamefont{G.}~\bibnamefont{Amaratunga}},
  \bibnamefont{and} \bibinfo{author}{\bibfnamefont{F.}~\bibnamefont{Udrea}},
  \bibinfo{journal}{Int. J. Mod. Phys. B} \textbf{\bibinfo{volume}{15}},
  \bibinfo{pages}{125} (\bibinfo{year}{2001}).

\bibitem[{\citenamefont{Bertoni}(2007)}]{Bertoni:2007-1}
\bibinfo{author}{\bibfnamefont{A.}~\bibnamefont{Bertoni}}, \bibinfo{journal}{J.
  Comput. Electron.} \textbf{\bibinfo{volume}{6}}, \bibinfo{pages}{67}
  (\bibinfo{year}{2007}).

\bibitem[{\citenamefont{Roussel et~al.}(2017)\citenamefont{Roussel, Cabart,
  F{\`e}ve, Thibierge, and Degiovanni}}]{Roussel:2016-2}
\bibinfo{author}{\bibfnamefont{B.}~\bibnamefont{Roussel}},
  \bibinfo{author}{\bibfnamefont{C.}~\bibnamefont{Cabart}},
  \bibinfo{author}{\bibfnamefont{G.}~\bibnamefont{F{\`e}ve}},
  \bibinfo{author}{\bibfnamefont{E.}~\bibnamefont{Thibierge}},
  \bibnamefont{and}
  \bibinfo{author}{\bibfnamefont{P.}~\bibnamefont{Degiovanni}},
  \bibinfo{journal}{Physica Status Solidi B} \textbf{\bibinfo{volume}{254}},
  \bibinfo{pages}{16000621} (\bibinfo{year}{2017}).

\bibitem[{\citenamefont{Le~Sueur et~al.}(2010)\citenamefont{Le~Sueur,
  Altimiras, Gennser, Cavanna, Mailly, and Pierre}}]{LeSueur:2010-1}
\bibinfo{author}{\bibfnamefont{H.}~\bibnamefont{Le~Sueur}},
  \bibinfo{author}{\bibfnamefont{C.}~\bibnamefont{Altimiras}},
  \bibinfo{author}{\bibfnamefont{U.}~\bibnamefont{Gennser}},
  \bibinfo{author}{\bibfnamefont{A.}~\bibnamefont{Cavanna}},
  \bibinfo{author}{\bibfnamefont{D.}~\bibnamefont{Mailly}}, \bibnamefont{and}
  \bibinfo{author}{\bibfnamefont{F.}~\bibnamefont{Pierre}},
  \bibinfo{journal}{Phys. Rev. Lett.} \textbf{\bibinfo{volume}{105}},
  \bibinfo{pages}{056803} (\bibinfo{year}{2010}).

\bibitem[{\citenamefont{Marguerite et~al.}(2016)\citenamefont{Marguerite,
  Cabart, Wahl, Roussel, Freulon, Ferraro, Grenier, Berroir, Pla{\c c}ais,
  Jonckheere et~al.}}]{Marguerite:2016-1}
\bibinfo{author}{\bibfnamefont{A.}~\bibnamefont{Marguerite}},
  \bibinfo{author}{\bibfnamefont{C.}~\bibnamefont{Cabart}},
  \bibinfo{author}{\bibfnamefont{C.}~\bibnamefont{Wahl}},
  \bibinfo{author}{\bibfnamefont{B.}~\bibnamefont{Roussel}},
  \bibinfo{author}{\bibfnamefont{V.}~\bibnamefont{Freulon}},
  \bibinfo{author}{\bibfnamefont{D.}~\bibnamefont{Ferraro}},
  \bibinfo{author}{\bibfnamefont{C.}~\bibnamefont{Grenier}},
  \bibinfo{author}{\bibfnamefont{J.-M.} \bibnamefont{Berroir}},
  \bibinfo{author}{\bibfnamefont{N.}~\bibnamefont{Pla{\c c}ais}},
  \bibinfo{author}{\bibfnamefont{T.}~\bibnamefont{Jonckheere}},
  \bibnamefont{et~al.}, \bibinfo{journal}{Phys. Rev. B}
  \textbf{\bibinfo{volume}{94}}, \bibinfo{pages}{115311}
  (\bibinfo{year}{2016}).

\bibitem[{\citenamefont{Freulon et~al.}(2015)\citenamefont{Freulon, Marguerite,
  Berroir, Pla{\c c}ais, Cavanna, Jin, and F{\`e}ve}}]{Freulon:2015-1}
\bibinfo{author}{\bibfnamefont{V.}~\bibnamefont{Freulon}},
  \bibinfo{author}{\bibfnamefont{A.}~\bibnamefont{Marguerite}},
  \bibinfo{author}{\bibfnamefont{J.}~\bibnamefont{Berroir}},
  \bibinfo{author}{\bibfnamefont{B.}~\bibnamefont{Pla{\c c}ais}},
  \bibinfo{author}{\bibfnamefont{A.}~\bibnamefont{Cavanna}},
  \bibinfo{author}{\bibfnamefont{Y.}~\bibnamefont{Jin}}, \bibnamefont{and}
  \bibinfo{author}{\bibfnamefont{G.}~\bibnamefont{F{\`e}ve}},
  \bibinfo{journal}{Nature Communications} \textbf{\bibinfo{volume}{6}},
  \bibinfo{pages}{6854} (\bibinfo{year}{2015}).

\bibitem[{\citenamefont{Tewari et~al.}(2016)\citenamefont{Tewari, Roulleau,
  Grenier, Portier, Cavanna, Gennser, Mailly, and Roche}}]{Tewari:2016-1}
\bibinfo{author}{\bibfnamefont{S.}~\bibnamefont{Tewari}},
  \bibinfo{author}{\bibfnamefont{P.}~\bibnamefont{Roulleau}},
  \bibinfo{author}{\bibfnamefont{C.}~\bibnamefont{Grenier}},
  \bibinfo{author}{\bibfnamefont{F.}~\bibnamefont{Portier}},
  \bibinfo{author}{\bibfnamefont{A.}~\bibnamefont{Cavanna}},
  \bibinfo{author}{\bibfnamefont{U.}~\bibnamefont{Gennser}},
  \bibinfo{author}{\bibfnamefont{D.}~\bibnamefont{Mailly}}, \bibnamefont{and}
  \bibinfo{author}{\bibfnamefont{P.}~\bibnamefont{Roche}},
  \bibinfo{journal}{Phys. Rev. B} \textbf{\bibinfo{volume}{93}},
  \bibinfo{pages}{035420} (\bibinfo{year}{2016}).

\bibitem[{\citenamefont{Levkivskyi and Sukhorukov}(2008)}]{Levkivskyi:2008-1}
\bibinfo{author}{\bibfnamefont{I.}~\bibnamefont{Levkivskyi}} \bibnamefont{and}
  \bibinfo{author}{\bibfnamefont{E.}~\bibnamefont{Sukhorukov}},
  \bibinfo{journal}{Phys. Rev. {\bf B}} \textbf{\bibinfo{volume}{78}},
  \bibinfo{pages}{045322} (\bibinfo{year}{2008}).

\bibitem[{\citenamefont{Slobodeniuk et~al.}(2016)\citenamefont{Slobodeniuk,
  Idrisov, and Sukhorukov}}]{Slobodeniuk:2016-1}
\bibinfo{author}{\bibfnamefont{A.~O.} \bibnamefont{Slobodeniuk}},
  \bibinfo{author}{\bibfnamefont{E.~G.} \bibnamefont{Idrisov}},
  \bibnamefont{and} \bibinfo{author}{\bibfnamefont{E.~V.}
  \bibnamefont{Sukhorukov}}, \bibinfo{journal}{Phys. Rev. B}
  \textbf{\bibinfo{volume}{93}}, \bibinfo{pages}{035421}
  (\bibinfo{year}{2016}).

\bibitem[{\citenamefont{F{\`e}ve et~al.}(2007)\citenamefont{F{\`e}ve, Mah{\'e},
  Berroir, Kontos, Pla{\c c}ais, Glattli, Cavanna, Etienne, and
  Jin}}]{Feve:2007-1}
\bibinfo{author}{\bibfnamefont{G.}~\bibnamefont{F{\`e}ve}},
  \bibinfo{author}{\bibfnamefont{A.}~\bibnamefont{Mah{\'e}}},
  \bibinfo{author}{\bibfnamefont{J.}~\bibnamefont{Berroir}},
  \bibinfo{author}{\bibfnamefont{T.}~\bibnamefont{Kontos}},
  \bibinfo{author}{\bibfnamefont{B.}~\bibnamefont{Pla{\c c}ais}},
  \bibinfo{author}{\bibfnamefont{D.C.}~\bibnamefont{Glattli}},
  \bibinfo{author}{\bibfnamefont{A.}~\bibnamefont{Cavanna}},
  \bibinfo{author}{\bibfnamefont{B.}~\bibnamefont{Etienne}}, \bibnamefont{and}
  \bibinfo{author}{\bibfnamefont{Y.}~\bibnamefont{Jin}},
  \bibinfo{journal}{Science} \textbf{\bibinfo{volume}{316}},
  \bibinfo{pages}{1169} (\bibinfo{year}{2007}).

\bibitem[{\citenamefont{Hohls et~al.}(2012)\citenamefont{Hohls, Welker, Leicht,
  Fricke, Kaestner, Mirovsky, M\"uller, Pierz, Siegner, and
  Schumacher}}]{Hohls:2012-1}
\bibinfo{author}{\bibfnamefont{F.}~\bibnamefont{Hohls}},
  \bibinfo{author}{\bibfnamefont{A.~C.} \bibnamefont{Welker}},
  \bibinfo{author}{\bibfnamefont{C.}~\bibnamefont{Leicht}},
  \bibinfo{author}{\bibfnamefont{L.}~\bibnamefont{Fricke}},
  \bibinfo{author}{\bibfnamefont{B.}~\bibnamefont{Kaestner}},
  \bibinfo{author}{\bibfnamefont{P.}~\bibnamefont{Mirovsky}},
  \bibinfo{author}{\bibfnamefont{A.}~\bibnamefont{M\"uller}},
  \bibinfo{author}{\bibfnamefont{K.}~\bibnamefont{Pierz}},
  \bibinfo{author}{\bibfnamefont{U.}~\bibnamefont{Siegner}}, \bibnamefont{and}
  \bibinfo{author}{\bibfnamefont{H.~W.} \bibnamefont{Schumacher}},
  \bibinfo{journal}{Phys. Rev. Lett.} \textbf{\bibinfo{volume}{109}},
  \bibinfo{pages}{056802} (\bibinfo{year}{2012}).

\bibitem[{\citenamefont{Waldie et~al.}(2015)\citenamefont{Waldie, See,
  Kashcheyevs, Griffiths, Farrer, Jones, Ritchie, Janssen, and
  Kataoka}}]{Waldie:2015-1}
\bibinfo{author}{\bibfnamefont{J.}~\bibnamefont{Waldie}},
  \bibinfo{author}{\bibfnamefont{P.}~\bibnamefont{See}},
  \bibinfo{author}{\bibfnamefont{V.}~\bibnamefont{Kashcheyevs}},
  \bibinfo{author}{\bibfnamefont{J.~P.} \bibnamefont{Griffiths}},
  \bibinfo{author}{\bibfnamefont{I.}~\bibnamefont{Farrer}},
  \bibinfo{author}{\bibfnamefont{G.~A.~C.} \bibnamefont{Jones}},
  \bibinfo{author}{\bibfnamefont{D.~A.} \bibnamefont{Ritchie}},
  \bibinfo{author}{\bibfnamefont{T.~J. B.~M.} \bibnamefont{Janssen}},
  \bibnamefont{and} \bibinfo{author}{\bibfnamefont{M.}~\bibnamefont{Kataoka}},
  \bibinfo{journal}{Phys. Rev. B} \textbf{\bibinfo{volume}{92}},
  \bibinfo{pages}{125305} (\bibinfo{year}{2015}).

\bibitem[{\citenamefont{Dubois et~al.}(2013{\natexlab{a}})\citenamefont{Dubois,
  Jullien, Grenier, Degiovanni, Roulleau, and Glattli}}]{Dubois:2013-1}
\bibinfo{author}{\bibfnamefont{J.}~\bibnamefont{Dubois}},
  \bibinfo{author}{\bibfnamefont{T.}~\bibnamefont{Jullien}},
  \bibinfo{author}{\bibfnamefont{C.}~\bibnamefont{Grenier}},
  \bibinfo{author}{\bibfnamefont{P.}~\bibnamefont{Degiovanni}},
  \bibinfo{author}{\bibfnamefont{P.}~\bibnamefont{Roulleau}}, \bibnamefont{and}
  \bibinfo{author}{\bibfnamefont{D.~C.} \bibnamefont{Glattli}},
  \bibinfo{journal}{Phys. Rev. B} \textbf{\bibinfo{volume}{88}},
  \bibinfo{pages}{085301} (\bibinfo{year}{2013}{\natexlab{a}}).

\bibitem[{\citenamefont{Fletcher et~al.}(2013)\citenamefont{Fletcher, See,
  Howe, Pepper, Giblin, Griffiths, Jones, Farrer, Ritchie, Janssen
  et~al.}}]{Fletcher:2013-1}
\bibinfo{author}{\bibfnamefont{J.~D.} \bibnamefont{Fletcher}},
  \bibinfo{author}{\bibfnamefont{P.}~\bibnamefont{See}},
  \bibinfo{author}{\bibfnamefont{H.}~\bibnamefont{Howe}},
  \bibinfo{author}{\bibfnamefont{M.}~\bibnamefont{Pepper}},
  \bibinfo{author}{\bibfnamefont{S.~P.} \bibnamefont{Giblin}},
  \bibinfo{author}{\bibfnamefont{J.~P.} \bibnamefont{Griffiths}},
  \bibinfo{author}{\bibfnamefont{G.~A.~C.} \bibnamefont{Jones}},
  \bibinfo{author}{\bibfnamefont{I.}~\bibnamefont{Farrer}},
  \bibinfo{author}{\bibfnamefont{D.~A.} \bibnamefont{Ritchie}},
  \bibinfo{author}{\bibfnamefont{T.~J. B.~M.} \bibnamefont{Janssen}},
  \bibnamefont{et~al.}, \bibinfo{journal}{Phys. Rev. Lett.}
  \textbf{\bibinfo{volume}{111}}, \bibinfo{pages}{216807}
  (\bibinfo{year}{2013}).

\bibitem[{\citenamefont{Hermelin et~al.}(2011)\citenamefont{Hermelin, Takada,
  Yamamoto, Tarucha, Wieck, Saminadayar, B\"{a}erle, and
  Meunier}}]{Hermelin:2011-1}
\bibinfo{author}{\bibfnamefont{S.}~\bibnamefont{Hermelin}},
  \bibinfo{author}{\bibfnamefont{S.}~\bibnamefont{Takada}},
  \bibinfo{author}{\bibfnamefont{M.}~\bibnamefont{Yamamoto}},
  \bibinfo{author}{\bibfnamefont{S.}~\bibnamefont{Tarucha}},
  \bibinfo{author}{\bibfnamefont{A.}~\bibnamefont{Wieck}},
  \bibinfo{author}{\bibfnamefont{L.}~\bibnamefont{Saminadayar}},
  \bibinfo{author}{\bibfnamefont{C.}~\bibnamefont{B\"{a}erle}},
  \bibnamefont{and} \bibinfo{author}{\bibfnamefont{T.}~\bibnamefont{Meunier}},
  \bibinfo{journal}{Nature} \textbf{\bibinfo{volume}{477}},
  \bibinfo{pages}{435} (\bibinfo{year}{2011}).

\bibitem[{\citenamefont{Ott and Moskalets}(2014)}]{Ott:2014-1}
\bibinfo{author}{\bibfnamefont{J.}~\bibnamefont{Ott}} \bibnamefont{and}
  \bibinfo{author}{\bibfnamefont{M.}~\bibnamefont{Moskalets}}
  (\bibinfo{year}{2014}), \bibinfo{note}{arXiv:1404.0185}.

\bibitem[{\citenamefont{Kashcheyevs and Samuelsson}(2017)}]{Kashcheyevs:2017-1}
\bibinfo{author}{\bibfnamefont{V.}~\bibnamefont{Kashcheyevs}} \bibnamefont{and}
  \bibinfo{author}{\bibfnamefont{P.}~\bibnamefont{Samuelsson}},
  \bibinfo{journal}{Phys. Rev. B} \textbf{\bibinfo{volume}{95}},
  \bibinfo{pages}{245424} (\bibinfo{year}{2017}).

\bibitem[{\citenamefont{Misiorny et~al.}(2018)\citenamefont{Misiorny, F\`eve,
  and Splettstoesser}}]{Misiorny:2018-1}
\bibinfo{author}{\bibfnamefont{M.}~\bibnamefont{Misiorny}},
  \bibinfo{author}{\bibfnamefont{G.}~\bibnamefont{F\`eve}}, \bibnamefont{and}
  \bibinfo{author}{\bibfnamefont{J.}~\bibnamefont{Splettstoesser}},
  \bibinfo{journal}{Phys. Rev. B} \textbf{\bibinfo{volume}{97}},
  \bibinfo{pages}{075426} (\bibinfo{year}{2018}).

\bibitem[{\citenamefont{Marguerite et~al.}(2017)\citenamefont{Marguerite,
  Roussel, Bisognin, Cabart, Kumar, Berroir, Bocquillon, Pla{\c c}ais, Cavanna,
  Gennser et~al.}}]{Marguerite:2017-1}
\bibinfo{author}{\bibfnamefont{A.}~\bibnamefont{Marguerite}},
  \bibinfo{author}{\bibfnamefont{B.}~\bibnamefont{Roussel}},
  \bibinfo{author}{\bibfnamefont{R.}~\bibnamefont{Bisognin}},
  \bibinfo{author}{\bibfnamefont{C.}~\bibnamefont{Cabart}},
  \bibinfo{author}{\bibfnamefont{M.}~\bibnamefont{Kumar}},
  \bibinfo{author}{\bibfnamefont{J.}~\bibnamefont{Berroir}},
  \bibinfo{author}{\bibfnamefont{E.}~\bibnamefont{Bocquillon}},
  \bibinfo{author}{\bibfnamefont{B.}~\bibnamefont{Pla{\c c}ais}},
  \bibinfo{author}{\bibfnamefont{A.}~\bibnamefont{Cavanna}},
  \bibinfo{author}{\bibfnamefont{U.}~\bibnamefont{Gennser}},
  \bibnamefont{et~al.} (\bibinfo{year}{2017}), \bibinfo{note}{submitted to
  PRL}.

\bibitem[{\citenamefont{Johnson et~al.}(2017)\citenamefont{Johnson, Fletcher,
  Humphreys, See, Griffiths, Jones, Farrer, Ritchie, Pepper, Janssen
  et~al.}}]{Johnson:2017-1}
\bibinfo{author}{\bibfnamefont{N.}~\bibnamefont{Johnson}},
  \bibinfo{author}{\bibfnamefont{J.~D.} \bibnamefont{Fletcher}},
  \bibinfo{author}{\bibfnamefont{D.~A.} \bibnamefont{Humphreys}},
  \bibinfo{author}{\bibfnamefont{P.}~\bibnamefont{See}},
  \bibinfo{author}{\bibfnamefont{J.~P.} \bibnamefont{Griffiths}},
  \bibinfo{author}{\bibfnamefont{G.~A.~C.} \bibnamefont{Jones}},
  \bibinfo{author}{\bibfnamefont{I.}~\bibnamefont{Farrer}},
  \bibinfo{author}{\bibfnamefont{D.~A.} \bibnamefont{Ritchie}},
  \bibinfo{author}{\bibfnamefont{M.}~\bibnamefont{Pepper}},
  \bibinfo{author}{\bibfnamefont{T.~J. B.~M.} \bibnamefont{Janssen}},
  \bibnamefont{et~al.}, \bibinfo{journal}{Applied Physics Letters}
  \textbf{\bibinfo{volume}{110}}, \bibinfo{pages}{102105}
  (\bibinfo{year}{2017}).

\bibitem[{\citenamefont{Degiovanni et~al.}(2009)\citenamefont{Degiovanni,
  Grenier, and F{\`e}ve}}]{Degio:2009-1}
\bibinfo{author}{\bibfnamefont{P.}~\bibnamefont{Degiovanni}},
  \bibinfo{author}{\bibfnamefont{C.}~\bibnamefont{Grenier}}, \bibnamefont{and}
  \bibinfo{author}{\bibfnamefont{G.}~\bibnamefont{F{\`e}ve}},
  \bibinfo{journal}{Phys. Rev. {\bf B}} \textbf{\bibinfo{volume}{80}},
  \bibinfo{pages}{241307(R)} (\bibinfo{year}{2009}).

\bibitem[{\citenamefont{Ferraro
  et~al.}(2014{\natexlab{a}})\citenamefont{Ferraro, Wahl, Rech, Jonckheere, and
  Martin}}]{Ferraro:2014-1}
\bibinfo{author}{\bibfnamefont{D.}~\bibnamefont{Ferraro}},
  \bibinfo{author}{\bibfnamefont{C.}~\bibnamefont{Wahl}},
  \bibinfo{author}{\bibfnamefont{J.}~\bibnamefont{Rech}},
  \bibinfo{author}{\bibfnamefont{T.}~\bibnamefont{Jonckheere}},
  \bibnamefont{and} \bibinfo{author}{\bibfnamefont{T.}~\bibnamefont{Martin}},
  \bibinfo{journal}{Phys. Rev. B} \textbf{\bibinfo{volume}{89}},
  \bibinfo{pages}{075407} (\bibinfo{year}{2014}{\natexlab{a}}).

\bibitem[{\citenamefont{Altimiras et~al.}(2010)\citenamefont{Altimiras,
  Le~Sueur, Gennser, Cavanna, Mailly, and Pierre}}]{Altimiras:2010-2}
\bibinfo{author}{\bibfnamefont{C.}~\bibnamefont{Altimiras}},
  \bibinfo{author}{\bibfnamefont{H.}~\bibnamefont{Le~Sueur}},
  \bibinfo{author}{\bibfnamefont{U.}~\bibnamefont{Gennser}},
  \bibinfo{author}{\bibfnamefont{A.}~\bibnamefont{Cavanna}},
  \bibinfo{author}{\bibfnamefont{D.}~\bibnamefont{Mailly}}, \bibnamefont{and}
  \bibinfo{author}{\bibfnamefont{F.}~\bibnamefont{Pierre}},
  \bibinfo{journal}{Phys. Rev. Lett.} \textbf{\bibinfo{volume}{105}},
  \bibinfo{pages}{226804} (\bibinfo{year}{2010}).

\bibitem[{\citenamefont{Huynh et~al.}(2012)\citenamefont{Huynh, Portier,
  le~Sueur, Faini, Gennser, Mailly, Pierre, Wegschider, and
  Roche}}]{Huynh:2012-1}
\bibinfo{author}{\bibfnamefont{P.-A.} \bibnamefont{Huynh}},
  \bibinfo{author}{\bibfnamefont{F.}~\bibnamefont{Portier}},
  \bibinfo{author}{\bibfnamefont{H.}~\bibnamefont{le~Sueur}},
  \bibinfo{author}{\bibfnamefont{G.}~\bibnamefont{Faini}},
  \bibinfo{author}{\bibfnamefont{U.}~\bibnamefont{Gennser}},
  \bibinfo{author}{\bibfnamefont{D.}~\bibnamefont{Mailly}},
  \bibinfo{author}{\bibfnamefont{F.}~\bibnamefont{Pierre}},
  \bibinfo{author}{\bibfnamefont{W.}~\bibnamefont{Wegschider}},
  \bibnamefont{and} \bibinfo{author}{\bibfnamefont{P.}~\bibnamefont{Roche}},
  \bibinfo{journal}{Phys. Rev. Lett.} \textbf{\bibinfo{volume}{108}},
  \bibinfo{pages}{256802} (\bibinfo{year}{2012}).

\bibitem[{\citenamefont{Glauber}(1963)}]{Glauber:1963-1}
\bibinfo{author}{\bibfnamefont{R.}~\bibnamefont{Glauber}},
  \bibinfo{journal}{Phys. Rev.} \textbf{\bibinfo{volume}{130}},
  \bibinfo{pages}{2529} (\bibinfo{year}{1963}).

\bibitem[{\citenamefont{Grenier et~al.}(2011)\citenamefont{Grenier, Herv{\'e},
  Bocquillon, Parmentier, Pla{\c c}ais, Berroir, F{\`e}ve, and
  Degiovanni}}]{Degio:2010-4}
\bibinfo{author}{\bibfnamefont{C.}~\bibnamefont{Grenier}},
  \bibinfo{author}{\bibfnamefont{R.}~\bibnamefont{Herv{\'e}}},
  \bibinfo{author}{\bibfnamefont{E.}~\bibnamefont{Bocquillon}},
  \bibinfo{author}{\bibfnamefont{F.}~\bibnamefont{Parmentier}},
  \bibinfo{author}{\bibfnamefont{B.}~\bibnamefont{Pla{\c c}ais}},
  \bibinfo{author}{\bibfnamefont{J.}~\bibnamefont{Berroir}},
  \bibinfo{author}{\bibfnamefont{G.}~\bibnamefont{F{\`e}ve}}, \bibnamefont{and}
  \bibinfo{author}{\bibfnamefont{P.}~\bibnamefont{Degiovanni}},
  \bibinfo{journal}{New Journal of Physics} \textbf{\bibinfo{volume}{13}},
  \bibinfo{pages}{093007} (\bibinfo{year}{2011}).

\bibitem[{\citenamefont{Haack et~al.}(2011)\citenamefont{Haack, Moskalets,
  Splettstoesser, and B\"uttiker}}]{Haack:2011-1}
\bibinfo{author}{\bibfnamefont{G.}~\bibnamefont{Haack}},
  \bibinfo{author}{\bibfnamefont{M.}~\bibnamefont{Moskalets}},
  \bibinfo{author}{\bibfnamefont{J.}~\bibnamefont{Splettstoesser}},
  \bibnamefont{and}
  \bibinfo{author}{\bibfnamefont{M.}~\bibnamefont{B\"uttiker}},
  \bibinfo{journal}{Phys. Rev. B} \textbf{\bibinfo{volume}{84}},
  \bibinfo{pages}{081303} (\bibinfo{year}{2011}).

\bibitem[{\citenamefont{Haack et~al.}(2012)\citenamefont{Haack, Moskalets, and
  B\"uttiker}}]{Haack:2012-2}
\bibinfo{author}{\bibfnamefont{G.}~\bibnamefont{Haack}},
  \bibinfo{author}{\bibfnamefont{M.}~\bibnamefont{Moskalets}},
  \bibnamefont{and}
  \bibinfo{author}{\bibfnamefont{M.}~\bibnamefont{B\"uttiker}},
  \bibinfo{journal}{Phys. Rev. B} \textbf{\bibinfo{volume}{87}}
  (\bibinfo{year}{2012}).

\bibitem[{\citenamefont{Ferraro et~al.}(2013)\citenamefont{Ferraro, Feller,
  Ghibaudo, Thibierge, Bocquillon, F{\`e}ve, Grenier, and
  Degiovanni}}]{Ferraro:2013-1}
\bibinfo{author}{\bibfnamefont{D.}~\bibnamefont{Ferraro}},
  \bibinfo{author}{\bibfnamefont{A.}~\bibnamefont{Feller}},
  \bibinfo{author}{\bibfnamefont{A.}~\bibnamefont{Ghibaudo}},
  \bibinfo{author}{\bibfnamefont{E.}~\bibnamefont{Thibierge}},
  \bibinfo{author}{\bibfnamefont{E.}~\bibnamefont{Bocquillon}},
  \bibinfo{author}{\bibfnamefont{G.}~\bibnamefont{F{\`e}ve}},
  \bibinfo{author}{\bibfnamefont{C.}~\bibnamefont{Grenier}}, \bibnamefont{and}
  \bibinfo{author}{\bibfnamefont{P.}~\bibnamefont{Degiovanni}},
  \bibinfo{journal}{Phys. Rev. B} \textbf{\bibinfo{volume}{88}},
  \bibinfo{pages}{205303} (\bibinfo{year}{2013}).

\bibitem[{\citenamefont{Jullien et~al.}(2014)\citenamefont{Jullien, Roulleau,
  Roche, Cavanna, Jin, and Glattli}}]{Jullien:2014-1}
\bibinfo{author}{\bibfnamefont{T.}~\bibnamefont{Jullien}},
  \bibinfo{author}{\bibfnamefont{P.}~\bibnamefont{Roulleau}},
  \bibinfo{author}{\bibfnamefont{B.}~\bibnamefont{Roche}},
  \bibinfo{author}{\bibfnamefont{A.}~\bibnamefont{Cavanna}},
  \bibinfo{author}{\bibfnamefont{Y.}~\bibnamefont{Jin}}, \bibnamefont{and}
  \bibinfo{author}{\bibfnamefont{D.~C.} \bibnamefont{Glattli}},
  \bibinfo{journal}{Nature} \textbf{\bibinfo{volume}{514}},
  \bibinfo{pages}{603} (\bibinfo{year}{2014}).

\bibitem[{\citenamefont{Mah{\'e} et~al.}(2008)\citenamefont{Mah{\'e},
  Parmentier, F{\`e}ve, Berroir, Kontos, Cavanna, Etienne, Jin, Glattli, and
  Pla{\c c}ais}}]{Mahe:2008-1}
\bibinfo{author}{\bibfnamefont{A.}~\bibnamefont{Mah{\'e}}},
  \bibinfo{author}{\bibfnamefont{F.}~\bibnamefont{Parmentier}},
  \bibinfo{author}{\bibfnamefont{G.}~\bibnamefont{F{\`e}ve}},
  \bibinfo{author}{\bibfnamefont{J.}~\bibnamefont{Berroir}},
  \bibinfo{author}{\bibfnamefont{T.}~\bibnamefont{Kontos}},
  \bibinfo{author}{\bibfnamefont{A.}~\bibnamefont{Cavanna}},
  \bibinfo{author}{\bibfnamefont{B.}~\bibnamefont{Etienne}},
  \bibinfo{author}{\bibfnamefont{Y.}~\bibnamefont{Jin}},
  \bibinfo{author}{\bibfnamefont{D. C.}~\bibnamefont{Glattli}}, \bibnamefont{and}
  \bibinfo{author}{\bibfnamefont{B.}~\bibnamefont{Pla{\c c}ais}},
  \bibinfo{journal}{Journal of Low Temperature Physics}
  \textbf{\bibinfo{volume}{153}}, \bibinfo{pages}{339} (\bibinfo{year}{2008}).

\bibitem[{\citenamefont{Mah{\'e} et~al.}(2010)\citenamefont{Mah{\'e},
  Parmentier, Bocquillon, Berroir, Glattli, Kontos, Pla{\c c}ais, F{\`e}ve,
  Cavanna, and Jin}}]{Mahe:2010-1}
\bibinfo{author}{\bibfnamefont{A.}~\bibnamefont{Mah{\'e}}},
  \bibinfo{author}{\bibfnamefont{F.}~\bibnamefont{Parmentier}},
  \bibinfo{author}{\bibfnamefont{E.}~\bibnamefont{Bocquillon}},
  \bibinfo{author}{\bibfnamefont{J.}~\bibnamefont{Berroir}},
  \bibinfo{author}{\bibfnamefont{D. C.}~\bibnamefont{Glattli}},
  \bibinfo{author}{\bibfnamefont{T.}~\bibnamefont{Kontos}},
  \bibinfo{author}{\bibfnamefont{B.}~\bibnamefont{Pla{\c c}ais}},
  \bibinfo{author}{\bibfnamefont{G.}~\bibnamefont{F{\`e}ve}},
  \bibinfo{author}{\bibfnamefont{A.}~\bibnamefont{Cavanna}}, \bibnamefont{and}
  \bibinfo{author}{\bibfnamefont{Y.}~\bibnamefont{Jin}},
  \bibinfo{journal}{Phys. Rev. B} \textbf{\bibinfo{volume}{82}},
  \bibinfo{pages}{201309} (\bibinfo{year}{2010}).

\bibitem[{\citenamefont{Roussel et~al.}()\citenamefont{Roussel, Cabart,
  F{\`e}ve, and Degiovanni}}]{Roussel:2017-1}
\bibinfo{author}{\bibfnamefont{B.}~\bibnamefont{Roussel}},
  \bibinfo{author}{\bibfnamefont{C.}~\bibnamefont{Cabart}},
  \bibinfo{author}{\bibfnamefont{R.}~\bibnamefont{Bisongnin}},
  \bibinfo{author}{\bibfnamefont{G.}~\bibnamefont{F{\`e}ve}}, \bibnamefont{and}
  \bibinfo{author}{\bibfnamefont{P.}~\bibnamefont{Degiovanni}},
  \bibinfo{note}{in preparation}.

\bibitem[{\citenamefont{Dubois et~al.}(2013{\natexlab{b}})\citenamefont{Dubois,
  Jullien, Portier, Roche, Cavanna, Jin, Wegscheider, Roulleau, and
  Glattli}}]{Dubois:2013-2}
\bibinfo{author}{\bibfnamefont{J.}~\bibnamefont{Dubois}},
  \bibinfo{author}{\bibfnamefont{T.}~\bibnamefont{Jullien}},
  \bibinfo{author}{\bibfnamefont{F.}~\bibnamefont{Portier}},
  \bibinfo{author}{\bibfnamefont{P.}~\bibnamefont{Roche}},
  \bibinfo{author}{\bibfnamefont{A.}~\bibnamefont{Cavanna}},
  \bibinfo{author}{\bibfnamefont{Y.}~\bibnamefont{Jin}},
  \bibinfo{author}{\bibfnamefont{W.}~\bibnamefont{Wegscheider}},
  \bibinfo{author}{\bibfnamefont{P.}~\bibnamefont{Roulleau}}, \bibnamefont{and}
  \bibinfo{author}{\bibfnamefont{D. C.}~\bibnamefont{Glattli}},
  \bibinfo{journal}{Nature} \textbf{\bibinfo{volume}{502}},
  \bibinfo{pages}{659} (\bibinfo{year}{2013}{\natexlab{b}}).

\bibitem[{\citenamefont{Levitov et~al.}(1996)\citenamefont{Levitov, Lee, and
  Lesovik}}]{Levitov:1996-1}
\bibinfo{author}{\bibfnamefont{L.}~\bibnamefont{Levitov}},
  \bibinfo{author}{\bibfnamefont{H.}~\bibnamefont{Lee}}, \bibnamefont{and}
  \bibinfo{author}{\bibfnamefont{G.}~\bibnamefont{Lesovik}},
  \bibinfo{journal}{J. Math. Phys.} \textbf{\bibinfo{volume}{37}},
  \bibinfo{pages}{4845} (\bibinfo{year}{1996}).

\bibitem[{\citenamefont{Keeling et~al.}(2006)\citenamefont{Keeling, Klich, and
  Levitov}}]{Keeling:2006-1}
\bibinfo{author}{\bibfnamefont{J.}~\bibnamefont{Keeling}},
  \bibinfo{author}{\bibfnamefont{I.}~\bibnamefont{Klich}}, \bibnamefont{and}
  \bibinfo{author}{\bibfnamefont{L.}~\bibnamefont{Levitov}},
  \bibinfo{journal}{Phys. Rev. Lett.} \textbf{\bibinfo{volume}{97}},
  \bibinfo{pages}{116403} (\bibinfo{year}{2006}).

\bibitem[{\citenamefont{Grenier et~al.}(2013)\citenamefont{Grenier, Dubois,
  Jullien, Roulleau, Glattli, and Degiovanni}}]{Grenier:2013-1}
\bibinfo{author}{\bibfnamefont{C.}~\bibnamefont{Grenier}},
  \bibinfo{author}{\bibfnamefont{J.}~\bibnamefont{Dubois}},
  \bibinfo{author}{\bibfnamefont{T.}~\bibnamefont{Jullien}},
  \bibinfo{author}{\bibfnamefont{P.}~\bibnamefont{Roulleau}},
  \bibinfo{author}{\bibfnamefont{D.~C.} \bibnamefont{Glattli}},
  \bibnamefont{and}
  \bibinfo{author}{\bibfnamefont{P.}~\bibnamefont{Degiovanni}},
  \bibinfo{journal}{Phys. Rev. B} \textbf{\bibinfo{volume}{88}},
  \bibinfo{pages}{085302} (\bibinfo{year}{2013}).

\bibitem[{\citenamefont{Landau}(1957)}]{Landau:1957-1}
\bibinfo{author}{\bibfnamefont{L.}~\bibnamefont{Landau}},
  \bibinfo{journal}{Sov. Phys. JETP} \textbf{\bibinfo{volume}{5}},
  \bibinfo{pages}{101 } (\bibinfo{year}{1957}).

\bibitem[{\citenamefont{Devoret et~al.}(1990)\citenamefont{Devoret, Esteve,
  Grabert, Ingold, Pothier, and Urbina}}]{Devoret:1990-1}
\bibinfo{author}{\bibfnamefont{M.}~\bibnamefont{Devoret}},
  \bibinfo{author}{\bibfnamefont{D.}~\bibnamefont{Esteve}},
  \bibinfo{author}{\bibfnamefont{H.}~\bibnamefont{Grabert}},
  \bibinfo{author}{\bibfnamefont{G.-L.} \bibnamefont{Ingold}},
  \bibinfo{author}{\bibfnamefont{H.}~\bibnamefont{Pothier}}, \bibnamefont{and}
  \bibinfo{author}{\bibfnamefont{C.}~\bibnamefont{Urbina}},
  \bibinfo{journal}{Phys. Rev. Lett.} \textbf{\bibinfo{volume}{64}},
  \bibinfo{pages}{1824} (\bibinfo{year}{1990}).

\bibitem[{\citenamefont{Girvin et~al.}(1990)\citenamefont{Girvin, Glazman,
  Jonson, Penn, and Stiles}}]{Girvin:1990-1}
\bibinfo{author}{\bibfnamefont{S.}~\bibnamefont{Girvin}},
  \bibinfo{author}{\bibfnamefont{L.}~\bibnamefont{Glazman}},
  \bibinfo{author}{\bibfnamefont{M.}~\bibnamefont{Jonson}},
  \bibinfo{author}{\bibfnamefont{D.}~\bibnamefont{Penn}}, \bibnamefont{and}
  \bibinfo{author}{\bibfnamefont{M.}~\bibnamefont{Stiles}},
  \bibinfo{journal}{Phys. Rev. Lett.} \textbf{\bibinfo{volume}{64}},
  \bibinfo{pages}{3183} (\bibinfo{year}{1990}).

\bibitem[{\citenamefont{Ferraro
  et~al.}(2014{\natexlab{b}})\citenamefont{Ferraro, Roussel, Cabart, Thibierge,
  F\`eve, Grenier, and Degiovanni}}]{Ferraro:2014-2}
\bibinfo{author}{\bibfnamefont{D.}~\bibnamefont{Ferraro}},
  \bibinfo{author}{\bibfnamefont{B.}~\bibnamefont{Roussel}},
  \bibinfo{author}{\bibfnamefont{C.}~\bibnamefont{Cabart}},
  \bibinfo{author}{\bibfnamefont{E.}~\bibnamefont{Thibierge}},
  \bibinfo{author}{\bibfnamefont{G.}~\bibnamefont{F\`eve}},
  \bibinfo{author}{\bibfnamefont{C.}~\bibnamefont{Grenier}}, \bibnamefont{and}
  \bibinfo{author}{\bibfnamefont{P.}~\bibnamefont{Degiovanni}},
  \bibinfo{journal}{Phys. Rev. Lett.} \textbf{\bibinfo{volume}{113}},
  \bibinfo{pages}{166403} (\bibinfo{year}{2014}{\natexlab{b}}).

\bibitem[{\citenamefont{Zurek et~al.}(1993)\citenamefont{Zurek, Habib, and
  Paz}}]{Zurek:1993-1}
\bibinfo{author}{\bibfnamefont{W.}~\bibnamefont{Zurek}},
  \bibinfo{author}{\bibfnamefont{S.}~\bibnamefont{Habib}}, \bibnamefont{and}
  \bibinfo{author}{\bibfnamefont{J.}~\bibnamefont{Paz}},
  \bibinfo{journal}{Phys. Rev. Lett.} \textbf{\bibinfo{volume}{70}},
  \bibinfo{pages}{1187 } (\bibinfo{year}{1993}).

\bibitem[{\citenamefont{Safi and Schulz}(1995{\natexlab{a}})}]{Safi:1995-1}
\bibinfo{author}{\bibfnamefont{I.}~\bibnamefont{Safi}} \bibnamefont{and}
  \bibinfo{author}{\bibfnamefont{H.}~\bibnamefont{Schulz}},
  \bibinfo{journal}{Phys. Rev. B} \textbf{\bibinfo{volume}{52}},
  \bibinfo{pages}{R1740} (\bibinfo{year}{1995}{\natexlab{a}}).

\bibitem[{\citenamefont{Safi and Schulz}(1995{\natexlab{b}})}]{Safi:1995-2}
\bibinfo{author}{\bibfnamefont{I.}~\bibnamefont{Safi}} \bibnamefont{and}
  \bibinfo{author}{\bibfnamefont{H.}~\bibnamefont{Schulz}}, in
  \emph{\bibinfo{booktitle}{Quantum Transport in Semiconductor Submicron
  Structures}}, edited by
  \bibinfo{editor}{\bibfnamefont{B.}~\bibnamefont{Kramer}}
  (\bibinfo{publisher}{Kluwer Academic Press, Dordrecht},
  \bibinfo{year}{1995}{\natexlab{b}}), p. \bibinfo{pages}{159}.

\bibitem[{\citenamefont{Safi}(1999)}]{Safi:1999-1}
\bibinfo{author}{\bibfnamefont{I.}~\bibnamefont{Safi}}, \bibinfo{journal}{Eur.
  Phys. J. {\bf D}} \textbf{\bibinfo{volume}{12}}, \bibinfo{pages}{451}
  (\bibinfo{year}{1999}).

\bibitem[{\citenamefont{Degiovanni et~al.}(2010)\citenamefont{Degiovanni,
  Grenier, F\`eve, Altimiras, le~Sueur, and Pierre}}]{Degio:2010-1}
\bibinfo{author}{\bibfnamefont{P.}~\bibnamefont{Degiovanni}},
  \bibinfo{author}{\bibfnamefont{C.}~\bibnamefont{Grenier}},
  \bibinfo{author}{\bibfnamefont{G.}~\bibnamefont{F\`eve}},
  \bibinfo{author}{\bibfnamefont{C.}~\bibnamefont{Altimiras}},
  \bibinfo{author}{\bibfnamefont{H.}~\bibnamefont{le~Sueur}}, \bibnamefont{and}
  \bibinfo{author}{\bibfnamefont{F.}~\bibnamefont{Pierre}},
  \bibinfo{journal}{Phys. Rev. B} \textbf{\bibinfo{volume}{81}},
  \bibinfo{pages}{121302(R)} (\bibinfo{year}{2010}).

\bibitem[{\citenamefont{Cauer}(1926)}]{Cauer:1926}
\bibinfo{author}{\bibfnamefont{W.}~\bibnamefont{Cauer}},
  \bibinfo{journal}{Archiv f{\"u}r Elektrotechnik}
  \textbf{\bibinfo{volume}{17}}, \bibinfo{pages}{355} (\bibinfo{year}{1926}).

\bibitem[{\citenamefont{Brune}(1931)}]{Brune:1931}
\bibinfo{author}{\bibfnamefont{O.}~\bibnamefont{Brune}}, \bibinfo{journal}{J.
  Math. and Phys.} \textbf{\bibinfo{volume}{10}}, \bibinfo{pages}{191}
  (\bibinfo{year}{1931}).

\bibitem[{\citenamefont{Bocquillon
  et~al.}(2013{\natexlab{a}})\citenamefont{Bocquillon, Freulon, Berroir,
  Degiovanni, Pla{\c c}ais, Cavanna, Jin, and F{\`e}ve}}]{Bocquillon:2013-2}
\bibinfo{author}{\bibfnamefont{E.}~\bibnamefont{Bocquillon}},
  \bibinfo{author}{\bibfnamefont{V.}~\bibnamefont{Freulon}},
  \bibinfo{author}{\bibfnamefont{J.}~\bibnamefont{Berroir}},
  \bibinfo{author}{\bibfnamefont{P.}~\bibnamefont{Degiovanni}},
  \bibinfo{author}{\bibfnamefont{B.}~\bibnamefont{Pla{\c c}ais}},
  \bibinfo{author}{\bibfnamefont{A.}~\bibnamefont{Cavanna}},
  \bibinfo{author}{\bibfnamefont{Y.}~\bibnamefont{Jin}}, \bibnamefont{and}
  \bibinfo{author}{\bibfnamefont{G.}~\bibnamefont{F{\`e}ve}},
  \bibinfo{journal}{Nature Communications} \textbf{\bibinfo{volume}{4}},
  \bibinfo{pages}{1839} (\bibinfo{year}{2013}{\natexlab{a}}).

\bibitem[{\citenamefont{Petkovi{\'c} et~al.}(2014)\citenamefont{Petkovi{\'c},
  Williams, and Glattli}}]{Petkovic:2014-1}
\bibinfo{author}{\bibfnamefont{I.}~\bibnamefont{Petkovi{\'c}}},
  \bibinfo{author}{\bibfnamefont{F.~I.~B.} \bibnamefont{Williams}},
  \bibnamefont{and} \bibinfo{author}{\bibfnamefont{D.~C.}
  \bibnamefont{Glattli}}, \bibinfo{journal}{Journal of Physics D: Applied
  Physics} \textbf{\bibinfo{volume}{47}}, \bibinfo{pages}{094010}
  (\bibinfo{year}{2014}).

\bibitem[{\citenamefont{Hashisaka et~al.}(2017)\citenamefont{Hashisaka, Hiyama,
  Akiho, Muraki, and Fujisawa}}]{Hashisaka:2017-1}
\bibinfo{author}{\bibfnamefont{M.}~\bibnamefont{Hashisaka}},
  \bibinfo{author}{\bibfnamefont{N.}~\bibnamefont{Hiyama}},
  \bibinfo{author}{\bibfnamefont{T.}~\bibnamefont{Akiho}},
  \bibinfo{author}{\bibfnamefont{K.}~\bibnamefont{Muraki}}, \bibnamefont{and}
  \bibinfo{author}{\bibfnamefont{T.}~\bibnamefont{Fujisawa}},
  \bibinfo{journal}{Nature Physics} \textbf{\bibinfo{volume}{advance online
  publication}},  (\bibinfo{year}{2017}).

\bibitem[{\citenamefont{Inoue et~al.}(2014)\citenamefont{Inoue, Grivnin, Ofek,
  Neder, Heiblum, Umansky, and Mahalu}}]{Inoue:2013-1}
\bibinfo{author}{\bibfnamefont{H.}~\bibnamefont{Inoue}},
  \bibinfo{author}{\bibfnamefont{A.}~\bibnamefont{Grivnin}},
  \bibinfo{author}{\bibfnamefont{N.}~\bibnamefont{Ofek}},
  \bibinfo{author}{\bibfnamefont{I.}~\bibnamefont{Neder}},
  \bibinfo{author}{\bibfnamefont{M.}~\bibnamefont{Heiblum}},
  \bibinfo{author}{\bibfnamefont{V.}~\bibnamefont{Umansky}}, \bibnamefont{and}
  \bibinfo{author}{\bibfnamefont{D.}~\bibnamefont{Mahalu}},
  \bibinfo{journal}{Phys. Rev. Lett.} \textbf{\bibinfo{volume}{112}},
  \bibinfo{pages}{166801} (\bibinfo{year}{2014}).

\bibitem[{\citenamefont{Pr{\^e}tre et~al.}(1996)\citenamefont{Pr{\^e}tre,
  Thomas, and B{\"u}ttiker}}]{Pretre:1996-1}
\bibinfo{author}{\bibfnamefont{A.}~\bibnamefont{Pr{\^e}tre}},
  \bibinfo{author}{\bibfnamefont{H.}~\bibnamefont{Thomas}}, \bibnamefont{and}
  \bibinfo{author}{\bibfnamefont{M.}~\bibnamefont{B{\"u}ttiker}},
  \bibinfo{journal}{Phys. Rev. {\bf B}} \textbf{\bibinfo{volume}{54}},
  \bibinfo{pages}{8130} (\bibinfo{year}{1996}).

\bibitem[{\citenamefont{Christen and B{\"u}ttiker}(1996)}]{Christen:1996-1}
\bibinfo{author}{\bibfnamefont{T.}~\bibnamefont{Christen}} \bibnamefont{and}
  \bibinfo{author}{\bibfnamefont{M.}~\bibnamefont{B{\"u}ttiker}},
  \bibinfo{journal}{Phys. Rev. {\bf B}} \textbf{\bibinfo{volume}{53}},
  \bibinfo{pages}{2064} (\bibinfo{year}{1996}).

\bibitem[{\citenamefont{Ingold and Nazarov}(1992)}]{Ingold:1992-1}
\bibinfo{author}{\bibfnamefont{G.-L.} \bibnamefont{Ingold}} \bibnamefont{and}
  \bibinfo{author}{\bibfnamefont{Y.}~\bibnamefont{Nazarov}},
  \emph{\bibinfo{title}{Single charge tunneling}} (\bibinfo{publisher}{Plenum
  Press, New York}, \bibinfo{year}{1992}), vol. \bibinfo{volume}{294} of
  \emph{\bibinfo{series}{NATO ASI Series B}}, chap. \bibinfo{chapter}{Charge
  tunneling rates in ultrasmall junctions}, pp. \bibinfo{pages}{21--107}.

\bibitem[{\citenamefont{Bocquillon
  et~al.}(2013{\natexlab{b}})\citenamefont{Bocquillon, Freulon, Berroir,
  Degiovanni, Pla{\c c}ais, Cavanna, Jin, and F{\`e}ve}}]{Bocquillon:2013-1}
\bibinfo{author}{\bibfnamefont{E.}~\bibnamefont{Bocquillon}},
  \bibinfo{author}{\bibfnamefont{V.}~\bibnamefont{Freulon}},
  \bibinfo{author}{\bibfnamefont{J.}~\bibnamefont{Berroir}},
  \bibinfo{author}{\bibfnamefont{P.}~\bibnamefont{Degiovanni}},
  \bibinfo{author}{\bibfnamefont{B.}~\bibnamefont{Pla{\c c}ais}},
  \bibinfo{author}{\bibfnamefont{A.}~\bibnamefont{Cavanna}},
  \bibinfo{author}{\bibfnamefont{Y.}~\bibnamefont{Jin}}, \bibnamefont{and}
  \bibinfo{author}{\bibfnamefont{G.}~\bibnamefont{F{\`e}ve}},
  \bibinfo{journal}{Science} \textbf{\bibinfo{volume}{339}},
  \bibinfo{pages}{1054} (\bibinfo{year}{2013}{\natexlab{b}}).

\bibitem[{\citenamefont{Ol'khovskaya et~al.}(2008)\citenamefont{Ol'khovskaya,
  Splettstoesser, Moskalets, and B{\"u}ttiker}}]{Olkhovskaya:2008-1}
\bibinfo{author}{\bibfnamefont{S.}~\bibnamefont{Ol'khovskaya}},
  \bibinfo{author}{\bibfnamefont{J.}~\bibnamefont{Splettstoesser}},
  \bibinfo{author}{\bibfnamefont{M.}~\bibnamefont{Moskalets}},
  \bibnamefont{and}
  \bibinfo{author}{\bibfnamefont{M.}~\bibnamefont{B{\"u}ttiker}},
  \bibinfo{journal}{Phys. Rev. Lett.} \textbf{\bibinfo{volume}{101}},
  \bibinfo{pages}{166802} (\bibinfo{year}{2008}).

\bibitem[{\citenamefont{Volkov and Mikhailov}(1988)}]{Volkov:1988-1}
\bibinfo{author}{\bibfnamefont{V.}~\bibnamefont{Volkov}} \bibnamefont{and}
  \bibinfo{author}{\bibfnamefont{S.}~\bibnamefont{Mikhailov}},
  \bibinfo{journal}{Sov. Phys. JETP} \textbf{\bibinfo{volume}{67}},
  \bibinfo{pages}{1639} (\bibinfo{year}{1988}).

\bibitem[{\citenamefont{Kumada et~al.}(2011)\citenamefont{Kumada, Kamata, and
  Fujisawa}}]{Kumada:2011-1}
\bibinfo{author}{\bibfnamefont{N.}~\bibnamefont{Kumada}},
  \bibinfo{author}{\bibfnamefont{H.}~\bibnamefont{Kamata}}, \bibnamefont{and}
  \bibinfo{author}{\bibfnamefont{T.}~\bibnamefont{Fujisawa}},
  \bibinfo{journal}{Phys. Rev. B} \textbf{\bibinfo{volume}{84}},
  \bibinfo{pages}{045314} (\bibinfo{year}{2011}).

\bibitem[{\citenamefont{Kumada et~al.}(2014)\citenamefont{Kumada, Roulleau,
  Roche, Hashisaka, Hibino, Petkovi\ifmmode~\acute{c}\else \'{c}\fi{}, and
  Glattli}}]{Kumada:2014-1}
\bibinfo{author}{\bibfnamefont{N.}~\bibnamefont{Kumada}},
  \bibinfo{author}{\bibfnamefont{P.}~\bibnamefont{Roulleau}},
  \bibinfo{author}{\bibfnamefont{B.}~\bibnamefont{Roche}},
  \bibinfo{author}{\bibfnamefont{M.}~\bibnamefont{Hashisaka}},
  \bibinfo{author}{\bibfnamefont{H.}~\bibnamefont{Hibino}},
  \bibinfo{author}{\bibfnamefont{I.}~\bibnamefont{Petkovi\ifmmode~\acute{c}\else
  \'{c}\fi{}}}, \bibnamefont{and} \bibinfo{author}{\bibfnamefont{D.~C.}
  \bibnamefont{Glattli}}, \bibinfo{journal}{Phys. Rev. Lett.}
  \textbf{\bibinfo{volume}{113}}, \bibinfo{pages}{266601}
  (\bibinfo{year}{2014}).

\bibitem[{\citenamefont{Wei et~al.}(2017)\citenamefont{Wei, van~der Sar,
  Sanchez-Yamagishi, Watanabe, Taniguchi, Jarillo-Herrero, Halperin, and
  Yacoby}}]{Wei:2017-1}
\bibinfo{author}{\bibfnamefont{D.~S.} \bibnamefont{Wei}},
  \bibinfo{author}{\bibfnamefont{T.}~\bibnamefont{van~der Sar}},
  \bibinfo{author}{\bibfnamefont{J.~D.} \bibnamefont{Sanchez-Yamagishi}},
  \bibinfo{author}{\bibfnamefont{K.}~\bibnamefont{Watanabe}},
  \bibinfo{author}{\bibfnamefont{T.}~\bibnamefont{Taniguchi}},
  \bibinfo{author}{\bibfnamefont{P.}~\bibnamefont{Jarillo-Herrero}},
  \bibinfo{author}{\bibfnamefont{B.~I.} \bibnamefont{Halperin}},
  \bibnamefont{and} \bibinfo{author}{\bibfnamefont{A.}~\bibnamefont{Yacoby}},
  \bibinfo{journal}{Science Advances} \textbf{\bibinfo{volume}{3}},
  \bibinfo{pages}{e1700600} (\bibinfo{year}{2017}).

\bibitem[{\citenamefont{Wahl et~al.}(2014)\citenamefont{Wahl, Rech, Jonckheere,
  and Martin}}]{Wahl:2013-1}
\bibinfo{author}{\bibfnamefont{C.}~\bibnamefont{Wahl}},
  \bibinfo{author}{\bibfnamefont{J.}~\bibnamefont{Rech}},
  \bibinfo{author}{\bibfnamefont{T.}~\bibnamefont{Jonckheere}},
  \bibnamefont{and} \bibinfo{author}{\bibfnamefont{T.}~\bibnamefont{Martin}},
  \bibinfo{journal}{Phys. Rev. Lett.} \textbf{\bibinfo{volume}{112}},
  \bibinfo{pages}{046802} (\bibinfo{year}{2014}).

\bibitem[{\citenamefont{Berg et~al.}(2009)\citenamefont{Berg, Oreg, Kim, and
  von Oppen}}]{Berg:2009-1}
\bibinfo{author}{\bibfnamefont{E.}~\bibnamefont{Berg}},
  \bibinfo{author}{\bibfnamefont{Y.}~\bibnamefont{Oreg}},
  \bibinfo{author}{\bibfnamefont{E.-A.} \bibnamefont{Kim}}, \bibnamefont{and}
  \bibinfo{author}{\bibfnamefont{F.}~\bibnamefont{von Oppen}},
  \bibinfo{journal}{Phys. Rev. Lett.} \textbf{\bibinfo{volume}{102}},
  \bibinfo{pages}{236402} (\bibinfo{year}{2009}).

\bibitem[{\citenamefont{Kamata et~al.}(2014)\citenamefont{Kamata, Kumada,
  Hashisaka, Muraki, and Fujisawa}}]{Kamata:2014-1}
\bibinfo{author}{\bibfnamefont{H.}~\bibnamefont{Kamata}},
  \bibinfo{author}{\bibfnamefont{N.}~\bibnamefont{Kumada}},
  \bibinfo{author}{\bibfnamefont{M.}~\bibnamefont{Hashisaka}},
  \bibinfo{author}{\bibfnamefont{K.}~\bibnamefont{Muraki}}, \bibnamefont{and}
  \bibinfo{author}{\bibfnamefont{T.}~\bibnamefont{Fujisawa}},
  \bibinfo{journal}{Nature Nanotechnology} \textbf{\bibinfo{volume}{9}},
  \bibinfo{pages}{177} (\bibinfo{year}{2014}).

\bibitem[{\citenamefont{Talyanskii et~al.}(1997)\citenamefont{Talyanskii,
  Shilton, Pepper, Smith, Ford, Linfield, Ritchie, and
  Jones}}]{Talyanskii:1997-1}
\bibinfo{author}{\bibfnamefont{V.~I.} \bibnamefont{Talyanskii}},
  \bibinfo{author}{\bibfnamefont{J.~M.} \bibnamefont{Shilton}},
  \bibinfo{author}{\bibfnamefont{M.}~\bibnamefont{Pepper}},
  \bibinfo{author}{\bibfnamefont{C.~G.} \bibnamefont{Smith}},
  \bibinfo{author}{\bibfnamefont{C.~J.~B.} \bibnamefont{Ford}},
  \bibinfo{author}{\bibfnamefont{E.~H.} \bibnamefont{Linfield}},
  \bibinfo{author}{\bibfnamefont{D.~A.} \bibnamefont{Ritchie}},
  \bibnamefont{and} \bibinfo{author}{\bibfnamefont{G.~A.~C.}
  \bibnamefont{Jones}}, \bibinfo{journal}{Phys. Rev. B}
  \textbf{\bibinfo{volume}{56}}, \bibinfo{pages}{15180} (\bibinfo{year}{1997}).

\bibitem[{\citenamefont{Kamata et~al.}(2010)\citenamefont{Kamata, Ota, Muraki,
  and Fujisawa}}]{Kamata:2010-1}
\bibinfo{author}{\bibfnamefont{H.}~\bibnamefont{Kamata}},
  \bibinfo{author}{\bibfnamefont{T.}~\bibnamefont{Ota}},
  \bibinfo{author}{\bibfnamefont{K.}~\bibnamefont{Muraki}}, \bibnamefont{and}
  \bibinfo{author}{\bibfnamefont{T.}~\bibnamefont{Fujisawa}},
  \bibinfo{journal}{Phys. Rev. B} \textbf{\bibinfo{volume}{81}},
  \bibinfo{pages}{085329} (\bibinfo{year}{2010}).

\bibitem[{\citenamefont{Petkovi\ifmmode~\acute{c}\else \'{c}\fi{}
  et~al.}(2013)\citenamefont{Petkovi\ifmmode~\acute{c}\else \'{c}\fi{},
  Williams, Bennaceur, Portier, Roche, and Glattli}}]{Petkovic:2013-1}
\bibinfo{author}{\bibfnamefont{I.}~\bibnamefont{Petkovi\ifmmode~\acute{c}\else
  \'{c}\fi{}}}, \bibinfo{author}{\bibfnamefont{F.~I.~B.}
  \bibnamefont{Williams}},
  \bibinfo{author}{\bibfnamefont{K.}~\bibnamefont{Bennaceur}},
  \bibinfo{author}{\bibfnamefont{F.}~\bibnamefont{Portier}},
  \bibinfo{author}{\bibfnamefont{P.}~\bibnamefont{Roche}}, \bibnamefont{and}
  \bibinfo{author}{\bibfnamefont{D.~C.} \bibnamefont{Glattli}},
  \bibinfo{journal}{Phys. Rev. Lett.} \textbf{\bibinfo{volume}{110}},
  \bibinfo{pages}{016801} (\bibinfo{year}{2013}).

\bibitem[{\citenamefont{Forster}(1924)}]{Foster:1924-1}
\bibinfo{author}{\bibfnamefont{R.}~\bibnamefont{Forster}},
  \bibinfo{journal}{Bell System Tech. J.} \textbf{\bibinfo{volume}{3}},
  \bibinfo{pages}{259} (\bibinfo{year}{1924}).

\bibitem[{\citenamefont{Neuenhahn and Marquardt}(2008)}]{Neuenhahn:2008-1}
\bibinfo{author}{\bibfnamefont{C.}~\bibnamefont{Neuenhahn}} \bibnamefont{and}
  \bibinfo{author}{\bibfnamefont{F.}~\bibnamefont{Marquardt}},
  \bibinfo{journal}{New Journal of Physics} \textbf{\bibinfo{volume}{10}},
  \bibinfo{pages}{115018} (\bibinfo{year}{2008}).

\bibitem[{\citenamefont{Neuenhahn and Marquardt}(2009)}]{Neuenhahn:2009-1}
\bibinfo{author}{\bibfnamefont{C.}~\bibnamefont{Neuenhahn}} \bibnamefont{and}
  \bibinfo{author}{\bibfnamefont{F.}~\bibnamefont{Marquardt}},
  \bibinfo{journal}{Phys. Rev. Lett.} \textbf{\bibinfo{volume}{102}},
  \bibinfo{pages}{046806} (\bibinfo{year}{2009}).

\end{thebibliography}

\end{document}